\newcommand{\tr}{{\rm tr} \,}
\newcommand{\arctanh}{{\rm arctanh} \,}
\newcommand{\arcsinh}{{\rm arcsinh} \,}
\newcommand{\Li}{{\rm Li_2 }\,}

\newcommand{\pslash}{\FMslash p}
\newcommand{\lslash}{\FMslash l}

\newcommand{\Qslash}{\FMslash Q}
\newcommand{\Pslash}{\FMslash P}

\newcommand{\wslash}{\FMslash w}

\newcommand{\Kslash}{\FMslash K}
\newcommand{\partialslash}{\FMslash \partial}

\documentclass{elsart}
\usepackage{feynmp,epsfig}
\begin{fmffile}{teq}
\begin{document}
GSI-Preprint-99-18, June 1999 (revised version)
\begin{frontmatter}
\title{Effective chiral theory of nucleon-nucleon scattering}
\author[GSI]{Matthias Lutz}
\address[GSI]{GSI, 64220 Darmstadt, Germany}
\centerline{E-mail: m.lutz@gsi.de}

\begin{abstract}
We present a new chiral expansion scheme for the nucleon-nucleon scattering amplitude 
which preserves unitarity exactly. Our effective field theory builds on the power counting 
rules for 2-nucleon reducible diagrams proposed in \cite{lutz}. We evaluate the leading order 
terms of the isospin one scattering amplitude and elaborate in detail on the $^1S_0$ phase 
shift. Our chiral description of the $^1S_0$-phase shift does compete in quality with modern 
phenomenological nucleon-nucleon potentials. We describe elastic and inelastic scattering
quantitatively up to laboratory energies of $E_{\rm lab} \simeq 600$ MeV.
\end{abstract}
\end{frontmatter}


\section{Introduction}

The nucleon-nucleon scattering problem has a long history in nuclear physics \cite{Watson}.
Nowadays there exist phenomenological nucleon-nucleon potentials 
\cite{Machleidt,Nim,Paris,Bochum} which successfully describe scattering data to high precision. These 
potentials are commonly constructed applying meson-exchange phenomenology. 
A serious drawback of the traditional approach is, however, the lack of a systematic scheme  
in the sense that there does not exist a well defined procedure to improve the potential. 
One has not yet identified a small expansion parameter which guides the construction of 
the potential nor has one resolved the renormalization problem. Results depend explicitly 
on ad-hoc formfactors. This will be an interesting issue when the data base on nucleon-nucleon 
scattering is further improved in the near future, in particular with upcoming experiments 
at the proton cooler facility COSY in J\"ulich.

A first attempt to systematically solve the problem of nucleon-nucleon scattering 
and construct a bridge to QCD was made by Weinberg \cite{Weinberg1}. He suggested 
to derive a nucleon-nucleon potential in time-ordered chiral perturbation theory 
\cite{Weinberg1}. The heavy meson-exchange contributions are, in accordance with 
chiral symmetry, systematically absorbed in local two-nucleon interaction vertices.
This program was first carried out by Ordonez, Ray and  van~Kolck 
\cite{bira1,bira2} who computed nucleon phase shifts from the solution of a cutoff 
regularized Schr\"odinger equation feeded properly with the chiral potential. 
More recent attempts within the potential approach can be found in \cite{Park,Epelbaoum,review}. 
The chiral potential scheme is, however, still plagued with serious shortcomings. The 
systematic renormalization of the chiral potential scheme remains an open problem. This 
leads again to a regularization scheme dependence of physical results. Though in certain 
cases the regularization scheme dependence is found to be weak the reason for which  
awaits a systematic explanation. 

In \cite{lutz} the author suggested to apply a relativistic version of chiral
perturbation theory ($\chi $PT) with power counting rules also for two nucleon 
reducible diagrams. The proposed scheme is practicable since the 'difficult' non-local 
pion dynamics is found to be perturbative. Non perturbative effects are generated 
by properly renormalized local two-nucleon vertices. As a result the s-wave scattering 
amplitude is the sum of a pole term ( representing the deuteron bound state in the spin 
triplet channel or the pseudo-bound state in the spin singlet channel ) and a smooth 
remainder which is evaluated perturbatively in the pion dynamics. 
The question whether pions are perturbative in the nucleon scattering problem has been 
addressed also in \cite{Gegelia,Steele}. Our scheme in \cite{lutz} shows similarities with 
a framework proposed recently by Kaplan, Savage and Wise \cite{KSW} and questioned by 
Cohen and Hansen \cite{Cohen-Hansen}. Starting with the static one-pion exchange potential 
the authors \cite{KSW} arrive at the power counting rules for reducible diagrams suggested in 
\cite{lutz}. However, as will be shown in detail the implementation of the generalized 
chiral counting rules of \cite{lutz} in \cite{KSW} is in contradiction to the low energy 
structure of the nucleon-nucleon scattering amplitude. In particular unitarity is strongly 
violated. The s-wave scattering amplitude acquires a more complicated structure.  
Its representation in terms of a pole term and a smooth remainder accessed in terms
of perturbative pions fails at unexpectedly small momenta. We will demonstrate that the 
failure of this scheme \cite{lutz} in the spin singlet channel is caused by a second 
non-perturbative phenomenon present in the subthreshold scattering amplitude. The purpose 
of this work is to develop a consistent effective field theoretic application of the chiral 
power counting rules proposed in \cite{lutz} in accordance with the low energy structure 
of the scattering amplitude. 

We choose the relativistic version of the chiral Lagrangian since it offers on the one hand 
a manifest covariant framework and on the other hand it is the natural scheme to describe 
particle production processes like $pp\rightarrow pp\,\pi^0, pn\,\pi^+$. A further important 
novel ingredient of our scheme is 
a reorganization of the local two-body interaction vertices. Rather than expanding the 
interaction terms around zero three-momenta we propose to expand around a finite value. 
The physical motivation for this reorganization follows from the empirical observation that 
for example the $^1S_0$ partial wave  shows a zero at intermediate energies 
$E_{lab.}\simeq 280$ MeV. A novel systematic approximation scheme for the solution of the  
Bethe-Salpeter equation consistent with the chiral counting rules and unitarity is presented. 

Our work is subdivided into four parts. In section two we discuss the chiral counting 
rules as they emerge from the relativistic chiral Lagrangian and introduce our 
expansion scheme for the nucleon-nucleon scattering amplitude. In section three we 
present analytic results of explicit evaluations at subleading orders. 
By evaluating the old scheme of \cite{lutz} at three-loop level it is demonstrated 
that indeed it fails at momenta $p > m_\pi $. Many technical 
details of this calculation can be found in a series of appendices. 
In section four we extend our scheme to describe the pion production process and 
confront our results with the empirical scattering phase shift. 
We obtain a quantitative description of elastic and inelastic s-wave nucleon-nucleon 
scattering in the spin singlet channel up to $E_{lab}\simeq $ 600 MeV.

\section{Discussion of relativistic chiral power counting rules}

The chiral Lagrangian is constructed conveniently in terms of 
Weinberg's stereographic coordinates \cite{Weinberg2}. The buildings blocks are 
the pionic four vector $D_\mu $ with
\begin{eqnarray}
D_{\mu }^c &=& \frac{\partial_{\mu} \pi^c}{1+\lambda^2 \pi^2 }
\end{eqnarray}
and the relativistic nucleon field $N$. The coupling $\lambda $ is identified with the 
chiral limit value of the pion decay constant
$\lambda^{-1} = 2\, f_{\pi } \simeq 186 \,{\rm MeV} $. The chiral Lagrangian is now the 
infinite sum of all isoscalar terms formed with any number of the pion field $D_\mu $,  
nucleon fields $N$ and their respective covariant derivatives with:
\begin{eqnarray}
{\mathcal D}_{\mu } N &&=
\left( \partial_{\mu }+i\, \frac{\lambda^2 }{1+\lambda^2 \pi^2 }
\vec \tau \cdot \left( \vec \pi \times \partial_{\mu } \vec \pi \right)
\right) N \; ,
\nonumber\\
{\mathcal D}_{\mu } D_\nu^{a} &&=
\partial_{\mu } D_\nu^{a}
+2\, \lambda \, \pi^a\, \left( \vec D_\nu \cdot \vec D_\mu \right)
-2\, \lambda \, \left( \vec \pi \cdot \vec D_\nu \right) D_\mu^a  \; .
\end{eqnarray}
Further terms are to be added in order to describe the explicit chiral symmetry breaking pattern
due to small current quark masses. Such terms do not add further complications to our 
discussion and therefore are not considered explicitly here. 
Weinberg's chiral power counting rules \cite{Weinberg1,Weinberg2} induce a systematic truncation 
of this infinite hierarchy. At a given order only a finite number of Feynman diagrams derived 
from a finite number of interaction terms need to be considered. Since we will generalize 
Weinberg's counting rules we briefly recall his scheme. 

For a given irreducible Feynman diagram one introduces the number of internal nucleon lines, 
$I_N$, the number of internal pion lines, $I_\pi $, the number of loops, $L$
and the number, $V_i$, of vertices of type $i$ with $d_i$
'small' derivatives involved. Here one calls a derivative small if it acts on the pion field. 
The chiral counting rule for a given Feynman diagram results if the nucleon propagator 
is given the chiral power $-1$, the pion propagator $-2$, a small derivative the power $1$ 
and the loop momentum the power $4$. The usefulness of the counting rules 
(\ref{counting-rules-1}) follows by means of the topological relations
$L=I_\pi +I_N-\sum_i V_i +1 $ and $2\, I_N +E_N =\sum V_i \, n_i $
where the number of external nucleon lines $E_N$ and $n_i$,  
the number of nucleon fields in an interaction term of type $i$ is introduced. 
The well know result 
\begin{eqnarray}
\nu &=& 4\, L -I_N -2\, I_\pi + \sum_i \, V_i \,d_i 
\nonumber\\
&=& 2-{\textstyle {1\over2}}\, E_N + 2\, L
+\sum_i V_i \left( d_i +{\textstyle {1\over2}}\, n_i-2 \right) 
\label{counting-rules-1}
\end{eqnarray}
shows that for $d_i +{\textstyle {1\over2}}\, n_i-2\geq 0 $, higher loop diagrams 
are more and more suppressed since they carry larger and larger chiral powers. 
We emphasize that the counting rule (\ref{counting-rules-1}) is
not applicable for multi-nucleon reducible diagrams which one encounters for example 
in the iteration of the Bethe-Salpeter equation for the nucleon-nucleon scattering amplitude. 

There are two problems inherent with the relativistic approach. First, any 
covariant derivative acting on the nucleon field produces the large nucleon mass 
and therefore must be assigned the minimal chiral power zero. Thus an infinite
tower of interaction terms needs to be evaluated at given finite
chiral order \cite{Gasser,Krause}. 
Second, the straightforward evaluation of Feynman diagrams involving relativistic 
nucleon propagators generates positive powers of the large nucleon mass \cite{Gasser}. 
Therefore the chiral power counting rule (\ref{counting-rules-1}) is spoiled. The 
relativistic scheme appeared impractical and the heavy mass formulation of $\chi $PT, 
which overcomes both problems by performing a non relativistic $1/m$ expansion at the 
level of the Lagrangian density, was developed \cite{Manohar} and applied extensively in 
the one nucleon sector \cite{Bernard}. In the two-nucleon sector, however, the heavy baryon 
mass formulation of $\chi $PT is not applicable due to the ill-defined  two-nucleon reducible 
diagrams \cite{Weinberg1}. Weinberg circumvented those problems by deriving chiral power 
counting rules for time-ordered perturbation theory \cite{Weinberg1}.

In this work we follow a different path and work directly with the relativistic form of the 
chiral Lagrangian \cite{lutz}. This has the advantage that it is possible 
to control covariance at each step of the calculation. We outline how the two aforementioned 
pertinent problems can be overcome within the framework of the relativistic chiral Lagrangian. 
We expect our relativistic scheme to be equivalent to the
more familiar heavy fermion formulation of chiral perturbation
theory in the one-nucleon sector. Rather than performing the $1/m$ expansion at the level 
of the Lagrangian density we suggest to work out this expansion explicitly
at the level of individual relativistic Feynman diagrams. Therewith we avoid the heavy baryon
chiral Lagrangian with its known artifacts in the two-nucleon sector.

It is convenient to consider the chiral Lagrangian (at an intermediate stage) to 
represent a finite cutoff theory \cite{Weinberg1}. The finite cutoff, $\Lambda \ll m$, is 
required to restrict the nucleon virtuality inside a given loop diagram such as to justify 
the $1/m$ expansion. Obviously a finite cutoff has to be introduced with great care not 
to break any chiral Ward identity. Alternatively one may  apply dimensional 
regularization \cite{Tang,Becher}. We emphasize that if dimensional regularization is applied 
one has to {\it first} expand in $1/m$ and then perform the loop integration. The two steps 
do {\it not} commute here. This is equivalent to first performing the $1/m$ expansion 
and then applying the large cutoff limit in a renormalized finite cutoff theory. 
This procedure  suppresses contributions from loop momenta larger than the nucleon mass
which are the source for both the conflict of relativistic diagrams with the counting rule 
(\ref{counting-rules-1}) and the appearance of multiple powers of the nucleon 
mass \cite{Gasser}. 

It remains to demonstrate that the relativistic chiral Lagrangian leads to
a practical scheme in the sense that at a given chiral order $Q$ there are only a
finite number of diagrams to be considered. Here it is convenient to 
generalize the notation of (\ref{counting-rules-1}) and denote with
$d_i$ the chiral order of an interaction vertex of type $i$. We will show
that by an appropriate regrouping of interaction terms in the chiral Lagrangian 
there are only a finite number of renormalized interaction 
vertices of given chiral order.  
Consider a covariant derivative ${\mathcal D}_\mu (x) $ acting on a
nucleon field $N(x)$. We need to construct a suitable 'counter interaction'
leading to a 'small' renormalized vertex. This crucial rearrangement is
illustrated at hand of the simplest case when the Lorentz index $\mu$
is saturated by another covariant derivative acting on the same nucleon field. 
In this case the counter interaction is readily found
\begin{eqnarray}
{\mathcal D}^\mu (x)\, {\mathcal D}_\mu(x) \, N(x) &&\rightarrow
\left({\mathcal D}^\mu (x)\, {\mathcal D}_\mu (x) +m^2 \right)\, N(x)
\sim Q \;.
\label{eq8}
\end{eqnarray}
The 'rearranged' vertex is proportional to the virtuality of
the nucleon, $(\partial_\mu \, \partial^\mu +m^2)\,N$, which is of 'minimal' chiral 
order $Q$. It is important to observe that this counter interaction is part of the chiral 
Lagrangian by construction.  
In the more general case where there are $2\,k$ derivatives acting on a nucleon field 
$({\mathcal D}^\mu (x)\, {\mathcal D}_\mu(x))^k \,N(x)$ an appropriate 
rearrangement leads to a renormalized structure 
$({\mathcal D}^\mu (x)\, {\mathcal D}_\mu(x)+m^2)^k \,N(x)$. 
The total suppression factor  of the interaction vertex $d_i=d_i^{(\pi)}+d_i^{(N)}$ is 
the sum of the number of derivatives acting on pion fields $d_i^{(\pi)}$, and the 
degree of virtuality of the nucleon field $d_i^{(N)}=k$. Note that a similar construction 
is applicable if covariant derivatives acting on different nucleon fields meet.

We are left to consider the case in which the Lorentz index $\mu $
of the covariant derivative ${\mathcal D}_\mu $ is contracted
either with the index of a Dirac matrix or the 
index of a derivative acting on a pion field. In the latter case 
vertices are well behaved since they are already suppressed by the pion 
derivative in $\partial_\mu  \pi $. Of
course, here, one has to give the nucleon derivative the chiral
power $Q^{0}$. The 'large' nucleon mass is not 'hazardous'
if one identifies the natural scale of any coupling constant $\bar
\Lambda $ to be of the same order as the nucleon mass. With $\bar
\Lambda \simeq m_\rho $ this seems indeed reasonable.
The remaining case in which the nucleon derivative couples to the 
index of a Dirac matrix in a vertex does not generate an infinite tower of
interaction terms since at a given number of nucleon fields $n_i$,
the number of available indices provided by Dirac matrices
is at most $n_i$. The degree of nucleon virtuality can be determined upon
inspection case by case.

The above arguments show how to organize the infinite tower of
chiral interaction terms such that there is always only a finite
number of interaction terms of a given chiral power. The counting 
rule (\ref{counting-rules-1}) applies where $d_i$ is identified with 
the sum of number of derivatives acting on pion fields and the degree of 
nucleon virtuality of the vertex as discussed above. Thus at
any given chiral order only a finite number of diagrams need to be
evaluated. We note that in the relativistic scheme it is convenient 
to use the notion of 'minimal' chiral power of a Feynman diagram since 
a given diagram leads to an infinite tower of terms with increasing chiral
powers upon expansion. The minimal chiral power is given by the counting rule 
(\ref{counting-rules-1}).

\subsection{2-Particle reducible diagrams}

In this section we discuss the 2-particle reducible diagrams for which 'standard' 
chiral power counting rules are not applicable \cite{lutz,Weinberg1}. 
In our relativistic framework one is therefore bound to consider
the Bethe-Salpeter equation for the nucleon-nucleon scattering. 
This is analogous to Weinberg's approach \cite{Weinberg1} which considers the 
Lippmann-Schwinger equation with a suitable potential expanded according to chiral 
power counting rules. 

The nucleon-nucleon scattering amplitude $T$, is given in terms of the 
Bethe-Salpeter kernel $K$
\begin{eqnarray}
\parbox{10mm}{
\begin{fmfgraph*}(10,10)
\fmfleftn{l}{2}
\fmfrightn{r}{2}
\fmfforce{(0.8w,0.2h)}{T1}
\fmfforce{(0.8w,0.8h)}{T2}
\fmfforce{(0.2w,0.8h)}{T3}
\fmfforce{(0.2w,0.2h)}{T4}
\fmfforce{(1.0w,0.0h)}{r1}
\fmfforce{(1.0w,1.0h)}{r2}
\fmfforce{(0.0w,1.0h)}{l2}
\fmfforce{(0.0w,0.0h)}{l1}
\fmfpolyn{empty,label=${T}$,smooth,pull=1.4}{T}{4}
\fmf{fermion}{r1,T1}
\fmf{fermion}{r2,T2}
\fmf{fermion}{T3,l2}
\fmf{fermion}{T4,l1}
\end{fmfgraph*}}
\,\,\,=\,\,\,
2\,\,\,
\parbox{10mm}{
\begin{fmfgraph*}(10,10)
\fmfleftn{l}{2}
\fmfrightn{r}{2}
\fmfforce{(0.8w,0.2h)}{T1}
\fmfforce{(0.8w,0.8h)}{T2}
\fmfforce{(0.2w,0.8h)}{T3}
\fmfforce{(0.2w,0.2h)}{T4}
\fmfforce{(1.0w,0.0h)}{r1}
\fmfforce{(1.0w,1.0h)}{r2}
\fmfforce{(0.0w,1.0h)}{l2}
\fmfforce{(0.0w,0.0h)}{l1}
\fmfpolyn{empty,label=${K}$,smooth,pull=1.4}{T}{4}
\fmf{fermion}{r1,T1}
\fmf{fermion}{r2,T2}
\fmf{fermion}{T3,l2}
\fmf{fermion}{T4,l1}
\end{fmfgraph*}}
\,\,\,+\,\,\,
\parbox{30mm}{
\begin{fmfgraph*}(30,10)
\fmfleftn{l}{2}
\fmfrightn{r}{2}
\fmfforce{(0.0w,1.0h)}{l2}
\fmfforce{(0.0w,0.0h)}{l1}
\fmfforce{(0.3w,0.2h)}{T1}
\fmfforce{(0.3w,0.8h)}{T2}
\fmfforce{(0.1w,0.8h)}{T3}
\fmfforce{(0.1w,0.2h)}{T4}
\fmfpolyn{empty,label=${K}$,smooth,pull=1.4}{T}{4}
\fmfforce{(0.9w,0.2h)}{V1}
\fmfforce{(0.9w,0.8h)}{V2}
\fmfforce{(0.7w,0.8h)}{V3}
\fmfforce{(0.7w,0.2h)}{V4}
\fmfpolyn{empty,label=$T$,smooth,pull=1.4}{V}{4}
\fmfforce{(1.0w,0.0h)}{r1}
\fmfforce{(1.0w,1.0h)}{r2}
\fmf{fermion}{T3,l2}
\fmf{fermion}{T4,l1}
\fmf{fermion,right=.15}{V3,T2}
\fmf{fermion,left=.15}{V4,T1}
\fmf{fermion}{r1,V1}
\fmf{fermion}{r2,V2}
\end{fmfgraph*}}
\;\;=\;\;\frac{2}{1-K\, G}\, K
\label{bs1}
\end{eqnarray}
where for convenience we use operator notation in the last expression. 
The scattering amplitude $T$ and interaction kernel $K$ of (\ref{bs1}) are antisymmetric 
under exchange of the ingoing or outging nucleons. The factor '2' in (\ref{bs1}) leads to  
the proper treatment of exchange diagrams. The interaction kernel
\begin{eqnarray}
\parbox{10mm}{
\begin{fmfgraph*}(10,10)
\fmfleftn{l}{2}
\fmfrightn{r}{2}
\fmfforce{(0.8w,0.2h)}{T1}
\fmfforce{(0.8w,0.8h)}{T2}
\fmfforce{(0.2w,0.8h)}{T3}
\fmfforce{(0.2w,0.2h)}{T4}
\fmfforce{(1.0w,0.0h)}{r1}
\fmfforce{(1.0w,1.0h)}{r2}
\fmfforce{(0.0w,1.0h)}{l2}
\fmfforce{(0.0w,0.0h)}{l1}
\fmfpolyn{empty,label=${K}$,smooth,pull=1.4}{T}{4}
\fmf{fermion}{r1,T1}
\fmf{fermion}{r2,T2}
\fmf{fermion}{T3,l2}
\fmf{fermion}{T4,l1}
\end{fmfgraph*}}
\,\,\,=\,\,\,
\parbox{10mm}{
\begin{fmfgraph*}(10,10)
\fmfleftn{l}{2}
\fmfrightn{r}{2}
\fmfforce{(1.0w,0.0h)}{r1}
\fmfforce{(1.0w,1.0h)}{r2}
\fmfforce{(0.0w,1.0h)}{l2}
\fmfforce{(0.0w,0.0h)}{l1}
\fmfforce{(0.5w,0.5h)}{o}
\fmfv{d.sh=square,d.f=full,d.si=3thick}{o}
\fmf{fermion}{r1,o}
\fmf{fermion}{r2,o}
\fmf{fermion}{o,l2}
\fmf{fermion}{o,l1}
\end{fmfgraph*}}
\,\,\,+\,\,\,
\parbox{10mm}{
\begin{fmfgraph*}(10,10)
\fmfleftn{l}{2}
\fmfrightn{r}{2}
\fmfforce{(1.0w,0.0h)}{r1}
\fmfforce{(1.0w,1.0h)}{r2}
\fmfforce{(0.0w,1.0h)}{l2}
\fmfforce{(0.0w,0.0h)}{l1}
\fmfforce{(0.5w,0.2h)}{od}
\fmfforce{(0.5w,0.8h)}{ou}
\fmfv{d.sh=circle,d.f=full,d.si=2thick}{od,ou}
\fmf{fermion}{r1,od}
\fmf{fermion}{r2,ou}
\fmf{fermion}{ou,l2}
\fmf{fermion}{od,l1}
\fmf{boson}{ou,od}
\end{fmfgraph*}}
\,\,\,+\,\,\dots \; .
\label{kernel1}
\end{eqnarray}
is the sum of two-particle irreducible diagrams deduced from the relativistic chiral 
Lagrangian.
To leading zeroth order it is the sum of local interaction terms and the one-pion exchange
diagram displayed in (\ref{kernel1}) with a wavy line representing the relativistic
pion propagator. The object $G$ in (\ref{bs1}) represents the relativistic two-nucleon 
propagator. Let us investigate the Bethe-Salpeter equation (\ref{bs1}) perturbatively. 
In the following we introduce the appropriate minimal chiral power $\nu$ for $K\,G\,K$ 
with respect to a selected part of the interaction kernel $K$ carrying the minimal chiral 
power $\nu_K$. The once iterated Bethe-Salpeter kernel is
\begin{eqnarray}
K\, G\, K
\sim
\int
\frac{d^4l }{(2\pi )^4} \frac{2\,m\,K(l)}
{\Big({\textstyle {1\over 2}}\,w-l\Big)^2-m^2+i\, \epsilon }
\frac{2\,m\,K(l)}{\Big({\textstyle {1\over 2}}\,w+l\Big)^2-m^2 +i\, \epsilon }
\label{bs2}
\end{eqnarray}
with $w^2 = s$. In the center of mass frame with $\vec w=0 $ the Mandelstam variable 
$s=4\,m^2+4\,p^2$ is given by the relative momentum $p$ of the nucleons.  It is instructive 
to study first the s-channel spectral density, $\rho(s)$, of this contribution. It typically 
receives strength from the pinch singularity generated by the two-nucleon propagators 
in (\ref{bs2}). In the center of mass frame one finds:
\begin{eqnarray}
\rho(s) = \Im \,\Big( K\, G\, K \Big) \sim  \frac{p}{\sqrt{m^2+p^2}}\,\frac{m^2}{4 \, \pi }
\,\int \,
\frac{d \Omega_{\vec l} }{4\, \pi }\, K(\vec l\, )\, K(\vec l\,)\,
\Big|_{l_0 =0,\, |\vec l | = p } \; .
\label{bs4}
\end{eqnarray}
Expression (\ref{bs4}) can now be used to introduce the minimal chiral power $\nu$
of the s-channel spectral density
\begin{eqnarray}
\nu = 1+ 2\,\nu_K
\label{bs5}
\end{eqnarray}
in terms of the minimal chiral power $\nu_K$ of the Bethe-Salpeter kernel
$K$ given by (\ref{counting-rules-1}). In (\ref{bs5}) we exploit 
that the elastic nucleon-nucleon phase space leads to the small momentum $p$ 
in the spectral density. Note that the angle average in (\ref{bs4}) does not affect the chiral 
power. If one assigns $K\sim 1/(4 f^2_\pi)$, as suggested by 
the one-pion exchange contribution to the kernel, and identifies $m\simeq 4\pi f_\pi $ 
the dimensionless expansion parameter $p/(4 f_\pi) $ arises. This indicates convergence 
for small momenta $p < 350$ MeV. Note, however, that if the kernel $K(p)$ in (\ref{bs4}) 
vanishes for large momenta $p$ we expect convergence possibly also for $p>350$ MeV. This
is typically the case if the spectral function $\rho(s)$ leads to a finite loop
$\Re \,K\,G\,K$ (see (\ref{bs8},\ref{ren-condition-1})). We then expect the dimensionless 
parameter $m_\pi/(4\,f_\pi) \simeq 0.4$ since the falloff in $K(l)$ must be driven by the
pion mass. It is instructive to compare the chiral power (\ref{bs5}) with the 'naive' chiral 
power, $2+2\,\nu_K$, implied by the standard counting rule (\ref{counting-rules-1}).
This confirms that the counting rule (\ref{counting-rules-1}) must not be applied to 
all parts of two-nucleon reducible diagrams. We point out that inelastic 
contributions to $\rho(s)$, for example from the one-pion production cut, can be treated 
in full analogy to the two-nucleon cut. The chiral power of the contribution to the spectral 
density follows upon inspection of the production phase space. In fact here the counting rule 
(\ref{counting-rules-1}) leads to the correct minimal chiral power (see appendix 
B.4 and B.5 for explicit examples). 

To have a useful scheme we also need to derive counting rules for
the real part of the scattering amplitude. Here we invoke causality which relates 
the real part of a given Feynman diagram to its imaginary part 
by means of a dispersion relation\footnote{The iterated Bethe-Salpeter kernel 
does not necessarily satisfy a s-channel dispersion relation. 
However, we point out that only the part which provides strength for the s-channel 
spectral density requires special attention and modified power counting rules. Obviously
a part which does not generate strength for the s-channel spectral
density has {\it no} pinch singularity leading to a s-channel cut so that
standard counting rules apply. Note that for a given contribution one has to specify whether 
the dispersion relation holds at fixed Mandelstam variable $t$
or $u$. This technical detail, however, does not affect our argument.}
\begin{eqnarray}
I(s) = \Re \Big( K \, G\, K \Big) = \frac{1}{\pi}\,{\mathcal P } 
\int_{4\, m^2}^{\infty }d\,\bar s\,
\frac{\rho(\bar s)}{\bar s -s } \; .
\label{bs7}
\end{eqnarray}
The dispersion relation (\ref{bs7}) need not to be finite and well defined. It may be 
ultraviolet divergent. However, one is free to consider the chiral Lagrangian as
a finite cutoff theory \cite{Weinberg1,LePage}. If the cutoff $\Lambda $ is
to be interpreted as a bound for the maximal virtuality of nucleons
one should first remove the nucleon rest mass.  Therefore we take the cutoff $\Lambda $ 
to restrict the momentum $p\leq \Lambda $ rather than the Mandelstam variable $s$ in the 
dispersion relation (\ref{bs7}):
\begin{eqnarray}
I(p, \Lambda ) = 
\frac{1}{\pi}\,{\mathcal P } 
\int_{4\, m^2}^{4\,(m^2+\Lambda^2)}d\,\bar s\,
\frac{\rho(\bar s)}{\bar s -s }=
\frac{1}{\pi}\,{\mathcal P } \int_{0}^{\Lambda^2  }d\,\bar p^2\,
\frac{\rho(4\,m^2+4\,\bar p^2)}{\bar p^2 -p^2 } \; .
\label{bs8}
\end{eqnarray}
Formally one counts $m_\pi/\Lambda  \sim \Lambda /m \sim Q$ with 
$m_\pi \ll \Lambda < m$ \cite{Weinberg1}.
The natural guess for the chiral power of the integral $I(p,\Lambda)$ would be
the chiral power of its imaginary part. However, the integral
(\ref{bs8}) as it stands does not lead to this chiral power yet. Though the separation
of scales is manifest in the spectral density $\rho(s) $ the cutoff scale $\Lambda $ 
prevents the conclusion that the integral behaves like a small scale to the power 
$1+2\,\nu_K $. We point out that after an appropriate  number of subtractions at 
$p^2=-\mu^2 < 0 $ with $\mu \sim m_\pi $ the counting rule (\ref{bs5}) can nevertheless be 
applied to the real part of the loop function. The subtraction is required to render 
the dispersion integral finite in the large cutoff limit: 
\begin{eqnarray}
I^{(n)}_S(p,\Lambda ;\mu ) =\frac{1}{\pi}\,{\mathcal P } \int_{0}^{\Lambda^2  }d\,\bar p^2\,
\frac{\rho(4\,m^2+4\,\bar p^2)}{\bar p^2 -p^2}
\left(\frac{p^2+\mu^2}{\bar p^2 +\mu^2}\right)^n \;.
\label{bs9}
\end{eqnarray}
The subtraction polynomial can always be absorbed into the contact 4-nucleon vertices.
Note that the performed subtractions do not necessarily
simplify our attempt to derive a counting rule since this procedure
introduces a new scale, $\mu $, the subtraction point. Nevertheless
we now can exploit the freedom to choose the subtraction point. If we
insist on a 'small' subtraction scale $\mu \sim m_\pi $ of the
order of the pion mass one can apply dimensional power counting
since the integral is finite in the large cutoff limit by construction.

The above arguments establish the chiral power of a once iterated Bethe-Salpeter kernel.
The general case of a n-times iterated kernel follows by complete induction provided that 
a suitable subtraction mechanism (see section 2.4) is implemented 
into the Bethe-Salpeter equation. We arrive at the desired chiral counting rule for 2-particle 
reducible diagrams: each pair of nucleon propagators which exhibit an s-channel
unitarity  cut receives the chiral power $-3$ in contrast to the 'naive'
power $-2$. Therefore the n-times iterated Bethe-Salpeter kernel receives the 
minimal chiral power
\begin{eqnarray}
\nu_K^{(n)} = n+n\,\nu_K \;.
\label{l-counting-rule}
\end{eqnarray}
We stress, first, that this counting rule is manifest {\it only if} one 
introduces a 'small' subtraction scale $\mu \sim m_\pi $ at the level of the Bethe-Salpeter
equation and second, that the subtraction scale is {\it independent}
of the intrinsic cutoff. We will refer to this counting rule
as the 'L'-counting rule since for the one-pion exchange interaction, which contributes 
at order zero to the interaction kernel, the chiral power of the n-th iterated interaction 
is simply given by the number of loops. In appendix B the reader may find explicit 
examples which confirm our L-counting rule (\ref{l-counting-rule}).  

We summarize what we derived. The {\it non-polynomial part} of the once iterated 
Bethe-Salpeter kernel has an excess of one chiral power as compared to the sum of
the chiral powers of the iterated Bethe-Salpeter kernels (see (\ref{bs5})). We point
out that this result, though obtained by means of dispersion
relations, holds also if a different technique for the evaluation
of the loop functions is applied. Since the chiral
power of the Bethe-Salpeter kernel starts at $\nu = 0$ our counting rules seem to 
suggest that the Bethe-Salpeter
kernel has to be iterated only a finite number of times at a given
chiral order $\nu $ for the scattering amplitude. Obviously this
can {\it not} be correct since there are well known
non-perturbative effects present in the scattering amplitude like
for example the deuteron bound state. This puzzle can be resolved
by realizing that the implicit assumption of any chiral power
counting scheme is the naturalness of all coupling constants
carrying non trivial dimensions. The characteristic scale is
assumed to be the mass of the lightest degree of freedom not
explicitly included in the effective chiral Lagrangian.

Therefore the crucial question to ask
is: {\it does the suggested subtraction scheme for the Bethe-Salpeter
equation affect the naturalness of the 4-nucleon coupling strengths?}

\subsection{Mutation of natural scale to anomalous scale}

In this section we address the question to what extent a subtraction scheme
affects the naturalness of the bare coupling function. The effect 
of our subtraction scheme is illustrated at hand of s-wave scattering in 
the isospin one channel. We consider here a theory without explicit pions defined 
by an infinite tower of separable interactions in the $^1S_0$-channel\footnote{Note 
that the terms displayed in (\ref{mut0}) are the only ones relevant for s-wave 
scattering in the spin singlet channel. Any further terms necessarily are on-shell 
equivalent to terms already included and lead to relativistic nucleonic tadpole diagrams upon 
unitary iteration in the Bethe-Salpeter equation. Such terms can be dropped in a 
consistent treatment (see appendix B.1 and (\ref{1/m-ren})). Note that the situation is 
different in a non-relativistic framework where one encounters a richer variety of interaction 
terms \cite{Kolck-1}.}:
\begin{eqnarray}
{\mathcal L}_{4N}(x) &=& \frac{1}{4}\sum_{n=0}^\infty g^{(2\,n)}\,
\partial^n \left(\bar N\,\gamma_5 \,C\,\, \tau \tau_2\,\bar N^t\right)
\partial^n \left(N^t\,\tau_2 \, \tau \, C^{-1}\gamma_5  \,N\right) 
\label{mut0}
\end{eqnarray}
with the charge conjugation matrix $C= i\,\gamma_0\,\gamma_2 $ and the isospin 
Pauli matrices $\tau_i$. The index contractions of the derivatives $\partial_\mu $ in 
(\ref{mut0}) are such that in momentum space the energy dependent 
coupling $g(s)$ arises:
\begin{eqnarray}
g(s) &=& \sum_{n=0}^\infty \,g^{(2\,n)}\,s^n 
=\sum_{n=0}^\infty \,g_{2\,n}\,\Big(s-4\,m^2\Big)^n 
\label{large-m-example}
\end{eqnarray}
with $p^2=s/4-m^2$. In (\ref{large-m-example}) we introduce renormalized couplings 
$g_{2\,n}$ according to the mandatory regrouping of interaction terms discussed in 
section 2. Note that in the derivation of Feynman rules from (\ref{mut0}) the 
troublesome time-derivatives can be treated naively since the form of the final 
effective Feynman vertex is dictated by covariance. At tree level the coupling $g_0$ 
can be interpreted in terms of the s-wave scattering lengths $a(^1S_0) $ with
$2\,m\,g_0 = -4\,\pi\, a(^1S_0)$. 
Recall that one may apply a phenomenological saturation model for the coupling function 
with $g(s)$ receiving contributions from heavy meson ($\sigma, \rho, \omega, ...$ ) 
exchanges. Such a model provides a bridge to the traditional meson-exchange picture 
of nucleon-nucleon scattering. It furthermore motivates the crucial assumption of 
any chiral expansion, namely, that all couplings are natural: in particular we expect 
$g_{n+2} \sim g_{n} /\Lambda_{nat.}^2 $. We identify here the natural scale 
$\Lambda_{nat.} $ with one half of the mass of the lightest meson not included explicitly 
in the theory. The factor $1/2$ follows since the t-channel meson-exchange diagram in a 
given partial wave suggests a convergence radius of half the meson mass for an expansion 
around $p^2=0$. In a theory with pions one identifies $\Lambda_{nat.}=m_\rho/2 $ and in a 
theory where pions are integrated out $\Lambda_{nat.}=m_\pi/2$. 
It is worth emphasizing that the leading term requires special attention since 
$g_0 \neq 1/\Lambda^2_{nat.}$. For example the vector meson $u$ and $t$ 
channel exchange diagrams predict 
$g_0\sim -g^2_{\rho NN}/(8\,m_\rho^2)-g^2_{\omega NN}/(8\,m_\omega^2)
\sim -1/(8\,f_\pi^2)$ with a typical meson-nucleon coupling constant 
$g_{\omega NN}\sim g_{\rho NN} \sim  m_\rho/(\sqrt{2}\,f_\pi )$ estimated with the 
KSFR relation. Therefore one expects $g_0 = \lambda^2\,\hat g_0 $ with 
the dimensionless parameter $\hat g_0$ of order unity.

We proceed and solve the Bethe-Salpeter equation for the nucleon-nucleon scattering 
amplitude $T$, with the tree-level  Bethe-Salpeter kernel of (\ref{mut0}). The equation 
is separable and therefore the solution can be written in terms of the divergent one-loop 
integral, $J(s,\Lambda )$, 
\begin{eqnarray}
T(p)&=& \frac{g(s)}{1- g(s)\,J (s) }\,
\left( \tau \,\tau_2
\,\gamma_5\,C^{-1} \right) \times
\Big( C\,\gamma_5 \,\tau_2\,\tau \Big)\; ,
\nonumber\\
J(s,\Lambda )&=&-i\,\tr \int \frac {d^4 l}{(2\pi)^4} \,C^{-1}\gamma_5
\,S_F\left(l+{\textstyle {1\over2}} w \right)\gamma_5 \,C
\,S^t_F\left(-l+{\textstyle {1\over2}} w \right) \; .
\label{mut2}
\end{eqnarray}
Here $S_F(p)=1/(\pslash-m +i\,\epsilon)$ is the relativistic nucleon propagator and $w^2 =4(m^2+p^2)$.
The loop integral may be evaluated by writing a dispersion
integral\footnote{We are aware that the introduction of a finite cutoff in
Feynman  diagrams requires special care in order not to break the
pertinent symmetries. For more details we refer to the appendix.}
\begin{eqnarray}
J(s,\Lambda ) =
\frac{1}{\pi} \int_{0}^{\Lambda^2 }
\frac {d\,\bar p^2\,\rho_J (\bar p )}{\bar p^2-p^2-i\,\epsilon}
\label{mut5}
\end{eqnarray}
where we restrict the relative momentum of the center of mass
system by the finite cutoff parameter $\Lambda $. The well defined
spectral density is:
\begin{eqnarray}
\rho_J (p )= \frac{m\,p }{2\pi}\,\sqrt{1+\frac{p^2}{m^2}}=
\frac{m\,p }{2\pi}\left(1 +\frac{p^2}{2\,m^2}+{\mathcal O}\left(\frac{p^4}{m^4}\right)\right)
\label{mut6} \; .
\end{eqnarray}
We expand the loop integral $J(s,\Lambda ) $ non relativistically in 
powers of $1/m$ by expanding its imaginary part (\ref{mut6}). Here we encounter 
a simple example how to systematically expand a Feynman diagram 
given by the relativistic chiral Lagrangian. Consider for example the loop function 
at zero momentum 
\begin{eqnarray}
J(4\,m^2,\Lambda) &=& \frac{1}{2\,\pi^2}\left(\Lambda\,\sqrt{m^2+\Lambda^2}
+m^2\,\arcsinh \left(\frac{\Lambda}{m}\right) \right) \;.
\label{mut-help}
\end{eqnarray}
A naive evaluation of the diagram with $\Lambda \gg m$ 
would generate a troublesome multiple power of the nucleon mass with 
$J \sim m^2\,\log \Lambda$ counteracting any chiral counting rule.
Equations (\ref{mut5}) and (\ref{mut-help}) exemplify the necessity of the intrinsic cutoff $\Lambda $ 
to be smaller than the nucleon mass such that the $1/m$ expansion is justified. 
Expanding (\ref{mut-help}) in $\Lambda/m$ reduces the leading power of $m$. This is also 
seen if the real part of the loop $J(s,\Lambda)$ is evaluated with the expanded 
spectral function (\ref{mut6}) by means of the dispersion integral (\ref{mut5}) :
\begin{eqnarray}
\Re \,J(s,\Lambda ) &=& 
\frac{m}{\pi^2}\, \Lambda \left( 1+\frac{\Lambda^2}{6\,m^2} +{\mathcal 
O}\left(\frac{\Lambda^4}{m^4}\right) \right) 
\nonumber\\
&+&\frac{m}{\pi^2} 
\frac{p^2}{\Lambda }\left(1 
+\frac{\Lambda^2}{2\,m^2}-\frac{\Lambda^4}{24\,m^4} +{\mathcal 
O}\left(\frac{\Lambda^6}{m^6}\right) 
\right) +{\mathcal O}\left(p^4 \right) \; .
\label{lin-div}
\end{eqnarray}
It is worth pointing out that the subleading terms in the $1/m$ expansion 
proportional to $(\Lambda/m)^n$ with $n>1$ can be evaluated 
with $\Lambda= m$. This follows since the $\Lambda/m$ expansion is controlled 
typically by square root singularities $\sqrt{m^2+p^2} $ for which even at 
$\Lambda/m=1$ the non-relativistic expansion (\ref{mut6}) is converging rapidly. 
This is nicely confirmed by the suppression factors $1/6$ and $1/12$ for the 
constant term and $p^2$-term in (\ref{lin-div}) respectively. 
Therefore, effectively we are left with only linear divergencies in $\Lambda $. In fact 
it is legitimate to drop all terms $(\Lambda/m)^n$ with $n>1$ since at given order 
$p^{2\,k}$ their effect can be compensated by a small change of $\Lambda $. This 
is particularly convenient since then nucleonic tadpole diagrams can be dropped 
systematically. With this simplified regularization scheme we avoid the 
problems one encounters in cutoff-regularized effective field theory. The ill-defined 
multiple powers of the cutoff are eliminated and need not to be canceled by the careful 
construction of the cutoff dependence of path integral measure \cite{cutoff-theory}.
For more details on our regularization procedure including or more rigorous definition
we refer the reader to appendix B. Here we point at a further important observation to 
be made. The presence of terms like $(p/\Lambda )^{2 k}$ necessarily leads to the 
identification $ \Lambda \simeq  \Lambda_{nat.} $. Only then such terms can be absorbed 
into contact interaction terms of the chiral Lagrangian. 

To leading order in the non-relativistic expansion the one-loop 
integral (\ref{mut5}) can be rendered finite by one subtraction at $s=4\,m^2-\mu^2$ 
\begin{eqnarray}
J_S(s,\Lambda ;\mu) = J(s,\Lambda )-J(4\,m^2-\mu^2, \Lambda  )
=\frac{1}{\pi} \int_{0}^{\Lambda^2 }
\frac {d\,\bar p^2\,\rho_J (\bar p)}{\bar p^2-p^2-i\,\epsilon}\,
\frac{ p^2+\mu^2 }
     {\bar p^2+\mu^2}
\label{mut7}
\end{eqnarray}
where the subtraction constant $J(4\,m^2-\mu^2, \Lambda  )$ can be absorbed into
the coupling function $g(s)=g(s,\Lambda)$. Note that subleading terms in the $1/m$
expansion require further subtractions. The subtracted loop (\ref{mut7}) is 
not identified with the renormalized loop. It is more convenient to deal with a manifest 
cutoff independent renormalized loop function. We introduce
$J_R(p,\mu ) = J_S(s , \infty ;\mu )$ and accordingly the renormalized 
coupling $g_R(p^2,\mu )$ with
\begin{equation}
T_{[^1S_0]}(p)=\frac{1}{g^{-1}(s, \Lambda)-J(s,\Lambda )} =
\frac{1}{ g_R^{-1}(p^2,\mu ) - J_R(p,\mu ) }\; .
\label{mut9}
\end{equation}
All dependence on the way in which the cutoff $\Lambda $ was 
introduced is absorbed into the bare coupling function $g(s,\Lambda)$. 
Consequently the choice of the cutoff scheme {\it does not} have any  
effect as long as the scheme does not break the pertinent symmetries. The leading 
order renormalized loop reads
\begin{eqnarray}
J_R(p,\mu ) =
\frac{m}{2\pi^2} \int_{0}^{\infty }
\frac {d\,\bar p^2\,\bar p }{\bar p^2-p^2-i\,\epsilon}\,
\frac{ p^2+\mu^2 }
     {\bar p^2+\mu^2}
=\frac{m}{2 \pi }\big( \mu   + i\,p\big)
\label{mut10}
\end{eqnarray}
where we dropped all $1/m$ correction terms in (\ref{mut10}). The leading order 
renormalized coupling function is:
\begin{eqnarray}
g_R^{-1}(p^2,\mu ) &=& g^{-1}(s, \Lambda) -
\frac{m}{2\pi^2} \int_{0}^{\Lambda^2 }
\frac {d\,\bar p^2\,\bar p }{\bar p^2+\mu^2}\,
+\frac{m}{2\pi^2} \int_{\Lambda^2 }^{\infty }
\frac {d\,\bar p^2\,\bar p }{\bar p^2-p^2}\,
\frac{ p^2+\mu^2 }
     {\bar p^2+\mu^2} 
\nonumber\\
&=& g^{-1}(s, \Lambda)-\frac{m}{\pi^2}
\left( \Lambda -\frac{\pi}{2}\, \mu  \right)
+\frac{m}{\pi^2} \frac{p^2}{\Lambda }\left(1 +{\mathcal
O}(p^2/\Lambda^2) \right) \;.
\label{mut12}
\end{eqnarray}
It is important to observe that for $p\leq \Lambda $ the contribution form the 
loop function to the renormalized coupling function can be expanded in powers 
of the small momentum $p$. 
Our result (\ref{mut12}) demonstrates that all cutoff dependence can safely be absorbed in the
bare coupling $g(s,\Lambda)$ provided that it is legitimate to identify the typical cutoff 
$\Lambda $ with the natural scale of the bare coupling function $\Lambda_{nat.}$. 
Equation (\ref{mut12}) exhibits an interesting effect. Starting with a
natural bare coupling function $g(s,\Lambda )$ it potentially generates
an anomalously large renormalized coupling 
$g_R(p^2=0, \mu )$, however, only if the subtraction scale
$\mu $ is taken to be sufficiently small as required by the L-counting rule. Then an 
attractive natural coupling $g^{-1}(4\,m^2,\Lambda )>0$ potentially cancels
the $m\, \Lambda $ term in (\ref{mut12}) predicting an anomalously large renormalized coupling 
$g_R(0, \mu)\sim 2\pi /(m\,\mu )$. We stress that this is a familiar phenomenon 
since it is long understood that an attractive system may dynamically
generate 'new' scales which are anomalously small (like for example the
binding energy of shallow bound state). What is intriguing here is the transparent
and simple mechanism provided by (\ref{mut12}) to naturally generate such a 'small'
scale  \cite{lutz,Beane}. If the cancellation in (\ref{mut12}) is manifest 
for a physical system the renormalized coupling $g_R(0,\mu )$ must be assigned an 
anomalous chiral power, e.g. the  'minimal' chiral power $-1$ at given subtraction 
scale with $\mu \sim m_\pi $. 

In (\ref{mut12}) we refrained from expanding the inverse bare coupling function 
$g^{-1}(s, \Lambda)$. 
We point out that it is necessary to discriminate two scenarios. In scenario I) 
the convergence radius of $g^{-1}(s, \Lambda )$ is set by the appropriate natural scale 
$\Lambda^{(I)}_{nat.}$ leading to a well behaved Taylor series in $p^2$. In scenario II) the 
convergence radius of $g^{-1}(s, \Lambda )$ is set by a zero of the bare coupling function at
$p^2=\mu_\sigma^2$ with $\mu_\sigma < \Lambda^{(I)}_{nat.}$. In this case it may be 
advantageous to expand the renormalized coupling around $p=\mu_\sigma $. If one insists 
on expanding around $p=0$ the theory breaks down at scale $p\leq \mu_\sigma < \Lambda^{(I)}_{nat.}$. 
In fact expanding around $p=\mu_\sigma $ further 
increases the convergence radius with $\Lambda^{(II)}_{nat.}=\Lambda^{(I)}_{nat.}+\mu_\sigma $ 
simply because the distance of the expansion point $p=\mu_\sigma $ to the closest subthreshold 
singularity at $p=\pm \,i\,\Lambda^{(I)}_{nat.}$ is increased. Note that here we implicitly
assume that $\mu_\sigma+\Lambda^{(II)}_{nat.}$ is smaller than the one-particle production cut 
with $2\,\mu^2_\sigma+\mu_\sigma\,\Lambda^{(I)}_{nat.} < m\,\Lambda^{(I)}_{nat.} $.
For example, in an effective theory which treats pions explicitly and where the 
$\rho$-meson is the lightest relevant degree-of-freedom that is integrated out, the optimal 
expansion point $p \simeq -m_\rho/4+(m\,m_\rho/4 +m_\rho^2/16)^{1/2}\simeq 275$ MeV leads 
to the maximal convergence radius or equivalently a natural scale of 
$\Lambda_{nat.}\simeq 660$ MeV.

We investigate the scattering amplitude $T_{[^1S_0]}(p)$ of (\ref{mut9}) in 
the two scenarios to leading order in the $1/m$ expansion
\begin{eqnarray}
T_{[^1S_0]}^{(I)}(p)&=& \left( g_{R,I}^{(0,-1)}(\mu)
+g_{R,I}^{(2,-1)}\,p^2+\cdots-\frac{m}{2\,\pi}\,\Big( \mu+i\,p\Big)\right)^{-1}\; ,
\nonumber\\
T_{[^1S_0]}^{(II)}(p)&=& \Bigg(\frac{g_{R,II}^{(-2,-1)}}{p^2-\mu_\sigma^2}
+g_{R,II}^{(0,-1)}(\mu )+g_{R,II}^{(2,-1)}\,\big( p^2-\mu_\sigma^2\big)
\nonumber\\
&&\;\;\;\;\;\;\;\;\;\;
+\cdots-\frac{m}{2\,\pi}\,\Big(\mu+i\,p \Big) \Bigg)^{-1}\;,
\label{two-scenarios}
\end{eqnarray}
where we implicitly assume that 
the bare coupling shows a simple zero with $g(\mu_\sigma^2)=0$ but $g'(\mu_\sigma^2)\neq 0$. 
In principal  a zero with higher degeneracy can not be excluded though it is less 
likely for a natural coupling function. Multiple zeros of 
$g(p^2,\Lambda )$ are difficult to reconcile with the naturalness assumption.
We emphasize that an effective field theory formulated with an explicit cutoff
may give rise to a similar behavior since here the inverse bare coupling is kept and not 
expanded. The zero at $p^2=\mu_\sigma^2$, however, is then necessarily a consequence of  
strongly correlated and fine-tuned parameters resulting from a fit to data. 
In our scheme (\ref{two-scenarios}) we avoid a strong correlation of parameters since 
the vanishing of the phase shift is already build in and controlled directly by $\mu_\sigma$.
There is a further important difference to our scheme: at order $Q^2$ the inverse 
scattering amplitude of scenario I) requires two parameters whereas scenario II) leads to 
four parameters. In other words scenario II) predicts the relevance of more operators 
at a given order.

It is instructive to explicitly work out the implications of the naturalness assumption
and discuss the low energy structure of $T_{[^1S_0]}(p)$ at 
momenta $p\ll \mu_\sigma $ in both scenarios. It leads to 
coupling functions of the form:
\begin{eqnarray}
g_{R,I}^{(0,-1)}(\mu) &=& -\frac{m}{\pi^2}\,\left( \Lambda^{(I)}_{nat.}-\frac{\pi}{2}\,\mu \right)+
\hat g_{R,I}^{(0,-1)}\,4\,f_\pi^2\; ,
\nonumber\\
g_{R,II}^{(-2,-1)} &=& \hat g_{R,II}^{(-2,-1)}\,4\,f_\pi^2\,
\left(\Lambda^{(II)}_{nat.}\right)^2 \; ,
\nonumber\\
g_{R,II}^{(0,-1)}(\mu) &=& -\frac{m}{\pi^2}\,
\left( \Lambda^{(II)}_{nat.}-\frac{\pi}{2}\,\mu \right)
+\hat g_{R,II}^{(0,-1)}\,4\,f_\pi^2
\label{nat-sat}
\end{eqnarray}
where the dimensionless quantities $\hat g_{R,II}^{(-2,-1)}$, $ \hat g_{R,II}^{(0,-1)}$ and 
$ \hat g_{R,I}^{(0,-1)}$ are of order unity. We point out that 
$g_{R,II}^{(0,-1)} \sim 4\,f_\pi^2 $ and 
$  g_{R,I}^{(0,-1)}(\mu )\sim 4\,f_\pi^2$ holds only for 
$\mu \simeq \Lambda^{(I)}_{nat.}$ and $\mu \simeq \Lambda^{(II)}_{nat.}$ respectively. 
The subleading coefficients $g_{R,I}^{(n,-1)}$ and $g_{R,II}^{(n,-1)}$ 
do not depend on the subtraction scale $\mu $ for $n>2$ to leading order in the $1/m$ 
expansion. Therefore, at $\mu=0$ the coefficient 
$-g_{R,II}^{(-2,-1)}/\mu_\sigma^2+g_{R,II}^{(0,-1)}(0)$ and also $g_{R,I}^{(0,-1)}(0)$ 
may be anomalously small leading to an unnatural large scattering length 
$2\pi\,a =-m\,T_{[^1S_0]}(0)$. The required cancellation is achieved with an attractive 
coupling $\hat g_{R,I}^{(0,-1)}(0)\simeq \Lambda^{(I)}_{nat.}/(\pi f_\pi) $ in scenario I) and with 
$\hat g_{R,II}^{(-2,-1)} \simeq - \mu_\sigma^2/(\pi f_\pi \,\Lambda^{(II)}_{nat.})$ in scenario II) 
where we identified $4\pi f_\pi \simeq m $ for simplicity.

An anomalously large scattering length is evidence for the presence of non-perturbative 
structures at small momenta. If the precise singular structure is known one may proceed 
and break up and further expand the denominator of (\ref{two-scenarios}) without loosing 
accuracy in the expansion. Note, however, that such an expansion 
will necessarily be in conflict with unitarity. In scenario I) a systematic expansion of 
the denominator is induced by 
identifying $g^{(0,-1)}_{R,I}(m_\pi) \sim Q$ and $g^{(2,-1)}_{R,I}(m_\pi)\,p^2 \sim Q^2$ with 
the typical small scale $Q$. This implies that the scattering amplitude $T_{[^1S_0]}(p)$ shows
either a shallow bound state or a pseudo-bound state \cite{lutz,KSW}. 
The perturbative inclusion of pion dynamics according to the L-counting rule is 
straight forward. If the scattering amplitude $T_{[^1S_0]}(p)$ exhibits a more complicated 
low-energy structure the scenario I) must be rejected as inefficient with an unnaturally small 
convergence radius. In scenario II) the relative importance of the two terms 
$g_{R,II}^{(-2,-1)}$ and $g_{R,II}^{(0,-1)}(\mu)$ depends crucially on the subtraction scale
$\mu$. For small subtraction scale $\mu\sim m_\pi $, required for a manifest realization of 
the $L$-counting rule (\ref{l-counting-rule}), both terms are equally important at $p\simeq 0$.
In fact the low-energy characteristics of scenario II) is much richer than that of scenario I). 
At 'leading' orders the inverse scattering amplitude may have zeros at three different 
complex momenta corresponding to the roots of a cubic equation. It is unclear, in particular 
when the pion dynamics is included, in which way to expand the denominator without loosing 
the homogenous convergence property of (\ref{two-scenarios}). In scenario II) it is therefore 
advantageous to keep unitarity exactly and refrain from any further expansion beyond 
(\ref{two-scenarios}).

\subsection{S-wave scattering in the spin singlet channel}

Before evaluating $T_{[^1S_0]}(p)$ to leading orders in the chiral
expansion we recall  its empirical behavior. A comparison of the structure of the 
empirical scattering amplitude with that of the amplitude in the 
two scenarios (\ref{two-scenarios}) provides strong indications to which scheme to choose.

The empirical scattering amplitude shows two striking phenomena: a pseudo-bound state close to 
threshold and a vanishing phase shift or equivalently a zero  of the scattering amplitude 
$T_{[^1S_0]}=0$ at $p\simeq 355$ MeV. At very low energies with
$E_{\mathrm{lab}}<5 $ MeV the partial wave amplitude $T_{[^1S_0]}(p)$ is dominated 
by a pseudo-bound state pole
\begin{eqnarray}
T_{[^1S_0]}(p) &=& \frac{8\,\pi}{\sqrt{s}}\,\frac{1}{p\,\cot
\,\delta(^1S_0) -i\,p}
\nonumber\\
&=&
\frac{4\,\pi }{m} \left( \frac{ i \, w}{p+i\, z}
-b^{(0)}_{\mathrm{eff}}-i\,b^{(1)}_{\mathrm{eff}}\,p\, 
+{\mathcal O}\left( \frac{p^2}{m_\pi^2}\right) \right)
\label{low-rep}
\end{eqnarray}
where we made the pseudo-bound state pole contribution explicit and
expanded the remainder in powers of $p$. The pole position $z$, its
residuum, $w$, and the leading background parameters $b^{(0,1)}_{\mathrm{eff}} $ are
predicted accurately by the effective range theory
\begin{eqnarray}
p\,\cot \,\delta =-\frac{1}{a}+\frac{1}{2}\,r\,p^2
+{\mathcal O}\left(p^4\right)
\label{eff-range}
\end{eqnarray}
in terms of the s-wave np-scattering length $a
\simeq -23.75 $ fm and effective range parameter $r\simeq 2.75$ fm.
One finds
\begin{eqnarray}
w^{-1} &&=\sqrt{1-2\,\frac{r}{a}} \simeq 0.90^{-1}
\; ,\;\;\;\;\;\;
z = \frac{1-w}{w\,r}\simeq 0.04\,{\mbox fm}^{-1} \; ,
\nonumber\\
b^{(0)}_{\mathrm{eff}} &&=a +\frac{r\, w^2}{1-w}
= -\frac{w}{z}\,\frac{1-w}{1+w} \simeq -1.18\, {\mbox fm} \; ,
\nonumber\\
b^{(1)}_{\mathrm{eff}} &&=\frac{w}{z^2}-a^2
= \frac{w}{z^2}\,\left(\frac{1-w}{1+w}\right)^2 \simeq 1.53\,{\mbox fm}^2 \; 
\label{aeff-def}
\end{eqnarray}
where we included also the prediction of the effective range theory for 
the next-to-leading moment in (\ref{low-rep}) for completeness.
Note that for example the effective scattering length $b^{(0)}_{\mathrm{eff}}$ is substantially 
suppressed as compared with the empirical scattering length $a
\simeq -23.75 $~fm\footnote{
It will turn out to be more convenient to work with the parameters
$w$ and $z$. The scattering length $a$ and effective range are
recovered as follows: $r= \frac{1-w}{w\,z} $ and
$a=-\frac{2\,w}{(w+1)z} $.}. 
We point out that the effective range formula (\ref{eff-range}) implies a second pole 
in the scattering amplitude at $p \simeq i\;0.77 $/fm:
\begin{eqnarray}
&&T_{[^1S_0]}(p) \rightarrow_{\mathrm{eff. range}} \frac{8\,\pi}{\sqrt{s}}\,
 \left( \frac{ i \, w}{p+i\, z} -\frac{ i \,
w}{p-i\, \bar z}
\right)\;,
\nonumber\\
&&\bar z = \frac{1+w}{1-w}\,z= \frac{1}{r}+{\mathcal O}\left(\frac{1}{a}\right)\;.
\label{barz-pole}
\end{eqnarray}
From a theoretical point of view it
is not obvious that the second pole at $p= i\,\bar z$ carries
direct physical significance since it may 'move' as more terms are
included in the expansion of the inverse $K$-matrix. In other words it could be
an effective singularity which accounts for non-polynomial contributions 
from the exchange of pions. Note, however, that from scattering 
theory one expects that the analyticity domain of the s-wave amplitude is bounded due to 
the one-pion exchange contribution at $\Im \,p < m_\pi/2 $ and $m_\pi/2 <\bar z$ numerically.

We now discriminate two alternative scenarios. Suppose first that the strength of the 
pole at $p= i\,\bar z$ effectively accounts for perturbative pionic cuts. Then one may 
expand the amplitude as indicated in (\ref{low-rep}). The chiral theory is expected to 
replace the polynomial remainder in (\ref{low-rep}) by smooth and calculable analytic 
functions which exhibit the proper pionic cuts. A natural starting point for a 
theoretical description is scenario I) of (\ref{two-scenarios}) with the 
denominator in $T^{(I)}_{[^1S_0]}(p)$ expanded according to the L-counting rule. This leads 
to a perturbative treatment of pions. 

Alternatively, suppose that the pole at $p= i\,\bar z$ has a physical significance in 
the sense that the exact solution of the problem shows an anomalously large subthreshold 
amplitude in the vicinity of $p\simeq i\,\bar z$. Then the expansion (\ref{low-rep}) 
turns useless at rather small momenta $p < \bar z $. Therefore an effective field theory 
description must incorporate that singularity properly. A promising starting point is now 
scenario II) of (\ref{two-scenarios}). We expand the inverse of $T^{(II)}_{[^1S_0]}$ and 
identify $\mu_\sigma $ with the empirical zero of the scattering phase. In this 
case we expect a significantly larger natural scale 
$\Lambda^{(II)}_{nat.}>\mu_\sigma \simeq 350$ MeV. In the next section we show in 
detail how the pion dynamics can be incorporated perturbatively according to the 
L-counting rule. 
 
In this paper we will argue that the second scenario is in fact realized in nature.
The success of the effective range theory, which describes the $^1S_0$ 
phase accurately up to momenta $p \simeq 150$ MeV, together with the L-counting rule already 
leads to the second scenario. Since the effective range theory can be viewed as an 
effective low energy theory where the pion degrees of freedom are integrated out, 
the second pole at $ p \simeq i\, 150$ MeV points either at the importance of unitary 
two-pion exchanges with a branch point at $p= i\,m_\pi $ or at the physical significance 
of the second pole. On the other hand the chiral L-counting rule predicts that the 
pionic cut at $p= i\,m_\pi$ is weak. From the fact that the effective range theory works well 
up to surprisingly large momenta $p \simeq \bar z \simeq m_\pi $ and the effective range 
theory prediction of a strong second pole at $p=i\,\bar z$ with residuum $w \simeq 1$ 
(see (\ref{barz-pole})) we then conclude that we have  choose the second scenario. This 
leads to an anomalously large subthreshold amplitude in the vicinity of $p\simeq i\,\bar z$
One arrives at a similar conclusion upon inspection of the typical size of the 
effective range parameter in both scenarios. Scenario II) yields a natural interpretation 
of  the rather large empirical effective range parameter  $1/\bar z \sim r \simeq 2.75$ fm 
also in a model without pions.  We return to this point in section 4 where we show that the 
empirical scattering amplitude can be rather nicely described by $T^{(II)}_{[^1S_0]}(p)$ 
parameterized in terms of the four leading parameters in (\ref{two-scenarios}).

\subsection{Solving the Bethe-Salpeter equation}

In the previous section we derived a chiral power counting rule for
appropriately subtracted 2-particle reducible diagrams generated by
the once iterated Bethe-Salpeter kernel. In order to show that
the counting rule (\ref{l-counting-rule}) holds for more than one iteration of
the Bethe-Salpeter kernel we now elaborate on how the subtraction scheme 
is implemented in the Bethe-Salpeter equation
\begin{eqnarray}
T = 2\,K + K\, G\, T \;.
\label{bs1b}
\end{eqnarray}
The interaction kernel $K$ is the sum of polynomial contributions from 2-nucleon 
contact interaction vertices, $K_{\sigma }$, which represents the exchange of 
heavy meson ($\sigma, \rho, \omega, ...$ ) and the term $K_\pi $, which 
represents 2-particle irreducible pion exchange diagrams: 
\begin{eqnarray}
\parbox{10mm}{
\begin{fmfgraph*}(10,10)
\fmfleftn{l}{2}
\fmfrightn{r}{2}
\fmfforce{(0.8w,0.2h)}{T1}
\fmfforce{(0.8w,0.8h)}{T2}
\fmfforce{(0.2w,0.8h)}{T3}
\fmfforce{(0.2w,0.2h)}{T4}
\fmfforce{(1.0w,0.0h)}{r1}
\fmfforce{(1.0w,1.0h)}{r2}
\fmfforce{(0.0w,1.0h)}{l2}
\fmfforce{(0.0w,0.0h)}{l1}
\fmfpolyn{empty,label=${K}$,smooth,pull=1.4}{T}{4}
\fmf{fermion}{r1,T1}
\fmf{fermion}{r2,T2}
\fmf{fermion}{T3,l2}
\fmf{fermion}{T4,l1}
\end{fmfgraph*}}
\,\,\,=\,\,\,
\parbox{10mm}{
\begin{fmfgraph*}(10,10)
\fmfleftn{l}{2}
\fmfrightn{r}{2}
\fmfforce{(1.0w,0.0h)}{r1}
\fmfforce{(1.0w,1.0h)}{r2}
\fmfforce{(0.0w,1.0h)}{l2}
\fmfforce{(0.0w,0.0h)}{l1}
\fmfforce{(0.5w,0.5h)}{o}
\fmfv{d.sh=square,d.f=full,d.si=3thick}{o}
\fmf{fermion}{r1,o}
\fmf{fermion}{r2,o}
\fmf{fermion}{o,l2}
\fmf{fermion}{o,l1}
\end{fmfgraph*}}
\,\,\,+\,\,\,
\parbox{10mm}{
\begin{fmfgraph*}(10,10)
\fmfleftn{l}{2}
\fmfrightn{r}{2}
\fmfforce{(1.0w,0.0h)}{r1}
\fmfforce{(1.0w,1.0h)}{r2}
\fmfforce{(0.0w,1.0h)}{l2}
\fmfforce{(0.0w,0.0h)}{l1}
\fmfforce{(0.5w,0.2h)}{od}
\fmfforce{(0.5w,0.8h)}{ou}
\fmfv{d.sh=circle,d.f=full,d.si=2thick}{od,ou}
\fmf{fermion}{r1,od}
\fmf{fermion}{r2,ou}
\fmf{fermion}{ou,l2}
\fmf{fermion}{od,l1}
\fmf{boson}{ou,od}
\end{fmfgraph*}}
\,\,\,+\,\,\dots
=K_{\sigma }+K_\pi \; .
\label{k-def}
\end{eqnarray}
Obviously the above separation of the interaction kernel $K$ is not
unique since the irreducible multi-pion loops may also
generate polynomial contact interaction terms.
We will elaborate on this issue in detail below. Let us introduce the amplitude $T_\pi $
\begin{eqnarray}
T_\pi = 2\, \Big( 1-K_\pi \, G \Big)^{-1} K_\pi
\label{}
\end{eqnarray}
which is the solution of the Bethe-Salpeter equation for $K_{\sigma }=0 $ and 
the auxiliary scattering amplitude $T_\sigma $: 
\begin{eqnarray}
\nonumber
\parbox{10mm}{
\begin{fmfgraph*}(10,10)
\fmfleftn{l}{2}
\fmfrightn{r}{2}
\fmfforce{(0.8w,0.2h)}{T1}
\fmfforce{(0.8w,0.8h)}{T2}
\fmfforce{(0.2w,0.8h)}{T3}
\fmfforce{(0.2w,0.2h)}{T4}
\fmfforce{(1.0w,0.0h)}{r1}
\fmfforce{(1.0w,1.0h)}{r2}
\fmfforce{(0.0w,1.0h)}{l2}
\fmfforce{(0.0w,0.0h)}{l1}
\fmfpolyn{empty,label=${ T_\sigma}$,smooth,pull=1.4}{T}{4}
\fmf{fermion}{r1,T1}
\fmf{fermion}{r2,T2}
\fmf{fermion}{T3,l2}
\fmf{fermion}{T4,l1}
\end{fmfgraph*}}
\,\,\,&=&\,\,\,
\frac{
2\,\,\,
\parbox{10mm}{
\begin{fmfgraph*}(10,10)
\fmfleftn{l}{2}
\fmfrightn{r}{2}
\fmfforce{(0.8w,0.2h)}{T1}
\fmfforce{(0.8w,0.8h)}{T2}
\fmfforce{(0.2w,0.8h)}{T3}
\fmfforce{(0.2w,0.2h)}{T4}
\fmfforce{(1.0w,0.0h)}{r1}
\fmfforce{(1.0w,1.0h)}{r2}
\fmfforce{(0.0w,1.0h)}{l2}
\fmfforce{(0.0w,0.0h)}{l1}
\fmfpolyn{empty,label=${K_\sigma }$,smooth,pull=1.4}{T}{4}
\fmf{fermion}{r1,T1}
\fmf{fermion}{r2,T2}
\fmf{fermion}{T3,l2}
\fmf{fermion}{T4,l1}
\end{fmfgraph*}}
} { \displaystyle {\huge 1}\,\,\, - \,\,\,
\parbox{10mm}{
\begin{fmfgraph*}(10,10)
\fmfleftn{l}{2}
\fmfrightn{r}{2}
\fmfforce{(0.8w,0.2h)}{T1}
\fmfforce{(0.8w,0.8h)}{T2}
\fmfforce{(0.2w,0.8h)}{T3}
\fmfforce{(0.2w,0.2h)}{T4}
\fmfforce{(1.0w,0.0h)}{r1}
\fmfforce{(1.0w,1.0h)}{r2}
\fmfforce{(0.0w,1.0h)}{l2}
\fmfforce{(0.0w,0.0h)}{l1}
\fmfpolyn{empty,label=${K_\sigma}$,smooth,pull=1.4}{T}{4}
\fmf{fermion}{r1,T1}
\fmf{fermion}{r2,T2}
\fmf{fermion}{T3,l2}
\fmf{fermion}{T4,l1}
\end{fmfgraph*}}
\,\,\, -\frac{ 1}{2}  \,\,\,
\parbox{30mm}{
\begin{fmfgraph*}(30,10)
\fmfleftn{l}{2}
\fmfrightn{r}{2}
\fmfforce{(0.0w,1.0h)}{l2}
\fmfforce{(0.0w,0.0h)}{l1}
\fmfforce{(0.3w,0.2h)}{T1}
\fmfforce{(0.3w,0.8h)}{T2}
\fmfforce{(0.1w,0.8h)}{T3}
\fmfforce{(0.1w,0.2h)}{T4}
\fmfpolyn{empty,label=${ T_\pi}$,smooth,pull=1.4}{T}{4}
\fmfforce{(0.9w,0.2h)}{V1}
\fmfforce{(0.9w,0.8h)}{V2}
\fmfforce{(0.7w,0.8h)}{V3}
\fmfforce{(0.7w,0.2h)}{V4}
\fmfpolyn{empty,label=${K_\sigma}$,smooth,pull=1.4}{V}{4}
\fmfforce{(1.0w,0.0h)}{r1}
\fmfforce{(1.0w,1.0h)}{r2}
\fmf{fermion}{T3,l2}
\fmf{fermion}{T4,l1}
\fmf{fermion,right=.15}{V3,T2}
\fmf{fermion,left=.15}{V4,T1}
\fmf{fermion}{r1,V1}
\fmf{fermion}{r2,V2}
\end{fmfgraph*}}
}
\nonumber\\
\nonumber\\
&=&2\,\left( 1-K_{\sigma }\,G- {\textstyle{1\over 
2}}\,K_{\sigma }\,G\, T_\pi\, G\right)^{-1}K_{\sigma } \; .
\label{}
\end{eqnarray}
The formal solution of the Bethe-Salpeter equation is easily expressed
in terms of these amplitudes $T_\sigma $ and $T_\pi $
\begin{eqnarray}\nonumber
\parbox{10mm}{
\begin{fmfgraph*}(10,10)
\fmfleftn{l}{2}
\fmfrightn{r}{2}
\fmfforce{(0.8w,0.2h)}{T1}
\fmfforce{(0.8w,0.8h)}{T2}
\fmfforce{(0.2w,0.8h)}{T3}
\fmfforce{(0.2w,0.2h)}{T4}
\fmfforce{(1.0w,0.0h)}{r1}
\fmfforce{(1.0w,1.0h)}{r2}
\fmfforce{(0.0w,1.0h)}{l2}
\fmfforce{(0.0w,0.0h)}{l1}
\fmfpolyn{empty,label=${T}$,smooth,pull=1.4}{T}{4}
\fmf{fermion}{r1,T1}
\fmf{fermion}{r2,T2}
\fmf{fermion}{T3,l2}
\fmf{fermion}{T4,l1}
\end{fmfgraph*}}
\,\,\,&=&\,\,\,
\,\,\,
\parbox{10mm}{
\begin{fmfgraph*}(10,10)
\fmfleftn{l}{2}
\fmfrightn{r}{2}
\fmfforce{(0.8w,0.2h)}{T1}
\fmfforce{(0.8w,0.8h)}{T2}
\fmfforce{(0.2w,0.8h)}{T3}
\fmfforce{(0.2w,0.2h)}{T4}
\fmfforce{(1.0w,0.0h)}{r1}
\fmfforce{(1.0w,1.0h)}{r2}
\fmfforce{(0.0w,1.0h)}{l2}
\fmfforce{(0.0w,0.0h)}{l1}
\fmfpolyn{empty,label=${T_\pi}$,smooth,pull=1.4}{T}{4}
\fmf{fermion}{r1,T1}
\fmf{fermion}{r2,T2}
\fmf{fermion}{T3,l2}
\fmf{fermion}{T4,l1}
\end{fmfgraph*}}
\,\,\,+\,\,\,\frac{ 1}{ 2}\,\,
\parbox{30mm}{
\begin{fmfgraph*}(30,10)
\fmfleftn{l}{2}
\fmfrightn{r}{2}
\fmfforce{(0.0w,1.0h)}{l2}
\fmfforce{(0.0w,0.0h)}{l1}
\fmfforce{(0.3w,0.2h)}{T1}
\fmfforce{(0.3w,0.8h)}{T2}
\fmfforce{(0.1w,0.8h)}{T3}
\fmfforce{(0.1w,0.2h)}{T4}
\fmfpolyn{empty,label=${ T_\pi}$,smooth,pull=1.4}{T}{4}
\fmfforce{(0.9w,0.2h)}{V1}
\fmfforce{(0.9w,0.8h)}{V2}
\fmfforce{(0.7w,0.8h)}{V3}
\fmfforce{(0.7w,0.2h)}{V4}
\fmfpolyn{empty,label=${ T_\sigma}$,smooth,pull=1.4}{V}{4}
\fmfforce{(1.0w,0.0h)}{r1}
\fmfforce{(1.0w,1.0h)}{r2}
\fmf{fermion}{T3,l2}
\fmf{fermion}{T4,l1}
\fmf{fermion,right=.15}{V3,T2}
\fmf{fermion,left=.15}{V4,T1}
\fmf{fermion}{r1,V1}
\fmf{fermion}{r2,V2}
\end{fmfgraph*}}
\,\,\,+\,\,\,\frac{ 1}{ 2}\,\,
\parbox{30mm}{
\begin{fmfgraph*}(30,10)
\fmfleftn{l}{2}
\fmfrightn{r}{2}
\fmfforce{(0.0w,1.0h)}{l2}
\fmfforce{(0.0w,0.0h)}{l1}
\fmfforce{(0.3w,0.2h)}{T1}
\fmfforce{(0.3w,0.8h)}{T2}
\fmfforce{(0.1w,0.8h)}{T3}
\fmfforce{(0.1w,0.2h)}{T4}
\fmfpolyn{empty,label=${ T_\sigma}$,smooth,pull=1.4}{T}{4}
\fmfforce{(0.9w,0.2h)}{V1}
\fmfforce{(0.9w,0.8h)}{V2}
\fmfforce{(0.7w,0.8h)}{V3}
\fmfforce{(0.7w,0.2h)}{V4}
\fmfpolyn{empty,label=${ T_\pi}$,smooth,pull=1.4}{V}{4}
\fmfforce{(1.0w,0.0h)}{r1}
\fmfforce{(1.0w,1.0h)}{r2}
\fmf{fermion}{T3,l2}
\fmf{fermion}{T4,l1}
\fmf{fermion,right=.15}{V3,T2}
\fmf{fermion,left=.15}{V4,T1}
\fmf{fermion}{r1,V1}
\fmf{fermion}{r2,V2}
\end{fmfgraph*}}
\\ \nonumber
\\ \nonumber
\,\,\,&+&\,\,\,\,\,\,
\parbox{10mm}{
\begin{fmfgraph*}(10,10)
\fmfleftn{l}{2}
\fmfrightn{r}{2}
\fmfforce{(0.8w,0.2h)}{T1}
\fmfforce{(0.8w,0.8h)}{T2}
\fmfforce{(0.2w,0.8h)}{T3}
\fmfforce{(0.2w,0.2h)}{T4}
\fmfforce{(1.0w,0.0h)}{r1}
\fmfforce{(1.0w,1.0h)}{r2}
\fmfforce{(0.0w,1.0h)}{l2}
\fmfforce{(0.0w,0.0h)}{l1}
\fmfpolyn{empty,label=${ T_\sigma}$,smooth,pull=1.4}{T}{4}
\fmf{fermion}{r1,T1}
\fmf{fermion}{r2,T2}
\fmf{fermion}{T3,l2}
\fmf{fermion}{T4,l1}
\end{fmfgraph*}}
\,\,\,+\,\,\,\frac{ 1}{4}\,\,
\parbox{45mm}{
\begin{fmfgraph*}(45,10)
\fmfleftn{l}{2}
\fmfrightn{r}{2}
\fmfforce{(0.0w,1.0h)}{l2}
\fmfforce{(0.0w,0.0h)}{l1}
\fmfforce{(0.23w,0.2h)}{T1}
\fmfforce{(0.23w,0.8h)}{T2}
\fmfforce{(0.1w,0.8h)}{T3}
\fmfforce{(0.1w,0.2h)}{T4}
\fmfpolyn{empty,label=${ T_\pi}$,smooth,pull=1.4}{T}{4}
\fmfforce{(0.56w,0.2h)}{M1}
\fmfforce{(0.56w,0.8h)}{M2}
\fmfforce{(0.43w,0.8h)}{M3}
\fmfforce{(0.43w,0.2h)}{M4}
\fmfpolyn{empty,label=${T_\sigma}$,smooth,pull=1.4}{M}{4}
\fmfforce{(0.9w,0.2h)}{V1}
\fmfforce{(0.9w,0.8h)}{V2}
\fmfforce{(0.76w,0.8h)}{V3}
\fmfforce{(0.76w,0.2h)}{V4}
\fmfpolyn{empty,label=${ T_\pi}$,smooth,pull=1.4}{V}{4}
\fmfforce{(1.0w,0.0h)}{r1}
\fmfforce{(1.0w,1.0h)}{r2}
\fmf{fermion}{T3,l2}
\fmf{fermion}{T4,l1}
\fmf{fermion,right=.15}{M3,T2}
\fmf{fermion,left=.15}{M4,T1}
\fmf{fermion,right=.15}{V3,M2}
\fmf{fermion,left=.15}{V4,M1}
\fmf{fermion}{r1,V1}
\fmf{fermion}{r2,V2}
\end{fmfgraph*}}
\\ \nonumber
\\
&=&T_\pi
+\Big( 1+{\textstyle{1\over 2}}\,T_\pi\,G \Big)\,T_\sigma \,
\Big( 1+{\textstyle{1\over 2}}\,G\, T_\pi \Big) \; .
\label{split-0}
\end{eqnarray}

Next we introduce a subtraction scheme for the Bethe-Salpeter 
equation. The subtracted kernels $K_{\pi ,S} $ and $K_{\sigma ,S}$ 
are defined in terms of an arbitrary and not yet specified 
subtraction interaction $K_S$ 
\begin{eqnarray}
K_{\pi ,S} &=& K_\pi + K_S \; ,
\nonumber\\
K_{\sigma ,S } &=& K_{\sigma } -K_S \; .
\label{split-1}
\end{eqnarray}
According to our construction the full interaction kernel $K $ and
consequently also the full scattering amplitude are manifestly independent
of the subtraction interaction $K_S$
\begin{eqnarray}
K= K_\pi+K_{\sigma }= K_{\pi, S}+K_{\sigma ,S} \; .
\label{split-2}
\end{eqnarray}
The subtracted interaction kernels induce the subtracted scattering
amplitudes  $T_{\pi , S}$ and $T_{\sigma ,S}$ with:
\begin{eqnarray}
T_{\pi ,S}[K_S] &=& 2\, \Big( 1-K_{\pi ,S} \, G \Big)^{-1} K_{\pi , S} \; ,
\nonumber\\
T_{\sigma,S}[K_S] &=& 2\,\Big(1-K_{\sigma ,S }\, G- {\textstyle{1\over 
2}}\,K_{\sigma ,S}\,G\,T_{\pi,S}[K_S]\,G \Big)^{-1} K_{\sigma ,S }  \; .
\label{split-3}
\end{eqnarray}
The full scattering amplitude remains formally invariant
\begin{eqnarray}
T=T_{\pi ,S}[K_S] +
\Big(1+{\textstyle{1\over 2}}\,T_{\pi , S}[K_S]\, G\Big) \, T_{\sigma ,S}[K_S]\,
\Big(1+{\textstyle{1\over 2}}\,G\, T_{\pi ,S}[K_S] \Big) \; .
\label{split-4}
\end{eqnarray}

One is now free to exploit the choice of the subtraction interaction  
$K_S$. Different choices for $K_S$ may increase or decrease the 
average interaction strength in $K_{\pi , S}$. Therefore one would 
expect that the non-perturbative part of the scattering amplitude 
like for example the deuteron bound state may sit either in 
$T_{\sigma ,S} $ or $T_{\pi ,S}$ depending on the choice of $K_S$. 
Suppose now that non-perturbative effects sit in $T_{\sigma ,S} $ then one may 
conclude that the amplitude $T_{\pi ,S}$ is perturbative and can be 
expanded in powers of $K_{\pi ,S }$. 
We point out that in this case also the amplitude $T_{\sigma ,S} $ need not to be
evaluated exactly. The perturbative character of the subtracted pion
iteration permits the expansion of the inverse amplitude $T_{\sigma ,S}  $ in
powers of the kernel $K_{\pi, S}$.
\begin{eqnarray}
T_{\sigma ,S}  &=& 2\,\Big(1-K_{\sigma ,S}\, G-K_{\sigma ,S} \,G\,K_{\pi 
,S}\, G + ... \Big)^{-1} K_{\sigma ,S } 
\nonumber\\
&=&2\,\Big(K_{\sigma ,S}^{-1}- G-\,G\,K_{\pi ,S}\, G + ... 
\Big)^{-1}  \; .
\label{tsigma}
\end{eqnarray}
Here we refrain form breaking up and further expanding the denominator in (\ref{tsigma})
since that requires further assumptions on the structures of non-perturbative 
effects in $T_{\sigma ,S}$. 

It is useful to outline our renormalization procedure. At 
given chiral order $n$, construct the local kernel $K_S=\bar K_S$ perturbatively such that
\begin{eqnarray}
T_{\pi ,S}[\bar K_S] =  \mbox{finite} + {\mathcal O} \left( Q^{n+1}\right) 
\;\;\;\&\;\;\;
T_{\pi ,S}[\bar K_S]\,G =  \mbox{finite} + {\mathcal O} \left( Q^{n+2}\right) 
\label{ren-condition-1}
\end{eqnarray}
holds. Note that $T_{\pi ,S}\,G$ can be defined via $T_{\pi ,S}\,G\,K_{\sigma ,S}$ if
$K_{\sigma ,S}$ is separable. We emphasize that the construction (\ref{ren-condition-1}) 
differs from conventional renormalization procedures insofar that the object $T_{\pi ,S}\,G$ 
is rendered finite by 'counter loops' rather then counter terms. In particular 
$T_{\pi ,S}$ and $T_{\pi ,S}\,G$ do not depend on any renormalization scale. As an 
immediate consequence note that for $T_{\pi ,S}$ and also the vertex loop functions 
$T_{\pi ,S}\,G$ the L-counting rule apply without a further subtraction 
procedure (see (\ref{bs9})). The only remaining divergencies sit 
in $T_{\sigma ,S}$ and can be absorbed into the local bare coupling $K_\sigma$:
\begin{eqnarray}
K_{\sigma ,S}^{-1}- G - G\,T_{\pi ,S}[\bar K_S]\,G 
= \mbox{finite} + {\mathcal O} \left( Q^{n+3}\right)\;.
\label{ren-condition-2}
\end{eqnarray}
Note that the divergencies in $K_{\sigma ,S}\,G\,T_{\pi ,S}\,G \,K_{\sigma ,S}$ are 'weak' 
since they are already tamed to a large extend by the renormalization 
condition (\ref{ren-condition-1}). Here a manifest realization of the L-counting rule 
requires a subtraction scale $\mu\sim m_\pi$. 

It is not obvious that the renormalization conditions (\ref{ren-condition-1}) and 
(\ref{ren-condition-2}) can be met. In fact our scheme requires that the Bethe-Salpeter 
interaction kernel $K$ is 'on-shell irreducible' if we insist on a separable counter 
interaction $K_S$. Basically we need a mechanism which absorbs systematically all 
divergencies not directly linked to a s-channel unitarity cut into $K_\sigma $ 
at the level of the interaction kernel. We  
decompose the kernel $K$ of (\ref{bs1b}) into an 'on-shell irreducible' part $\bar K $ 
and 'on-shell reducible' terms $K_L$, $K_R$ and $K_{LR}$ which vanish if evaluated with 
on-shell kinematics either in the incoming or outgoing channel
\begin{eqnarray}
K=\bar K+K_L+K_R+K_{LR} \;.
\label{}
\end{eqnarray}
Here $K_R$ ($K_L$) vanishes at on-shell kinematics for the incoming (outgoing) nucleons after 
multiplication with the incoming (outgoing) nucleon spinors 
by construction. The term $K_{LR}$ vanishes if evaluated with either incoming or outgoing 
on-shell kinematics. For example the s-channel spectral density of $K_R\,G\,\bar K$ or 
$\bar K\,G\,K_L$ is zero. A term like $K_R\,G$ would be renormalized according to 
(\ref{ren-condition-1}) by an appropriate counter loop $K_S\,G$ with a non-separable $K_S$. 
This would destroy the simple structure of $T_{\sigma ,S}$.  It is therefore advantageous to 
eliminate all on-shell reducible terms by introducing the on-shell equivalent effective 
interaction kernel $\hat K $ with
\begin{eqnarray}
\hat T&=&2\,\hat K+\hat K\,G\,\hat T =\frac{2}{1-K\,G}\,K \Bigg|_{\mbox{on-shell}} \;,
\nonumber\\
\hat K &=& \bar K+K_R\,G\,\bar K+\bar K\,G\,K_L
+ K_R\,G\,K_L+\bar K\,G\,K_{LR}\,G\,\bar K +\cdots \;.
\label{matching-xx}
\end{eqnarray}
By construction the effective interaction kernel $\hat K$ is on-shell irreducible 
and one therefore expects a non-zero s-channel spectral density of $\hat K \,G\,\hat K $. 
We point out that it is sufficient to work out the  matching in (\ref{matching-xx}) 
perturbatively since chiral power counting rules apply for $\hat K $
\footnote{Note that the introduction of the on-shell irreducible kernel via (\ref{matching-xx}) 
demonstrates that the infinite tower of local s-wave interaction terms in (\ref{mut0}) 
is effectively complete. Further s-wave interaction terms simply renormalize the coupling 
function  of (\ref{mut0}) according to (\ref{matching-xx}).}. In full 
analogy to (\ref{split-2}) the effective kernel $\hat K=\hat K_\pi +\hat K_\sigma $ is written 
as a sum  of non-local contributions $\hat K_\pi $ and local counter terms $\hat K_\sigma $.
Divergent terms in $\hat K $ can now be absorbed into the local interaction vertex 
$\hat K_\sigma $ of (\ref{k-def}) at the level of the interaction kernel.

We need to address a further aspect of our scheme: how does one renormalize divergencies 
that appear in subleading orders in the $1/m$ expansion. To this end it is convenient to 
introduce a partly renormalized two-nucleon propagator $\tilde G$:
\begin{eqnarray}
&&\tilde G(l;w) = \int_{4\,m^2}^{4\,(m^2+\Lambda^2)}
\frac{d\,\bar s}{\pi}\,\sqrt{\frac{\bar s}{w^2}}
\frac{-1}{\bar s-w^2-i\,\epsilon}\,\Delta S^{(+)}_F(l,w;\bar s)
\otimes \Delta S^{(-)}_F(l,w;\bar s) \; ,
\nonumber\\
&&\Delta S^{(\pm )}_F(l,w;\bar s) =
\,i\,\pi\, \delta \Big( l^2\pm l\cdot w+{\textstyle{1\over 4}}\,\bar s-m^2 \Big)
\,\Big(\pm \lslash + {\textstyle{1\over2}}\,\wslash +m \Big)\;
\label{tilde-G}
\end{eqnarray}
where we keep the cutoff parameter $\Lambda$ at this intermediate stage.
Replacing the two-nucleon propagator $G$ by the renormalized propagator $\tilde G$
leads to a renormalized interaction kernel $\tilde K$
\begin{eqnarray}
\hat T&=&2\,\hat K+\hat K\,G\,\hat T= 2\,\tilde K +\tilde K\,\tilde G\,\hat T \,,
\;\;\;
\tilde K = \frac{1}{1-\hat K\,(G-\tilde G)}\,\hat K\;.
\label{1/m-ren}
\end{eqnarray}
It is crucial to observe that by construction $\tilde K $ is free of s-channel unitarity cuts. 
Therefore $\tilde K $ can be evaluated in perturbation theory applying chiral power 
counting rules. The construction (\ref{1/m-ren}) defines a systematic prescription 
for absorbing ill-defined  power divergent terms $(\Lambda/m)^n$ with $n>1$ 
induced by the $1/m$ expansion into $K_\sigma$ (see (\ref{lin-div})).
Note that for a local vertex (\ref{mut0}) we would find $\tilde K = \hat K$ 
if we replaced $\sqrt{\bar s/w^2}\rightarrow 1$ in (\ref{tilde-G}). 

In the following we assume that the kernel $K$ in (\ref{bs1b}) is on-shell 
irreducible and $1/m$ renormalized  according to (\ref{matching-xx},\ref{1/m-ren}). 
We emphasize that the constructions (\ref{matching-xx}) and (\ref{1/m-ren}) are needed to 
systematically absorb a large class of divergencies into the local interaction kernel 
$K_\sigma$. Only after the effective interaction kernel has been properly identified, can 
one apply the renormalization conditions (\ref{ren-condition-1},\ref{ren-condition-2}). Note 
that in actual calculations it is not mandatory to construct the effective interaction 
explicitly. Since the pion-dynamics at given chiral order leads to a finite number of 
Feynman diagrams to be considered it is sufficient to evaluate those diagrams in terms 
of the original bare interaction kernel $K$. The construction (\ref{matching-xx}) 
and(\ref{1/m-ren}) then leads to a prescription for moving divergencies, which are not 
absorbed with (\ref{ren-condition-1},\ref{ren-condition-2}), into $K_\sigma$ (see appendix B). 

We turn to a problem with the proposed expansion: as it stands it violates unitarity. 
Here we anticipate an important result to be discussed in detail below, namely that such an 
expansion can be reconciled with unitarity if the subtraction interaction $K_S$ is 
generalized to allow for more complicated separable structures. Rather than evaluating 
for example $T_{\pi,S}[K_S]$ strictly at $K_S = \bar K_S$ as suggested by the 
renormalization condition (\ref{ren-condition-1}) we consider $T_{\pi,S}[K_S]$ at 
$K_S \neq \bar K_S$. It is crucial not to spoil the renormalization procedure 
(\ref{ren-condition-1},\ref{ren-condition-2}). Therefore the non-trivial part of $K_S$ 
is permitted only in the finite parts. More specifically we split for example 
$T_{\pi ,S}[K_S]$ in a finite part $T_{\pi ,S }^{fin.}[K_S]$ and a divergent part 
$T_{\pi ,S }^{div.}[K_S]$ and expand in $K_S$ around $\bar K_S$:
\begin{eqnarray}
T_{\pi ,S}[p^2;K_S] &=& T_{\pi ,S }^{fin.}[p^2,\bar \mu^2;K_S]
+T_{\pi ,S }^{div.}[p^2,\bar \mu^2;K_S]
\nonumber\\
&=& \sum^{\infty }_{n=0} \,\Big( T_{\pi, \,div.}^{(n)}(p^2,\bar \mu^2)
+T_{\pi, \,fin.}^{(n)}(p^2,\bar \mu^2)\Big) \,
\Big( K_S-\bar K_S\Big)^n\; ,
\nonumber\\
T_{\pi, \,fin.}^{(n)}(-\bar \mu^2,\bar \mu^2) &=& 0 \;\;\;\mbox{if}
 \;\;T_{\pi, \,div.}^{(n)}(p^2,\bar \mu^2) \neq 0
\label{split-fin-div-1}
\end{eqnarray}
where we introduce a necessary splitting point $\bar \mu $. In (\ref{split-fin-div-1}) the local 
kernel $\bar K_S$ is fixed according to the renormalization condition (\ref{ren-condition-1}).
We implicitly assume a projection onto a given partial wave. Thus the scattering amplitude 
is a function of only the relative momentum $p$ with $s=4\,(m^2+p^2) $.
Note for example that by construction $T_{\pi, \,div.}^{(0)}(p^2,\bar \mu^2) = 0$ and therefore 
$T_{\pi, \,fin.}^{(n)}(p^2,\bar \mu^2)$ is independent of the splitting point $\bar \mu$. 
Moreover in (\ref{split-fin-div-1}) we assume that the non-trivial part of the kernel 
$K_S$ is separable since only then the multiple powers of $( K_S-\bar K_S)^n$ are defined 
properly. The separation of the objects $T_{\pi ,S}\,G$ and $G\,T_{\pi ,S}$ into finite and 
divergent parts is performed in complete analogy with (\ref{split-fin-div-1}). The remaining 
objects $G\,T_{\pi ,S}\,G$ and $K_{\sigma ,S}$ require a separate treatment since they 
involve further divergencies which are not taken care of by $\bar K_S$:
\begin{eqnarray}
K_{\sigma ,S}^{-1} - G  &=& G\,T_{\pi ,S}[K_S]\,G
\nonumber\\
&+&\sum^{\infty }_{n=0} \,\Big( T_{\sigma, \,div.}^{(n,-1)}(p^2,\bar \mu^2)
+T_{\sigma, \,fin.}^{(n,-1)}(p^2,\bar \mu^2)\Big) \,
\Big( K_S-\bar K_S\Big)^n\; ,
\nonumber\\
T_{\sigma, \,fin.}^{(n,-1)}(-\bar \mu^2,\bar \mu^2) &=& 0 
\;\;\;\mbox{if} \;\;T_{\sigma, \,div.}^{(n,-1)}(p^2,\bar \mu^2) \neq 0
\; .
\label{split-fin-div-2}
\end{eqnarray}
Again by construction $T_{\sigma, \,div.}^{(0,-1)}(p^2,\bar \mu^2)=0$ and therefore 
$T_{\sigma, \,fin.}^{(0,-1)}(p^2,\bar \mu^2)$ is independent on the splitting point $\bar \mu$. 
Note that the splitting procedure in (\ref{split-fin-div-1}) and (\ref{split-fin-div-2}) 
must necessarily be correlated in order to ensure that the full scattering amplitude $T$ 
is independent of $K_S$ and thus on any power $( K_S-\bar K_S )^n$ as it should be according 
to our construction (\ref{split-1},\ref{split-2}). The independence of $T$ on $K_S-\bar K_S$
serves as a powerful argument that all divergent terms like 
$T_{\pi, \,div.}^{(n)}(p^2,\bar \mu^2)\,\Big( K_S-\bar K_S\Big)^n $ can be dropped. 
Note that any truncation of $T_{\pi,S}$ and $T_{\sigma,S}$ of the form (\ref{tsigma}) 
necessarily leads to a residual dependence of $T$ on $K_S$. This dependence 
of the approximate $T$ on $K_S-\bar K_S$ can then be exploited to restore unitarity.

We proceed and present an argument that the splitting point $\bar \mu^2$ must in fact be 
identified with the chiral limit value of a potential singularity present in $T$ at 
$p^2=-z_0^2=-\bar \mu^2$ in a given partial wave\footnote{In the spin triplet channel 
for example we would identify $z_0 = \sqrt{2\,m\,\epsilon_D-\epsilon_D^2}$ with the nucleon 
mass $m$ and the deuteron binding energy $\epsilon_D$ evaluated in the chiral limit with
$m_\pi =0$.}. Suppose we drop the non-trivial terms $(K_S-\bar K_S)^n$ for a moment since in 
the full expression for $T$ they cancels anyway. 
Then we expect the singularity to sit in $T^{(0,-1)}_{\sigma ,fin.}(-z_0^2)=0$ since 
$T^{(0)}_{\pi ,fin.}(p^2)$ is perturbative and independent on $\bar \mu$ by construction. 
Therefore, if we now include the terms $(K_S-\bar K_S)^n$, we must insist on 
$T^{(n,-1)}_{\sigma ,fin.}(-z_0^2,\bar \mu^2)=0$ also for $n>0$. If we do not insist 
on this condition the singularity at $p^2=-z_0^2$ must sit in $T^{(0)}_{\pi ,fin.}$ which 
contradicts our construction. Here we implicitly assume 
that the non-trivial part in $K_S$ is also perturbative. This will be confirmed later 
when this part is constructed explicitly. We now arrive at our result $\bar \mu^2=z_0^2 $ since
from (\ref{split-fin-div-1}) it follows that 
$T_{\sigma, \,div.}^{(n>0,-1)}(p^2,\bar \mu^2) \neq 0$ holds.

We emphasize that the local part of $\bar K_S$ is fixed by the renormalization condition 
(\ref{ren-condition-1}) whereas the non-trivial terms $(K_S-\bar K_S)^n$, which are  
included exclusively in the finite parts, are determined by the unitarity constraint. 
In our discussion of 
the separation of finite parts from infinite parts, we suppressed the chiral expansion 
of e.g. $K_S$ and $T_{\pi,S}$. It is implicitly understood that 
(\ref{split-fin-div-1},\ref{split-fin-div-2}) apply order by order
in the chiral expansion.

On a rather formal level the underlying mechanism leading to exact unitarity is quickly 
revealed. At given chiral order $n$ construct the non-trivial but separable part of $K_S$ 
such that
\begin{eqnarray}
\langle T_{\pi ,S} \rangle =0 +{\mathcal O} \left( Q^{n+1}\right)
\label{unitarity-construct}
\end{eqnarray}
holds, where the brackets refers to projection onto a given partial wave with on-shell 
kinematics. Insisting on a separable construction of $K_S$ leads 
to the crucial identities: 
\begin{eqnarray}
\langle T \rangle  &=& 2\,\frac{\langle 1+ \frac{1}{2}\,T_{\pi,S}\,G \rangle\,
\langle 1+ \frac{1}{2}\,G\,T_{\pi,S} \rangle+{\mathcal O} \left( Q^{n+2}\right)}
{\langle K^{-1}_{\sigma,S}-G
-\frac{1}{2}\,G\,T_{\pi,S}\,G\rangle +{\mathcal O} \left( Q^{n+3}\right)}
\;,
\nonumber\\
0 &=& \Im \,\langle 1+ \frac{1}{2}\,T_{\pi,S}\,G \rangle \,\langle
 1+ \frac{1}{2}\,G\,T_{\pi,S} \rangle  
+{\mathcal O} \left( Q^{n+2}\right)\;.
\label{ltr}
\end{eqnarray}
The unitarity condition
\begin{eqnarray}
\Im \,T^{-1} &=& -\frac{1}{2}\,\Im \,G
\label{gen-unitarity-1}
\end{eqnarray}
applied to (\ref{split-4}) reduces to:
\begin{eqnarray}
\Im \,\langle G+\frac{1}{2}\,G\,T_{\pi,S}\,G\rangle 
&=&\Big(\Im \,\langle G \rangle\Big) \,\Big(\Re \,\langle 1+ \frac{1}{2}\,T_{\pi,S}\,G \rangle\,
\langle 1+ \frac{1}{2}\,G\,T_{\pi,S} \rangle \Big) \;.
\label{gen-unitarity-2}
\end{eqnarray}
The unitarity condition (\ref{gen-unitarity-2}) is indeed fulfilled if expanded according
to chiral power counting rules. This is basically Cutkosky's  cutting 
rule \cite{Cutkosky}. The explicit technical realization of this mechanism will be 
presented in section 3.4. 

According to the chiral counting rules one expects that any non-perturbative structure must 
sit in the denominator of (\ref{ltr}). Thus any attempt to break up and expand further 
this denominator is subtle if the precise singularity structure is unknown. 
We suggest a scheme in which the numerator and denominator of 
(\ref{ltr}) are expanded according to the counting rule (\ref{l-counting-rule}) separately. 
This expansion has the great advantage that it respects unitarity exactly order by order 
in the chiral expansion. Note that if $T_{\pi , S} =0 $ the 
full solution of the Bethe-Salpeter equation is represented by $T_{\sigma ,S }$. In this sense 
one may say that unitarity leads to an optimized chiral expansion of the scattering amplitude.  
We stress that a perturbative $K_{\pi,S } $ kernel is predicted by chiral power counting 
rules provided that the subtraction interaction $\bar K_S$ is constructed properly. 
In particular all divergencies must be absorbed in $T_{\pi ,S}$ and $T_{\pi ,S}\,G$. A 
positive feature of this scheme is its practicality: the object $T_{\pi ,S}$ can 
be evaluated perturbatively and therefore, due to the separable nature of $K_S$, 
the scattering amplitude can be evaluated analytically.

\section{The nucleon-nucleon scattering amplitude}

In this section we derive explicit expressions for the leading chiral orders of the 
nucleon-nucleon scattering amplitude constructed according to the chiral counting scheme of 
section 2. We will present analytic results for the required loop functions and confirm the 
usefulness of our L-counting rule. 

We parameterize the on-shell scattering amplitude, $T(t,u)$, in terms of five isospin
zero amplitudes, $a_i^{(0)}(t,u)$, and five isospin one
amplitudes, $a_i^{(1)}(t,u)$,
\begin{equation}
T(t,u) ={\mathcal P}_0 \,\sum_{i=0}^5 a_i^{(0)} (t,u)\,  T_i
+ {\mathcal P}_1\,\sum_{i=0}^5 a_i^{(1)} (t,u)\,  T_i
\label{ampl1}
\end{equation}
defined with respect to the Dirac tensors
\begin{eqnarray}
 T_1 &=& \left(   \gamma_5\, C^{-1}\times C\,\gamma_5  \right)\; ,
\nonumber\\
 T_2 &=&  \left(  \gamma_{\alpha }\,C^{-1}\times C\,\gamma^{\alpha }  \right)\; ,
\nonumber\\
 T_3 &=& \left( C^{-1}  \times C \right)\; ,
\nonumber\\
 T_4 &=&  \left(  \gamma_{\alpha }\,C^{-1}\times C\,\gamma^{\alpha }  \right)
-{\textstyle {1\over2}}
\, \left( \sigma_{\alpha \beta }\,C^{-1} \times C \,\sigma^{\alpha \beta }  \right)\; ,
\nonumber\\
 T_5 &=& \left(  \gamma_5\, C^{-1} \times C \,\gamma_5  \right)
+\left(  \gamma_5 \gamma_{\alpha }\,C^{-1}\times C\, \gamma_5\gamma^{\alpha }  \right)
\label{ampl2}
\end{eqnarray}
with the charge conjugation matrix $C= i\gamma_0 \gamma_2 $ and the isospin projection 
operators
\begin{eqnarray}
{\mathcal P}_0 &=& {\textstyle {1\over2}}\left( \tau_2 \times
\tau_2\right)\; ,
\nonumber\\
{\mathcal P}_1 &=& {\textstyle {1\over2}}
\left( \vec \tau \tau_2 \times \tau_2\vec \tau \right) \,.
\label{ampl3}
\end{eqnarray}
For the precise definition of the tensor product $\times $ we refer to appendix A. 
The invariant amplitudes, $a_i^{(I)}(t,u)$, depend on the Mandelstam variables $t=(p_1-p_1')^2$ 
and $u=(p_1-p_2')^2$ where $p_{1,2}$ are the 4-momenta of the incoming nucleons and $p_{1,2}'$ 
those of the outgoing nucleons. The Dirac tensors $ T_i $ are constructed such that they 
are invariant modulo a sign upon flipping incoming or outgoing indices:
\begin{equation}
\left(  T_i \right)_{klmn} = (-1)^i \left(  T_i \right)_{klnm}
=(-1)^i \left(  T_i \right)_{lkmn} \; .
\label{ampl4}
\end{equation}
As a consequence one finds also a simple transformation property of the reduced amplitudes, 
$a^{(I)}_i(u,t)= (-1)^{i+I}\,a^{(I)}_i(t,u)$, under the interchange of incoming 
or the outgoing particles. Furthermore the tensors $ T_i$ are chosen 
such that the $ T_{i\ge3} $ are kinematically suppressed by one power of $u$ or $t$ if 
multiplied with on-shell spinors. 
In appendix A the invariant amplitudes $a_i^{(I)}(t,u) $ are
uniquely reconstructed in terms of the nucleon phase shifts.

In this work we focus on the isospin one channel and evaluate the
$a^{(1)}_i(u,t)$-amplitudes to leading orders in our chiral
expansion scheme. Results on the isospin zero amplitudes will be
reported on in a forthcoming publication. Here we elaborate in particular on the s-wave 
partial wave amplitude $T_{[^1S_0]}(p) $ which is parameterized uniquely
in terms of the s-wave phase shift $\delta_{[^1S_0]}(p)$. The s-wave amplitude
can be constructed in terms of our reduced amplitudes
\begin{eqnarray}
T_{[^1S_0]}(p) &=& \int \frac{d\,\Omega}{4\pi }\, T_{[^1S_0]}(u,t)
=\frac {4\, \pi}{\sqrt{s}}\,\frac{1}{i\,p}
\left(e^{2\,i\,\delta_{[^1S_0]} }-1 \right)
\label{s-def}
\end{eqnarray}
where
\begin{eqnarray}
T_{[^1S_0]}(u,t) &=&
a^{(1)}_1(u,t)-\frac{u-t}{s}\,a^{(1)}_4(u,t)-\frac{u+t}{s}\,a^{(1)}_5(u,t)
\label{s-proj}
\end{eqnarray}
and $t=-2\,p^2\,(1-\cos \theta ) $, $ u = -2\,p^2\,(1+\cos \theta )$ (see (A.5,A.8)).

We proceed and derive the leading terms of the scattering amplitude
(\ref{split-4}). The central interaction vertex relevant for our work is the pion-nucleon 
coupling vertex extracted from the chiral Lagrangian
\begin{eqnarray}
{\mathcal L} &=& g_A\,\lambda\;
\bar N \,\gamma_5 \,\gamma_\mu\,\Big(\partial^\mu \,\vec \pi \cdot \vec \tau\Big)\,N
= \frac{g_A\,m}{f_\pi}\,
\bar N \,i\,\gamma_5 \,\vec \pi \cdot \vec \tau \,N +\cdots
\label{pion-nucleon-coupling}
\end{eqnarray}
where the coupling constant $\lambda $ is identified with the pion decay constant
$\lambda^{-1} = 2\, f_{\pi } \simeq 186 \,{\rm MeV} $. For the axial vector coupling 
constant of the nucleon we take $g_A\simeq 1.26$. To the leading orders we will be working here 
the interaction kernel $K_\pi$ of (\ref{k-def}) is identified with the one-pion exchange 
contribution
\begin{eqnarray}
K_\pi &=& -\frac{g_A^2}{8\,f_\pi^2}\,
\frac{\gamma_5\,\Qslash \otimes \gamma_5\,\Qslash}{t-m_\pi^2} \,
\Big({\mathcal P}_1 -3\,{\mathcal P}_0 \Big) + \mbox{exchange \,term}
\label{kpi-def}
\end{eqnarray}
following from (\ref{pion-nucleon-coupling}) and $Q=p_1-p_1'$. The 
on-shell irreducible kernel $\hat K_\pi$ of (\ref{matching-xx}) follows with
\begin{eqnarray}
\hat K_\pi &=& \frac{g_A^2\,m^2}{2\,f_\pi^2}\,
\frac{\gamma_5\otimes \gamma_5}{t-m_\pi^2} \,
\Big({\mathcal P}_1 -3\,{\mathcal P}_0 \Big) 
\nonumber\\
&-&\frac{g_A^4\,m^2}{16\,f_\pi^4}
\left( 2\left(m_\pi^2-{\textstyle {t\over 2}}\right) I_{3,\pi}(t)+I_{2,\pi}(t)-2\,I_2(m^2) \right)
\Big( T_1 +T_2 \Big)\,\Big( {\mathcal P}_1+9\,{\mathcal P}_1 \Big)
\nonumber\\
&-& \frac{11\,g_A^4\,m^2}{384\,f_\pi^4}\,I_{1,\pi}\,
\Big( T_1 +T_2 \Big)\,\Big( {\mathcal P}_1+9\,{\mathcal P}_1 \Big)
+{\mathcal O} \left( Q^2 \right)+\mbox{exchange \,terms} \;.
\label{kpi-hat-def}
\end{eqnarray}
At leading orders it is given by the pion-nucleon vertex with 
pseudo-scalar coupling and contributions from scalar one-loop functions 
$I_{3,\pi}(t), I_{2,\pi}(t), I_{1,\pi}$ and $I_2(m^2)$ defined in appendix B. All we would like 
to point out here is that we find:
\begin{eqnarray}
\Big(K_\pi +K_\pi\,G\,K_\pi \Big) \Big|_{\mbox{on-shell}}
= \hat K_\pi +\hat K_\pi\,G\,\hat K_\pi +{\mathcal O} \left( Q^2 \right)
\label{}
\end{eqnarray}
by construction. Our result (\ref{kpi-hat-def}) clearly exemplifies the merit of 
transforming to the on-shell irreducible kernel: it permits to move the divergent 
pionic tadpole contribution $I_{1,\pi}$ into the local interaction vertex 
$K_\sigma$ of (\ref{k-def}) at the level of the interaction kernel. 
An interesting observation worked out in detail in appendix B.1 concerns the chiral power 
of the pionic tadpole diagram $I_{1,\pi}$. As is demonstrated explicitly a consistent $1/m$ 
expansion implies $I_{1,\pi}\sim Q$ rather than $I_{1,\pi}\sim Q^2$ as one might expect 
naively from (\ref{counting-rules-1}). Here this does not cause any complication since 
$I_{1,\pi}$ is absorbed into the bare coupling function. Note a further subtle property of 
the on-shell irreducible kernel $\hat K_\pi$: the scalar pion bubble loop $I_{2,\pi}(t)$ in 
(\ref{kpi-hat-def}), carries chiral power zero (see (\ref{i2-pi})). This does not contradict 
the chiral expansion since this term is canceled by a corresponding term in 
$\hat K_\pi\,G\,\hat K_\pi $ by construction. 

The kernel $K_\sigma $ was already given in (\ref{large-m-example}) in terms of the energy 
dependent coupling function $g(s)$. It is on-shell irreducible by construction. 
We further need to specify the subtraction interaction $K_{S}$ of (\ref{split-1})
and $\bar K_S$ of (\ref{ren-condition-1})
\begin{eqnarray}
K_S &=& \parbox{20mm}{
\begin{fmfgraph*}(10,10)
\fmfleftn{l}{2}
\fmfrightn{r}{2}
\fmf{fermion}{r2,o,l2}
\fmf{fermion}{r1,o,l1}
\fmfv{d.sh=c,d.filled=0,d.size=3thick}{o}
\end{fmfgraph*}}
= c(s)\,g_A^2\,\lambda^2\, T_1\,{\mathcal P}_1 \; ,\;\;
\bar K_S = \bar c(s)\,g_A^2\,\lambda^2\, T_1\,{\mathcal P}_1 \;,
\nonumber\\
K_{\sigma,S} &=& g_{\sigma,S}(s)\,\,\,\,\;\parbox{20mm}{
\begin{fmfgraph*}(10,10)
\fmfleftn{l}{2}
\fmfrightn{r}{2}
\fmf{fermion}{r2,o,l2}
\fmf{fermion}{r1,o,l1}
\fmfv{d.sh=c,d.filled=1,d.size=3thick}{o}
\end{fmfgraph*}}
= \Big( g(s)-c(s)\,g_A^2\,\lambda^2\Big)\, T_1\,{\mathcal P}_1
\label{barc-def}
\end{eqnarray}
appropriate for s-wave scattering in the spin singlet channel. Here we allow for a s-dependence 
of the subtraction function $c(s)$. The subtraction function $c(s)$ is on the one hand 
required to justify our previously introduced L-counting rule and on the other hand 
ultimately leads to unitary results for the scattering amplitude. 

The leading orders as they contribute to our amplitudes are readily derived. The amplitude
$T_{\pi,S}(s) $ of (\ref{split-3}) receives at chiral order $Q$ contributions
\begin{eqnarray}
T_{\pi,S}&=&
\parbox{15mm}{
\begin{fmfgraph}(15,10)
\fmfleftn{l}{2}
\fmfrightn{r}{2}
\fmf{fermion}{r2,o,l2}
\fmf{fermion}{r1,o,l1}
\fmfv{d.sh=c,d.filled=0,d.size=4thick}{o}
\end{fmfgraph}}+
\parbox{15mm}{
\begin{fmfgraph*}(15,10)
\fmfleftn{l}{2}
\fmfrightn{r}{2}
\fmf{fermion}{r2,u,l2}
\fmf{fermion}{r1,d,l1}
\fmfforce{(0.0w,0.0h)}{l1}
\fmfforce{(0.5w,0.0h)}{d}
\fmfforce{(1.0w,0.0h)}{r1}
\fmfforce{(0.0w,1.0h)}{l2}
\fmfforce{(0.5w,1.0h)}{u}
\fmfforce{(1.0w,1.0h)}{r2}
\fmf{boson}{d,u}
\fmfdot{u,d}
\end{fmfgraph*}}
+
\parbox{20mm}{
\begin{fmfgraph*}(20,10)
\fmfleftn{l}{2}
\fmfrightn{r}{2}
\fmf{fermion}{r2,u2,u1,l2}
\fmf{fermion}{r1,d2,d1,l1}
\fmfforce{(0.0w,0.0h)}{l1}
\fmfforce{(0.33w,0.0h)}{d1}
\fmfforce{(0.66w,0.0h)}{d2}
\fmfforce{(1.0w,0.0h)}{r1}
\fmfforce{(0.0w,1.0h)}{l2}
\fmfforce{(0.33w,1.0h)}{u1}
\fmfforce{(0.66w,1.0h)}{u2}
\fmfforce{(1.0w,1.0h)}{r2}
\fmf{boson}{d2,u2}
\fmf{boson}{d1,u1}
\fmfdot{u1,u2,d1,d2}
\end{fmfgraph*}
}
\nonumber\\ \nonumber\\
&+&
\parbox{30mm}{
\begin{fmfgraph*}(30,15)
\fmfleft{o}
\fmfrightn{r}{2}
\fmfleftn{l}{2}
\fmf{fermion}{r2,u1,o,l2}
\fmf{fermion}{r1,d1,o,l1}
\fmfforce{(0.3w,0.5h)}{o}
\fmffreeze
\fmf{boson}{d1,u1}
\fmfdot{d1,u1}
\fmfv{d.sh=c,d.filled=0,d.size=4thick}{o}
\end{fmfgraph*}}
+
\parbox{30mm}{
\begin{fmfgraph*}(30,15)
\fmfleft{o}
\fmfrightn{r}{2}
\fmfleftn{l}{2}
\fmf{fermion}{r2,o,u1,l2}
\fmf{fermion}{r1,o,d1,l1}
\fmfforce{(0.6w,0.5h)}{o}
\fmffreeze
\fmf{boson}{d1,u1}
\fmfdot{d1,u1}
\fmfv{d.sh=c,d.filled=0,d.size=4thick}{o}
\end{fmfgraph*}}
+{\mathcal O} \left(Q^2 \right)
\nonumber\\
&+& \mbox{exchange terms}
\label{ampl5-a}
\end{eqnarray}
from the subtraction vertex $K_S$, the one-pion exchange, the pion nucleon box and 
triangle diagrams induced by $K_S$. In (\ref{ampl5-a}) the wavy line represents the 
pion exchange $K_\pi$ of (\ref{kpi-def}) rather than $\hat K_\pi$ of (\ref{kpi-hat-def}) 
for convenience. Transforming back to the on-shell irreducible kernel $\hat K_\pi$ is
straightforward and only required for the renormalization process. The full scattering 
amplitude follows now from (\ref{split-4}). We
note that all reduced amplitudes $a^{(1)}_i(u,t)$ receive contributions from 
$T_{\pi ,S}$ whereas the amplitude $T_{\sigma ,S}$ affects only $a^{(1)}_1(u,t)$:
\begin{eqnarray}
a_1^{(1)}(u,t) &=& 2\,\,g_A^2\,\lambda^2\,c(s)
\left(1+2\,g_A^2\,\lambda^2\,V_{\pi }(s)\right)
+2\,g_A^4\,\lambda^4 \,c^2(s)\,J(s)
\nonumber\\
&+& g_A^4\,\lambda^4\,\Big( B_1(u,t)+B_1(t,u)\Big)
\nonumber\\
&+& \left(1+ g_A^2 \, \lambda^2\,
\,V_{\pi ,S }(s) +g_A^4\,\lambda^4\,V_{\pi \pi,S}(s) \right)^2
T_{\sigma ,S}(s)\; ,
\nonumber\\
a_2^{(1)}(u,t) &=& g_A^4\,\lambda^4\,\Big( B_2(u,t)-B_2(t,u)\Big)\; ,
\nonumber\\
a_3^{(1)}(u,t) &=&  m^2\, g_A^2
\,\lambda^2\left(
\frac{1}{t-m_{\pi }^2}+\frac{1}{u-m_{\pi }^2} \right)
+g_A^4\,\lambda^4\,\Big( B_3(u,t)+B_3(t,u)\Big)\; ,
\nonumber\\
a_4^{(1)}(u,t) &=& m^2\, g_A^2 \,\lambda^2 \left(
\frac{1}{u-m_{\pi }^2}-\frac{1}{t-m_{\pi }^2}\right)
+g_A^4\,\lambda^4\,\Big( B_4(u,t)-B_4(t,u)\Big)\; ,
\nonumber\\
a_5^{(1)}(u,t) &=& m^2 \,g_A^2 \,\lambda^2\left(
\frac{1}{t-m_{\pi }^2}+\frac{1}{u-m_{\pi }^2} \right)
+g_A^4\,\lambda^4\,\Big( B_5(u,t)+B_5(t,u)\Big)\;.
\nonumber\\
\label{ampl5}
\end{eqnarray}
The functions $B_i(u,t) $ represent the contribution from the pion-nucleon box
diagram in (\ref{ampl5-a}). The precise definition for all loop functions can be found in 
appendix B. The one-pion exchange contribution is made explicit in (\ref{ampl5}) and confirms
the transformation property of the amplitudes $a_i(t,u)$ under interchange of $t$ and $u$.
The vertex loop functions $V_{\pi^n,S}(s)$ in (\ref{ampl5}) are introduced with respect 
to the interaction kernel $K_{\pi , S}$ of (\ref{split-1}). 
The one and two-loop vertex functions $ V_{\pi ,S} (s)$ and $V_{\pi 
\pi ,S }(s) $ are then given by
\begin{eqnarray}
V_{\pi, S}(s)&=&\frac{1}{g_A^2\,\lambda^2} \left(
\parbox{30mm}{
\begin{fmfgraph*}(30,15)
\fmfleft{o}
\fmfrightn{r}{2}
\fmfleftn{l}{2}
\fmf{fermion}{r2,u1,o,l2}
\fmf{fermion}{r1,d1,o,l1}
\fmfforce{(0.3w,0.5h)}{o}
\fmffreeze
\fmf{boson}{d1,u1}
\fmfdot{d1,u1}
\fmfv{d.sh=c,d.filled=1,d.size=4thick}{o}
\end{fmfgraph*}}
+\parbox{30mm}{
\begin{fmfgraph*}(30,15)
\fmfleftn{l}{2}
\fmfrightn{r}{2}
\fmf{fermion}{ol,l2}
\fmf{fermion}{ol,l1}
\fmf{fermion}{r1,or}
\fmf{fermion}{r2,or}
\fmf{fermion,left=.7,tension=.5}{or,ol}
\fmf{fermion,right=.7,tension=.5}{or,ol}
\fmfv{d.sh=c,d.filled=1,d.size=4thick}{ol}
\fmfv{d.sh=c,d.filled=0,d.size=4thick}{or}
\end{fmfgraph*}}
\right)
\nonumber\\ \nonumber\\ 
&=& V_{\pi }(s) +c(s)\,J(s)\; ,
\nonumber\\ \nonumber\\
V_{\pi \pi,S}(s) &=&\frac{1}{g_A^4\,\lambda^4}\, 
\left( 
\parbox{30mm}{
\begin{fmfgraph*}(30,15)
\fmfleft{o}
\fmfrightn{r}{2}
\fmfleftn{l}{2}
\fmf{fermion}{r2,u2,u1,o,l2}
\fmf{fermion}{r1,d2,d1,o,l1}
\fmfforce{(0.3w,0.5h)}{o}
\fmffreeze
\fmf{boson}{d1,u1}
\fmf{boson}{d2,u2}
\fmfdot{d1,u1,d2,u2}
\fmfv{d.sh=c,d.filled=1,d.size=4thick}{o}
\end{fmfgraph*}}
+
\parbox{40mm}{
\begin{fmfgraph*}(40,15)
\fmfleft{o}
\fmfrightn{r}{2}
\fmfleftn{l}{2}
\fmf{fermion}{ol,l1}
\fmf{fermion}{ol,l2}
\fmf{fermion,left=.7,tension=.5}{o,ol}
\fmf{fermion,right=.7,tension=.5}{o,ol}
\fmf{fermion}{r2,u1,o}
\fmf{fermion}{r1,d1,o}
\fmfforce{(0.3w,0.5h)}{ol}
\fmfforce{(0.6w,0.5h)}{o}
\fmffreeze
\fmf{boson}{d1,u1}
\fmfdot{d1,u1}
\fmfv{d.sh=c,d.filled=0,d.size=4thick}{o}
\fmfv{d.sh=c,d.filled=1,d.size=4thick}{ol}
\end{fmfgraph*}}\right.
\nonumber\\  \nonumber
&&\;\;\;\;\;\;\;\;+\left.
\parbox{30mm}{
\begin{fmfgraph*}(30,15)
\fmfleftn{l}{2}
\fmfrightn{r}{2}
\fmf{fermion}{ol,l2}
\fmf{fermion}{ol,l1}
\fmf{fermion}{r1,or}
\fmf{fermion}{r2,or}
\fmf{phantom,left=.7,tension=.5,tag=1}{or,ol}
\fmf{phantom,left=.7,tension=.5,tag=2}{ol,or}
\fmfv{d.sh=c,d.filled=1,d.size=4thick}{ol}
\fmfv{d.sh=c,d.filled=0,d.size=4thick}{or}
\fmfposition
\fmfipath{p[]}
\fmfiset{p1}{vpath1(__or,__ol)}
\fmfiset{p2}{vpath2(__ol,__or)}
\fmfi{fermion}{subpath (0,length(p1)/2) of p1}
\fmfi{fermion}{subpath (length(p1)/2,length(p1)) of p1}
\fmfi{fermion}{subpath (length(p2),length(p2)/2) of p2}
\fmfi{fermion}{subpath (length(p2)/2,0) of p2}
\fmfi{boson}{point length(p1)/2 of p1 -- point length(p2)/2 of p2}
\fmfiv{d.sh=c,d.filled=1,d.size=2thick}{point length(p1)/2 of p1}
\fmfiv{d.sh=c,d.filled=1,d.size=2thick}{point length(p2)/2 of p2}
\end{fmfgraph*}}
+
\parbox{40mm}{
\begin{fmfgraph*}(40,15)
\fmfleftn{l}{2}
\fmfrightn{r}{2}
\fmf{fermion}{ol,l2}
\fmf{fermion}{ol,l1}
\fmf{fermion}{r1,or}
\fmf{fermion}{r2,or}
\fmf{fermion,left=.7,tension=.5}{or,om}
\fmf{fermion,right=.7,tension=.5}{or,om}
\fmf{fermion,left=.7,tension=.5}{om,ol}
\fmf{fermion,right=.7,tension=.5}{om,ol}
\fmfv{d.sh=c,d.filled=1,d.size=4thick}{ol}
\fmfv{d.sh=c,d.filled=0,d.size=4thick}{or,om}
\end{fmfgraph*}}
\right)
\nonumber\\\nonumber\\ 
&=&V_{\pi \pi }(s) +c(s)\left(J(s)\,V_{\pi
}(s)+J_{\pi }(s)\right) +c^2(s)\,J^2(s)
\label{svertexdef}
\end{eqnarray}
in terms of the vertex functions $V_{\pi }(s) $ and
$V_{\pi \pi }(s) $ defined with respect to the unsubtracted interaction kernel $K_\pi $ 
of (\ref{kpi-def}), the not yet specified subtraction function $c(s)$ and the  one-loop bubble 
function $J(s)$ introduced in (\ref{mut2}). The pole term in (\ref{ampl5})
\begin{eqnarray}
T_{\sigma ,S}(s) &=&\frac{2}{g_{\sigma,S}^{-1}(s)-J(s)-g_A^2\,\lambda^2\,J_{\pi,S}(s)
-g_A^4\,\lambda^4\,J_{\pi \pi,S}(s)+\; ...}
\label{ampl6}
\end{eqnarray}
is determined by the bubble  functions $J_{\pi^n,S}(s)$ introduced with respect 
to the interaction kernel $K_{\pi ,S}$ of (\ref{split-1}). 
Again the subtracted two and three-loop bubble integrals $J_{\pi ,S}(s) $ and $ J_{\pi
\pi ,S}(s) $ can be expressed as follows
\begin{eqnarray}
J_{\pi,S} (s ) &=& \frac{1}{g_A^2\,\lambda^2}\,\left(
\parbox{30mm}{
\begin{fmfgraph*}(30,15)
\fmfleftn{l}{2}
\fmfrightn{r}{2}
\fmf{fermion}{ol,l2}
\fmf{fermion}{ol,l1}
\fmf{fermion}{r1,or}
\fmf{fermion}{r2,or}
\fmf{phantom,left=.7,tension=.5,tag=1}{or,ol}
\fmf{phantom,left=.7,tension=.5,tag=2}{ol,or}
\fmfv{d.sh=c,d.filled=1,d.size=4thick}{ol}
\fmfv{d.sh=c,d.filled=1,d.size=4thick}{or}
\fmfposition
\fmfipath{p[]}
\fmfiset{p1}{vpath1(__or,__ol)}
\fmfiset{p2}{vpath2(__ol,__or)}
\fmfi{fermion}{subpath (0,length(p1)/2) of p1}
\fmfi{fermion}{subpath (length(p1)/2,length(p1)) of p1}
\fmfi{fermion}{subpath (length(p2),length(p2)/2) of p2}
\fmfi{fermion}{subpath (length(p2)/2,0) of p2}
\fmfi{boson}{point length(p1)/2 of p1 -- point length(p2)/2 of p2}
\fmfiv{d.sh=c,d.filled=1,d.size=2thick}{point length(p1)/2 of p1}
\fmfiv{d.sh=c,d.filled=1,d.size=2thick}{point length(p2)/2 of p2}
\end{fmfgraph*}}
+
\parbox{40mm}{
\begin{fmfgraph*}(40,15)
\fmfleftn{l}{2}
\fmfrightn{r}{2}
\fmf{fermion}{ol,l2}
\fmf{fermion}{ol,l1}
\fmf{fermion}{r1,or}
\fmf{fermion}{r2,or}
\fmf{fermion,left=.7,tension=.5}{or,om}
\fmf{fermion,right=.7,tension=.5}{or,om}
\fmf{fermion,left=.7,tension=.5}{om,ol}
\fmf{fermion,right=.7,tension=.5}{om,ol}
\fmfv{d.sh=c,d.filled=1,d.size=4thick}{ol,or}
\fmfv{d.sh=c,d.filled=0,d.size=4thick}{om}
\end{fmfgraph*}}
\right)
\nonumber\\\nonumber\\ 
&=& J_{\pi }(s) +c(s)\, J^2(s)\; ,
\nonumber\\\nonumber\\ 
J_{\pi \pi,S}(s) &=& \frac{1}{g_A^4\,\lambda^4}\,\left(
\parbox{40mm}{
\begin{fmfgraph*}(40,15)
\fmfleftn{l}{2}
\fmfrightn{r}{2}
\fmf{fermion}{ol,l2}
\fmf{fermion}{ol,l1}
\fmf{fermion}{r1,or}
\fmf{fermion}{r2,or}
\fmf{phantom,left=.7,tension=.5,tag=1}{or,ol}
\fmf{phantom,left=.7,tension=.5,tag=2}{ol,or}
\fmfv{d.sh=c,d.filled=1,d.size=4thick}{ol}
\fmfv{d.sh=c,d.filled=1,d.size=4thick}{or}
\fmfposition
\fmfipath{p[]}
\fmfiset{p1}{vpath1(__or,__ol)}
\fmfiset{p2}{vpath2(__ol,__or)}
\fmfi{fermion}{subpath (0,length(p1)/3) of p1}
\fmfi{fermion}{subpath (length(p1)/3,2length(p1)/3) of p1}
\fmfi{fermion}{subpath (2length(p1)/3,length(p1)) of p1}
\fmfi{fermion}{subpath (length(p2),2length(p2)/3) of p2}
\fmfi{fermion}{subpath (2length(p2),length(p2)/3) of p2}
\fmfi{fermion}{subpath (length(p2)/3,0) of p2}
\fmfi{boson}{point length(p1)/3 of p1 -- point 2length(p2)/3 of p2}
\fmfi{boson}{point 2length(p1)/3 of p1 -- point length(p2)/3 of p2}
\fmfiv{d.sh=c,d.filled=1,d.size=2thick}{point length(p1)/3 of p1}
\fmfiv{d.sh=c,d.filled=1,d.size=2thick}{point length(p2)/3 of p2}
\fmfiv{d.sh=c,d.filled=1,d.size=2thick}{point 2length(p1)/3 of p1}
\fmfiv{d.sh=c,d.filled=1,d.size=2thick}{point 2length(p2)/3 of p2}
\end{fmfgraph*}}
+
\parbox{45mm}{
\begin{fmfgraph*}(45,15)
\fmfleftn{l}{2}
\fmfrightn{r}{2}
\fmf{fermion}{ol,l2}
\fmf{fermion}{ol,l1}
\fmf{fermion}{r1,or}
\fmf{fermion}{r2,or}
\fmf{phantom,left=.7,tension=.5,tag=1}{or,om}
\fmf{phantom,right=.7,tension=.5,tag=2}{or,om}
\fmf{fermion,left=.7,tension=.5}{om,ol}
\fmf{fermion,right=.7,tension=.5}{om,ol}
\fmfv{d.sh=c,d.filled=1,d.size=4thick}{ol,or}
\fmfv{d.sh=c,d.filled=0,d.size=4thick}{om}
\fmfposition
\fmfipath{p[]}
\fmfiset{p1}{vpath1(__or,__om)}
\fmfiset{p2}{vpath2(__or,__om)}
\fmfi{fermion}{subpath (0,length(p1)/2) of p1}
\fmfi{fermion}{subpath (length(p1)/2,length(p1)) of p1}
\fmfi{fermion}{subpath (length(p2)/2,length(p2)) of p2}
\fmfi{fermion}{subpath (0,length(p2)/2) of p2}
\fmfi{boson}{point length(p1)/2 of p1 -- point length(p2)/2 of p2}
\fmfiv{d.sh=c,d.filled=1,d.size=2thick}{point length(p1)/2 of p1}
\fmfiv{d.sh=c,d.filled=1,d.size=2thick}{point length(p2)/2 of p2}
\end{fmfgraph*}}\right.
\nonumber\\  \nonumber
&&\;\;\;\;\;\;\;\;+\left.
\parbox{40mm}{
\begin{fmfgraph*}(40,15)
\fmfleftn{l}{2}
\fmfrightn{r}{2}
\fmf{fermion}{ol,l2}
\fmf{fermion}{ol,l1}
\fmf{fermion}{r1,or}
\fmf{fermion}{r2,or}
\fmf{fermion,left=.7,tension=.5}{or,om}
\fmf{fermion,right=.7,tension=.5}{or,om}
\fmf{phantom,left=.7,tension=.5,tag=1}{om,ol}
\fmf{phantom,right=.7,tension=.5,tag=2}{om,ol}
\fmfv{d.sh=c,d.filled=1,d.size=4thick}{ol,or}
\fmfv{d.sh=c,d.filled=0,d.size=4thick}{om}
\fmfposition
\fmfipath{p[]}
\fmfiset{p1}{vpath1(__om,__ol)}
\fmfiset{p2}{vpath2(__om,__ol)}
\fmfi{fermion}{subpath (0,length(p1)/2) of p1}
\fmfi{fermion}{subpath (length(p1)/2,length(p1)) of p1}
\fmfi{fermion}{subpath (length(p2)/2,length(p2)) of p2}
\fmfi{fermion}{subpath (0,length(p2)/2) of p2}
\fmfi{boson}{point length(p1)/2 of p1 -- point length(p2)/2 of p2}
\fmfiv{d.sh=c,d.filled=1,d.size=2thick}{point length(p1)/2 of p1}
\fmfiv{d.sh=c,d.filled=1,d.size=2thick}{point length(p2)/2 of p2}
\end{fmfgraph*}}
+
\parbox{50mm}{
\begin{fmfgraph*}(50,15)
\fmfleftn{l}{2}
\fmfrightn{r}{2}
\fmf{fermion}{ol,l2}
\fmf{fermion}{ol,l1}
\fmf{fermion}{r1,or}
\fmf{fermion}{r2,or}
\fmf{fermion,left=.7,tension=.5}{or,om}
\fmf{fermion,right=.7,tension=.5}{or,om}
\fmf{fermion,left=.7,tension=.5}{om,omr}
\fmf{fermion,right=.7,tension=.5}{om,omr}
\fmf{fermion,left=.7,tension=.5}{omr,ol}
\fmf{fermion,right=.7,tension=.5}{omr,ol}
\fmfv{d.sh=c,d.filled=1,d.size=4thick}{ol,or}
\fmfv{d.sh=c,d.filled=0,d.size=4thick}{om,omr}
\end{fmfgraph*}}
\right)
\nonumber\\ \nonumber\\ 
&=&J_{\pi \pi }(s)  +2\,c(s)\,J(s)\,J_{\pi }(s)
+c^2(s)\,J^3(s)
\label{sbubbledef}
\end{eqnarray}
where $ J_{\pi }(s) $ and $J_{\pi \pi }(s)$ are the 'unsubtracted' loop functions. 

We will demonstrate that our L-counting rule can be applied for the loop 
functions
\begin{eqnarray}
&&V_{\pi^n,S}(s)  \sim Q^n\;,\;\;\;\;\;\;\;\;\;\;\;\;\;
J_{\pi^n,S}(s)-J_{\pi^n,S}(s_0)\sim  Q^{n+1}
\label{L-count}
\end{eqnarray}
provided that $c=\bar c $ with $\bar c=1/2+{\mathcal O}\left(Q^2 \right)$ and 
the subtraction point $s_0 \sim 4\,m^2 $ sufficiently close to threshold.
According to the L-counting rule 
the leading chiral power of the vertex functions $V_{\pi^n,S}(s)$ and bubble functions 
$J_{\pi^n,S}(s)-J_{\pi^n,S}(s_0)$ is given by the number of independent loops. 
We return to the chiral powers of these loop functions in more detail below when we 
present explicit analytic results for all loop functions introduced here. 
For more details on the evaluation of the loop functions we refer the reader to appendix B.

We proceed and systematically rewrite the bare loop functions $J(p)$, $J_{\pi,S}(p)$ and 
$V_{\pi,S}(p) $ of (\ref{ampl5}) and (\ref{ampl6}) in terms of renormalized loop functions 
$J_R(p,z_0)$, $J_{\pi ,R}(p,z_0)$ and $ V_{\pi,R}(p)$. Note that from now on we consider 
all loop functions as a function of the small momentum $p$ rather than the Mandelstam 
variable $s=4\,(m^2+p^2)$. For simplicity we consider in this section only the leading term 
in the $1/m$ expansion and systematically drop correction terms. Technical details on the 
regularization scheme and on subleading terms in the $1/m$ expansion can be found in 
appendix B. We find:
\begin{eqnarray}
J(p)&=&J(-i\,z_0)+J_R(p,z_0) \;,
\;\;\,V_{\pi,S}(p) =V_{\pi ,R}(p)+\left(c-{\textstyle{1\over2}}\right) J(p)\;,
\nonumber\\
J_{\pi,S}(p) &=& J_{\pi,R}(p,z_0)+\bar J_{\pi}(-i\,z_0 )+
\left(c-{\textstyle{1\over2}}\right)J^2(p)  \; .
\label{loops-zero}
\end{eqnarray}
Note that here we also drop a contribution from the pion production cut in 
$J_{\pi,S}(p)$ (see (\ref{jpi-prod})). The subtraction point $z_0$ will 
ultimately be identified with the pseudo-bound state pole position in the chiral limit. 
If $z_0>0$ the subtraction point lies on the second Riemann sheet of $p=\sqrt{s/4-m^2}$. 
The renormalized bubble loop functions $J_R(p,z_0)$ and $J_{\pi,R}(p,z_0)$ vanish at 
$p=-i\,z_0$ by definition. The analytic expressions for the renormalized 
loop functions read: 
\begin{eqnarray}
J_R(p,z_0) &=&\frac{m}{2\,\pi} \Big( -z_0+i\,p
\Big) \;,\;\;\,
V_{\pi,R } (p) = 
\frac{m\,m_\pi}{8\,\pi}\,
\frac{m_\pi}{p}\,i\,\ln \left(1-i\,\frac{2\,p}{m_\pi}\right) \;,
\nonumber\\
J_{\pi ,R} (p,z_0)
&=&-2\,\frac{m^2\,m^2_\pi}{(4\,\pi)^2}\,
\ln \left(\frac{m_\pi-2\,i\,p}{m_\pi-2\,z_0 }\right)\; .
\label{1-2-a-result}
\end{eqnarray}
All divergencies sit in the one-loop and two-loop subtraction coefficients $J(-i\,z_0)$ 
and $\bar J_{\pi}(-i\,z_0)$ which may be estimated in terms of a cutoff parameter 
$\Lambda $ introduced in the s-channel dispersion relation (see (\ref{mut5})): 
\begin{eqnarray}
J(-i\,z_0 ) &\sim & \frac{m}{\pi}\,\Lambda \;,
\;\;\;
\bar J_{\pi}(-i\,z_0) \sim \frac{m^2\,m_\pi^2}{(4\,\pi)^2}\,\ln \left(
\frac{\Lambda^2}{m^2_\pi}\right) \;.
\label{subtr.-zero}
\end{eqnarray}
We point out that the vertex function $V_{\pi,S}(p)$ is rendered finite by the unique choice 
$c=1/2$ whereas the two-loop bubble $J_{\pi,S}(p)$ requires a further subtraction 
even at $c=1/2$. This reflects our renormalization procedure (\ref{ren-condition-1}) 
and (\ref{ren-condition-2}). The vertex function is renormalized by an appropriate 
'counter loop' leading to the renormalization point independent result $V_{\pi,R}(p)$. 
The bubble function $J_{\pi,S}(p)$ is renormalized in two steps. The counter loops
lead to a function $\bar J_\pi (p)$ which defines the renormalized loop 
modulo a subtraction $J_{\pi ,R} (p,z_0)=\bar J_{\pi}(p)-\bar J_{\pi}(-i\,z_0)$. 
This leads to a subtraction point dependent result for $J_{\pi ,R} (p,z_0)$. 
Note that here we are implicitly assuming that the on-shell irreducible interaction 
kernel $\hat K_\pi$ is applied (see appendix B.3). 

Consistency of our L-counting rule requires the subtraction constant 
$c$ to be rather close to one half. For example if $c$ deviates largely from 
$1/2$ the term proportional to $(c-{\textstyle{1\over 2}})\,J(-i\,z_0)\, J_R(p,z_0) $ in 
our expression for $J_{\pi ,S }$ must not be counted as chiral order two since 
$g_A^2\,\lambda^2\,J(-i\,z_0)  > 1 $. We count $(c-\bar c)\sim Q $ and therewith confirm 
$V_{\pi ,S}(p) \sim Q $ and $J_{\pi ,S}(p)- J_{\pi ,S}(-i\,z_0)\sim Q^2 $.
Note that according to (\ref{loops-zero}) $J_{\pi ,S}(p)$ carries in fact the leading 
chiral power $Q$. This result generalizes to the n-th order bubble $J_{\pi^n ,S}(p)$. 
The anomalous contribution to all orders is readily anticipated
\begin{eqnarray}
J_{\pi^n,S }(p)&=& \Big( c-{\textstyle{1\over 2}}\,\Big)^{n} \,J^{n+1}(-i\,z_0)
+{\mathcal O} \left( Q^{n+1}\right) \; .
\label{all-order}
\end{eqnarray}
and leads to  $J_{\pi^n ,S}(p) \sim Q^n$. 

We proceed and express the not yet specified two-loop function $V_{\pi\pi,S}(p)$ and three-loop 
function $J_{\pi \pi,S}(p) $ of (\ref{ampl5}) in terms of the renormalized loops 
$V_{\pi\pi,R}(p)$ and $J_{\pi \pi,R}(p,z_0)$
\begin{eqnarray}
V_{\pi\pi,S}(p) &=&V_{\pi \pi,R}(p)
+\left(c-{\textstyle{1\over 2}}\right) J(p)\,V_{\pi ,R}(p) 
+\left(c-{\textstyle{1\over 2}}\right)^2 J^2(p)
\nonumber\\
&+&\left(c-{\textstyle{1\over 2}}\right) \Big( \bar J_{\pi}(-i\,z_0)+J_{\pi ,R}(p,z_0)\Big)
\nonumber\\
J_{\pi \pi,S}(p) &=& 
J_{\pi \pi,R }(p,z_0)+\bar J_{\pi \pi}(-i\,z_0)
+\left(c-{\textstyle{1\over 2}}\right)^2 J^3(p)
\nonumber\\
&+&2\left(c-{\textstyle{1\over 2}}\right) J(p) 
\,\Big( \bar J_{\pi}(-i\,z_0)+J_{\pi ,R}(p,z_0) \Big)
\label{2-3-loop}
\end{eqnarray}
at leading order in the $1/m$ expansion. We suppress here subleading contributions from 
pion production cuts. All divergencies in (\ref{2-3-loop}) sit  in
the previously introduced power divergent one-loop bubble subtraction 
$J(-i\,z_0)$ and  the log-divergent two-loop bubble subtraction $J_{\pi}(-i\,z_0)$.
The three-loop subtraction constant $J_{\pi \pi}(-i\,z_0)$ is in fact finite:
\begin{eqnarray}
\bar J_{\pi\pi}(-i\,z_0) 
&=&\frac{m^3\,m_\pi^3}{(4\,\pi)^3}\,
\left(\ln \left(16 \right) 
+{\mathcal O}\left(\frac{z_0}{m_\pi} \right)\right)\; .
\label{}
\end{eqnarray}
The multiple powers of $J^n(-i\,z_0) \sim \Lambda^n$ in (\ref{2-3-loop}) 
reflect the presence of overlapping divergencies in the two-loop vertex and three-loop 
bubble functions. They are easily disentangled by proper loop subtractions and 
as shown in appendix B. Note that our result (\ref{2-3-loop}) confirms the anomalous 
part of $J_{\pi\pi,S}(p)$ as anticipated in (\ref{all-order}). 
It is straightforward to derive analytic expressions for the
renormalized loop functions. Again we find it most economic to deduce the analytic 
form of the loop functions from their imaginary parts by means of their known analytic
structure. We drop $1/m$ correction terms and present the leading 
renormalized loop functions
\begin{eqnarray}
V_{\pi \pi,R }(p)&=&
\frac{m^2\,m^2_\pi}{(4\,\pi)^2 }
\,\frac{m_\pi^2}{4\,p^2} \,
\Bigg(\Li \left(\frac{1}{2}+\frac{i\,p}{m_\pi}\right)
-2\, \Li \left(\frac{1}{2}\,\frac{m_\pi}{m_\pi-i\,p}\right)
\nonumber\\
&+&\Li \left(\frac{1}{2}\right)
-\ln^2 \left(\frac{m_\pi-2\,i\,p}{2\,m_\pi-2\,i\,p }\right)
-\ln \left(\frac{1}{2}-\frac{i\,p}{m_\pi } \right)\,\ln \left( 2\right)  \Bigg)
\Bigg)\; ,
\nonumber\\
\bar J_{\pi \pi }(p )&=&-i\,
\frac{m^3\,m^3_\pi}{(4\,\pi)^3 }
\,\frac{m_\pi}{p} \,
\Bigg(
2\,\Li \left(\frac{1}{2}\,\frac{m_\pi}{m_\pi-i\,p}\right) 
-\frac{i\,p}{m_\pi}\,\ln \left(16 \right)
\nonumber\\
&&\;\;\;\;\;\;\;\;\;\;\;\;\;\;\;\;\;\;\;\;
+\ln^2 \left(\frac{m_\pi-2\,i\,p}{2\,m_\pi-2\,i\,p }\right)
-\frac{\pi^2}{6} \Bigg) \Bigg)+\bar J_{\pi \pi }(0)
\label{2-3-a-result}
\end{eqnarray}
with  $J_{\pi \pi,R}(p,z)=\bar J_{\pi \pi }(p)-\bar J_{\pi \pi }(-i\,z)$
\footnote{ Here we recall the 
definition of the polylog function 
$$
\Li(x) = \int_x^0 dt\,\frac{\ln (1-t)}{t} = \sum_{k=1}^\infty \frac{x^k}{k^2} \; . 
$$}.
Again we confirm the anticipated chiral powers of the loop functions with
$V_{\pi\pi ,S}(p) \sim Q^2 $ and  $J_{\pi\pi,S}(p)- J_{\pi\pi ,S}(-i\,z_0)\sim Q^3 $. 
The analytic results (\ref{2-3-a-result}) explicitly demonstrate the presence of 
a branch point at $p=-i\,m_\pi$ in the two-loop vertex function 
$V_{\pi \pi,R}(p) $ and three-loop bubble function $J_{\pi \pi ,R}(p,z_0) $. Note that 
the polylog function $\Li (x)$ has one branch point at $x=1$.
It remains the contribution  from the pion-nucleon box diagram to the s-wave 
scattering amplitude. The relevant combination 
\begin{eqnarray}
B_1(u,t) -\frac{u-t}{4\,m^2}\, B_4(u,t)-\frac{u+t}{4\,m^2}\,B_5(u,t)
&=& B(u,t) -V_{\pi,R}(p)+\frac{1}{4}\, J(p)
\label{}
\end{eqnarray}
is conveniently  expressed in terms of the reduced pion-nucleon box function 
$B(u,t)$, the renormalized vertex loop $V_{\pi,R}(p)$ and $J(p)$. 
We derive the analytic expression
\begin{eqnarray}
B(u,t)&=&\frac{1}{4\,\pi
}\,\frac{m\,m_\pi^4}{\sqrt{-t\,b(u,t)}}
\Bigg( i\,\arctanh \left(\sqrt{\frac{t(u+t)}{b(u,t)}}\,\right)
\nonumber\\
&+&\frac{1}{2}\,\arctan
\left(  \frac{\sqrt{-t\,b(u,t) }}{4\,m_\pi^3}
-\frac{\sqrt{-u-t}}{m_\pi } \left(1-\frac {t}{4\,m_\pi^2} \right) \right)
\nonumber\\
&+&\frac{1}{2}\,\arctan
\left(  \frac{\sqrt{-t\,b(u,t) }}{4\,m_\pi^3}
+\frac{\sqrt{-u-t}}{m_\pi } \left(1-\frac {t}{4\,m_\pi^2} \right) \right)
\Bigg)\;,
\nonumber\\
b(u,t) &=&4\,m_\pi^4
-4\,m_\pi^2 \left(u+t\right) + t \left( u+t\right) \; .
\label{reduced-box}
\end{eqnarray}
The result (\ref{reduced-box}) confirms the expected chiral order of the reduced 
diagram with $
B(u,t) \sim Q $. 

Finally we collect all terms introduced in (\ref{ampl5}) and (\ref{ampl6}) relevant 
for s-wave scattering according to (\ref{s-proj})
\begin{eqnarray}
T^{}_{[^1S_0]}(u,t)&=&2\,g_A^2\,\lambda^2\left(c-{\textstyle{1\over 2}}\right)
\left(1+\left(c-{\textstyle{1\over 2}}\right) J(p)\right)
+4\,g_A^4\,\lambda^4\,\left(c-{\textstyle{1\over 2}}\right)
V_{\pi ,R}(p) 
\nonumber\\
&-&\frac{1}{2}\, g_A^2\,\lambda^2\,\Bigg(\frac{m_\pi^2}{u-m_\pi^2}+\frac{m_\pi^2}{t-m_\pi^2} 
\Bigg)+g_A^4\,\lambda^4\,\Big(  B(u,t) + B(t,u)\Big)
\nonumber\\
&+&2\,
\frac{\Big(1+g_A^2\,\lambda^2\,V_{\pi,S}(p)\Big)^2+g_A^2\,\lambda^2\,V_{\pi \pi,S}(p)}
{g_{\sigma,S}^{-1}(s)-J(p)-g_A^2\,\lambda^2\,J_{\pi,S}(p)
-g_A^4\,\lambda^4\,J_{\pi \pi,S}(p)}
\label{tpis-one}
\end{eqnarray}
at chiral order $Q$. According to the discussion in section 2 we refrain from 
expanding the denominator in (\ref{tpis-one}). Expression (\ref{tpis-one}) serves 
as a convenient starting point for any further expansion implied by scenario 
I and II of section 2.

It is worth pointing out a generic property of the renormalized 
loop functions $V_{\pi^n,R}(p)$ and $J_{\pi^n,R}(p,z_0)$. Since they satisfy a 
dispersion relation in the Mandelstam variable 
$s$ they are in fact analytic in $p$ except for multiple branch points at $p=-i\,k\,m_\pi/2$ with
$0\geq n \geq k $ . This convenient property can be applied to derive the analytic form 
of the loop function in terms of its imaginary part (see appendix B). 
The s-wave scattering amplitude $T_{[^1S_0]}(p)$ shows also branch points on the upper complex 
half plane with $p=+i\,k\,m_\pi/2$. These branch points are, however, exclusively induced 
by the s-wave projected multiple pion exchanges of $T_{\pi, S}$. 
We perform the angle average of the one-pion exchange contribution and $B(u,t)$  
according to (\ref{s-def}) 
\begin{eqnarray}
R_\pi(p) &=& -\frac{1}{4}\,\int \frac{d\,\Omega}{4\pi }\, 
\Bigg(\frac{m_\pi^2}{u-m_\pi^2}+\frac{m_\pi^2}{t-m_\pi^2} \Bigg)
=\frac{m_\pi^2}{8\,p^2}\,\ln \Bigg(
1+\frac{4\,p^2}{m_\pi^2} \Bigg) \;,
\nonumber\\
R_{\pi \pi}(p) &=&
\int \frac{d\,\Omega}{4\pi }\, B(u,t)
\nonumber\\
&=&\frac{m\,m_\pi}{8\,\pi}\,\frac{m_\pi^3}{8\,p^3}\,\Bigg(
i\,\ln^2 \left(\frac{1}{2}+\frac{i\,p}{m_\pi}\right)
-2\,i\,\Li\left(\frac{1}{2}+\frac{i\,p}{m_\pi}\right)
\nonumber\\
&&\;\;\;\;\;\;-2\,i\,\Li\left(\frac{1}{2}\,\frac{m_\pi}{m_\pi+i\,p}\right)
+2\,i\,\Li\left(\frac{1}{2}\,\frac{m_\pi}{m_\pi-i\,p}\right)
\nonumber\\
&&\;\;\;\;\;\; +2\,\ln
\left(\frac{m_\pi^2+4\,p^2}{4\,m_\pi^2+4\,p^2}\right)
\Bigg( \arctan \left( \frac{2\,p}{m_\pi}\right)-\arctan \left( \frac{p}{m_\pi}\right)
\Bigg)
\nonumber\\
&&\;\;\;\;\;\; +2\,i\,\ln (2)\,\ln
\left(\frac{1}{4}+\frac{p^2}{m^2_\pi } \right)
+2\,i\,\ln^2(2)+i\,\frac{\pi^2}{6}
\Bigg) \;.
\label{Rpipi}
\end{eqnarray}
The expressions (\ref{Rpipi}) confirm that indeed the function $R_\pi(p)$ shows branch 
points at $p=\pm \,i\, m_\pi/2$ and $R_{\pi \pi}(p) $ branch points at $p=\pm \,i\,m_\pi$ 
and $p=\pm \,i\,m_\pi/2$ as expected.

We conclude that the chiral L-counting rule derived by means of the s-channel dispersion 
relation in section 2 is confirmed explicitly for the loop functions considered here.
Let us shed some light on the convergence property of the chiral expansion. A convenient 
measure is the ratio of the one-to-two loop and two-to-three loop subtraction constants:
\begin{eqnarray}
&& g_A^2\,\lambda^2\,\frac{J_{\pi,S}(0) }{J(0)}
\simeq g_A^2\,\lambda^2\,\frac{m\,m_\pi }{4\,\pi}\,\frac{m_\pi}{\Lambda }
\,\ln \left(\frac{\Lambda^2 }{m_\pi^2} \right) \; ,
\nonumber\\
&& g_A^2\,\lambda^2\,\frac{J_{\pi \pi,S}(0) }{J_{\pi ,S}(0) }
\simeq  \,g_A^2\,\lambda^2 \,\frac{m\,m_\pi}{4\,\pi}
\,\frac{\ln \left( 4 \right)}{\ln \left(\frac{\Lambda }{m_\pi}\right)}
\label{ratio-1}
\end{eqnarray}
at $c=\bar c$. The typical cutoff $\Lambda \simeq 600 $ MeV implies reasonable small ratios
of about $0.27 $ and $0.35$ for the leading and subleading ratio respectively. 
Finally consider also the ratio of leading to subleading vertex functions:
\begin{eqnarray}
g_A^2\,\lambda^2\,\frac{V_{\pi \pi,R}(0) }{V_{\pi ,R}(0) }
&=&g_A^2\,\lambda^2 \,\frac{m\,m_\pi}{4\,\pi} \ln \left(2 \right)
\simeq 0.28
\label{ratio-2}
\end{eqnarray}
evaluated at $p^2=0$. All together we find good convergence behavior of the chiral expansion. 
The underlying expansion parameter appears to be roughly $1/3$. 
Before presenting further details of our chiral expansion scheme we compare the analytic 
results with a calculation by Fogel\cite{Fogel} obtained in the static pion limit.

\subsection{Static pion limit}

It is instructive to compare our result (\ref{ampl5}) with a corresponding calculation
performed with static pion propagators. As an example we consider here the 
one-pion exchange contributions $R_\pi(p)$ and $R_{\pi \pi}(p)$ (see (\ref{Rpipi})). 
The limit of static pions is accessed most economically by considering the Schr\"odinger 
equation of a Yukawa potential
\begin{eqnarray}
\psi''(r)+\Big( p^2+g_\pi\,m_\pi\,\frac{e^{-m_\pi \,r}}{r} \Big) \,\psi (r) = 0 \;.
\label{schrodinger}
\end{eqnarray}
with the coupling constant $g_\pi = g_A^2\,m_\pi\,m/(16 \pi f_\pi^{\,2})\simeq 0.45$. 
This problem was solved analytically to lowest orders in the coupling constant by 
Fogel \cite{Fogel}.  The s-wave scattering phase shift $\delta_{[^1S_0]}(p) $ 
\begin{eqnarray}
\tan \delta_{[^1S_0]}(p) = \frac{\Im \,f(p)}{\Re \,f(p)}
\label{}
\end{eqnarray}
is expressed in terms of the Jost function, $f(p)$, which
is the solution of (\ref{schrodinger}) with asymptotic behavior $\sim e^{-i\,p\,r}$ evaluated
at $r=0$. In \cite{Fogel} the Jost function is evaluated perturbatively 
\begin{eqnarray}
f(p) &=& 1+ \sum_{n=1}^\infty \,g^n_\pi\,f_{n}(p)
\label{}
\end{eqnarray}
with analytic results given for leading and subleading terms:
\begin{eqnarray}
f_1(p) &=& \frac{i\,m_\pi}{2\,p}\,\ln \left(1+2\frac{i\,p}{m_\pi} \right)\; ,
\nonumber\\
f_2(p) &=& -\frac{m_\pi^2}{4\,p^2}\,\Bigg( 
\frac{1}{2}\,\ln^2 \left( 1+2\frac{i\,p}{m_\pi}\right)+\frac{\pi^2}{12}
-\Li \left(\frac{2\,i\,p-m_\pi}{2\,i\,p+m_\pi} \right)
\nonumber\\
&&\;\;\;\;\;\;\;\;\;\;+2\,\Li \left(\frac{-m_\pi}{2\,i\,p+m_\pi} \right)
\Bigg) \;.
\label{}
\end{eqnarray}
It is found \cite{Fogel,Friman} that for physical coupling $g_\pi \simeq 0.45 $ the 
perturbative expansion of the Jost function is rapidly converging for momenta 
$2\,p>g_\pi\, m_\pi  \simeq 70$ MeV. The convergence property is in fact improving 
significantly as $p$ gets larger. The next moment $f_3(p)$ was evaluated by Friman \cite{Friman}.

We confront our result with the analytic expressions of Fogel by matching the s-wave 
scattering amplitude, $T_{[^1S_0]}(p)$, at leading and subleading order in the coupling 
constant $g_\pi $:
\begin{eqnarray}
T_{[^1S_0]}(p) &=& \frac{8\pi}{\sqrt{s}\,p}\,\frac{\Im \,f(p)}{\Re \,f(p)-i\,\Im\,f(p)}
\label{}
\end{eqnarray}
We identify
\begin{eqnarray}
R_\pi(p)  && \leftrightarrow  \frac{m_\pi}{2\,p}\,\Im\,f_1(p)\; ,
\nonumber\\
R_{\pi \pi}(p) &&\leftrightarrow \frac{m\,m_\pi^2}{8\,\pi\,p}\,\Bigg(
\Im \,f_2(p)-\Im\,f_1(p)\,\Big( \Re\,f_1(p) -i\,\Im\,f_1(p)\Big)
\Bigg)
\label{}
\end{eqnarray}
and conclude that, as to be expected, the terms agree identically.  
For a discussion and evaluation of $1/m$ correction terms as they evolve from relativistic
Feynman diagrams see appendix B.1.

\subsection{Expansion of effective coupling function}

In this section we will further prepare the ground to generalize (\ref{two-scenarios}) 
for the case in which the  pion dynamics is included explicitly in the theory. Starting 
from (\ref{tpis-one}) we will specify how to identify and expand the renormalized coupling 
function.

The scattering amplitude (\ref{tpis-one}) involves the subtracted 
coupling function $g_{\sigma,S}=g-g_A^2\,\lambda^2\,c$. Within our renormalization procedure 
(\ref{ren-condition-1}) and (\ref{ren-condition-2}), which holds for 
both scenarios I) and II), the loop subtraction coefficient $\bar c$ is determined by 
enforcing finite vertex functions $V_{\pi^n ,S}(p)$. At leading orders 
this is achieved with $\bar c=1/2+{\mathcal O}(Q^2)$ corresponding to the 'local' part of 
the one-pion exchange potential (see (\ref{tpis-one})). The remaining divergencies sit in 
the bubble loop functions $J_{\pi^n ,S}(s)$ (see (\ref{loops-zero},\ref{2-3-loop})) and can 
be absorbed into the bare coupling function $g(p,m_\pi)$ by a subtraction procedure. To this 
end it is convenient to introduce the auxiliary coupling function $g_H(s,m_\pi; c ) $, the 
basic building block in (\ref{tpis-one}), 
\begin{eqnarray}
g_H^{-1}(s,m_\pi; c ) &&= g_{\sigma ,S}^{-1}(s,m_\pi;c)
-J(-i\,z_0)
\nonumber\\
&& - g_A^2\,\lambda^2\,J_{\pi,
S}(-i\,z_0 )-g_A^4\,\lambda^4\,J_{\pi \pi ,S}(-i\,z_0)+\cdots   \;,
\nonumber\\
g_R(p,m_\pi)&&=g_H\left(s,m_\pi; \bar c\right)
\label{gren}
\end{eqnarray}
which then defines the renormalized coupling function if evaluated at $c=\bar c$. 
In (\ref{gren}) we consider the bubble loops $J_{\pi^n ,S}(p)$ as a function of the 
momentum $p$. The auxiliary coupling function $g_H(s,m_\pi; c )$ 
depends on the subtraction $c$ via the coupling function $g_{\sigma ,S}$ and via the 
explicit dependence of the bubble loop functions $J_{\pi^n, S}(p)$. The subtraction point 
$z_0=z(m_\pi=0)$ is identified with the pseudo-bound state pole position in the chiral limit. 

The bare coupling function $g(s,m^2_\pi)$ in $g_H(s,m_\pi;c)$ permits a Taylor expansion in 
$p^2 $  and $m_\pi^2 $. In scenario I) the bare coupling function $g(s,m^2_\pi)$ and the 
renormalized coupling function $g_R(p,m_\pi)$ are expanded as follows:
\begin{eqnarray}
g(s,m^2_\pi) &=& \lambda^2 \,\sum_{n,k}\,
\beta^{(2k)}_{\;2n}\,\Big(p^{2}+z^2_0\Big)^n\, m_\pi^{2k} \;,
\nonumber\\
g^{-1}_{R,I}\big(p,m_\pi\big) 
&=&\frac{m}{2\,\pi} \,\sum_{n,k} \,\zeta^{(k)}_{\;2n}\,\Big( p^2+z_0^2 \Big)^n\,m_\pi^k \;.
\label{natural}
\end{eqnarray}
Note that in (\ref{natural}) we slightly reorganized the expansion of the bare 
coupling function as compared to (\ref{large-m-example}). The function $g(s,m_\pi)$ represents 
an infinite hierarchy  of interaction terms of the chiral Lagrangian with increasing number 
of derivative couplings and increasing degree of explicit chiral symmetry breaking. 
The naturalness assumption leads to $\beta_{\,n}^{(m)}\sim \Lambda_{nat.}^{-n-m}$ and 
$\Lambda_{nat.} =\Lambda^{(I)}_{nat.}\simeq m_\rho/2 $ at given cutoff 
$\Lambda \simeq \Lambda_{nat.}$. 

The renormalized coupling constants $\zeta^{(k)}_{\;2n}$ can be expressed in terms of 
the bare parameters $\beta^{(2k)}_{\;2n}$ and the subtraction coefficients 
$J_{\pi^n,S}(-i\,z_0)$. We expand the inverse auxiliary coupling function 
$g_H^{-1}(s,m_\pi ;c)$ around the point $c=\bar c$ (see (\ref{split-fin-div-2})).
The leading terms read:
\begin{eqnarray}
\frac{\lambda^2}{g(s,m_\pi)-c\,g_A^2\,\lambda^2 }&=& 
\frac{1}{\beta^{(0)}_0-g_A^2/2}
-\frac{\beta^{(2)}_0\,m_\pi^2+\beta^{(0)}_2\,\Big(p^2+z_0^2 \Big)
\,}{\Big(\beta^{(0)}_0-g_A^2/2\Big)^2} 
\nonumber\\
&+&  \frac{g_A^2\,\Big(c-\bar c\Big)}{\left(\beta^{(0)}_0-g_A^2/2\right)^2 } 
+ \frac{g_A^4\,\Big( c-\bar c\Big)^2}
{\left(\beta^{(0)}_0-g_A^2/2\right)^3} 
\nonumber\\
&-&2\,g_A^2\,\Big(c-\bar c\Big)\, 
\frac{\beta_0^{(2)}\,m_\pi^2+\beta_2^{(0)}\,\Big(p^2+z_0^2\Big)}{\Big( 
\beta^{(0)}_0-g_A^2/2\Big)^3} +{\mathcal O} \Big(Q^3 \Big)  \; .
\label{exp-subtr.-coup}
\end{eqnarray}
where we count $c-\bar c \sim Q$. The divergent subtraction coefficients can now be absorbed 
into the bare parameters $\beta_{2n}^{(2k)}$. The linear divergence in $ J(-i\,z_0)$ is 
absorbed into the bare coupling $\beta^{(0)}_0$ and the log-divergence in 
$\bar J_{\pi }(-i\,z_0)$ is absorbed into $\beta^{(0)}_2$ (see (\ref{subtr.-zero})). The 
leading order renormalized coupling constants $\zeta_0^{(2)}, \zeta_2^{(0)}$ and 
$\zeta_0^{(3)}$ can now be identified
\begin{eqnarray}
\lambda^2\,J (-i\,z_0) &=& 
\frac{1}{\Big(\beta^{(0)}_0-g_A^2/2\Big)} \;\; ,\;\;\;\;\;
\zeta^{(0)}_2 =- \frac{2\,\pi\,\beta^{(0)}_2 /\lambda^2}
{m\,\Big(\beta^{(0)}_0-g_A^2/2\Big)^2} \;,
\nonumber\\
\zeta_0^{(2)} &=&-g_A^2\,\lambda^2\,\frac{2\,\pi}{m} \,\frac{\bar 
J_{\pi }(-i\,z_0)}{m_\pi^2 }-\frac{2\,\pi\,\beta^{(2)}_0 /\lambda^2}
{m\,\Big(\beta^{(0)}_0-g_A^2/2\Big)^2}\;,
\nonumber\\
\zeta_0^{(3)}&=&- 
g_A^2\,\lambda^2\,\frac{2\,\pi}{m}\,\frac{\bar J_{\pi \pi}(-i\,z_0)}{m_\pi^3} \;.
\label{ren-bare}
\end{eqnarray}
Note that the renormalized coupling function $g_R(p,m_\pi)$ includes non-analytic terms 
in the squared pion mass (e.g. $m_\pi^3\,\zeta_0^{(3)}$) from the bubble loop subtraction
coefficients $J_{\pi^n,S}(-i\,z_0)$ not present in the bare coupling function
$g(s,m_\pi^2)$. Moreover we point out that the inverse auxiliary coupling function 
$g^{-1}_H(s,m_\pi;c)$ and the inverse renormalized coupling function $g_R^{-1}(p,m_\pi)$
agree up to chiral order $Q^2$:  
\begin{eqnarray}
g_H^{-1}(s,m_\pi;c) &=&\zeta^{(0)}_2\,\frac{m}{2\,\pi} \,\Big( p^2+z_0^2 
\Big)+ \frac{m}{2\,\pi}\, 
\zeta^{(2)}_0\,m_\pi^2  +{\mathcal O} \left(Q^3 \right) \; .
\label{ghq-help}
\end{eqnarray} 
This follows since (\ref{ren-bare}) implies that the power divergent terms proportional 
to $(c-\bar c)^n$ with $n=1,2$ in (\ref{exp-subtr.-coup}) are canceled by matching 
contributions from $J_{\pi,S}(-i\,z_0)$ and  $J_{\pi \pi,S}(-i\,z_0)$ 
(see (\ref{loops-zero},\ref{all-order})). For the generalization of (\ref{ghq-help}), which 
does show an explicit dependence on $c-\bar c$, valid up to chiral order $Q^3$ see 
(\ref{gren-sub}).

We turn to scenario II). Most economically it may be considered as a particular resummation 
of scenario I) with:
\begin{eqnarray}
g^{-1}_{R,II}\big(p,m_\pi\big) 
&=& \frac{m}{2\,\pi}\,\Bigg(\xi_\sigma \,
\frac{p^2+z_0^2}{p^2-\mu_\sigma^2}
+\xi_{2}^{(0)} \,\Big(p^2+z_0^2\Big)  
\nonumber\\
&&\;\;\;\;\;\;\;\;\;\;\;\;+\xi_{0}^{(2)}\,m_\pi^2+\xi^{(3)}_{0}\,m_\pi^3\Bigg)
+{\mathcal O} \left(Q^4 \right) \; ,
\nonumber\\
\sum_{k=0}^\infty \,\zeta^{(k)}_{2\,n}\,m_\pi^k\, 
&=& -\xi_\sigma \,\Big(\mu_\sigma^2+z^2_0\Big)^{-n}
+\sum_{k=0}^\infty\,\xi^{(k)}_{2\,n}\,m_\pi^k \;
\label{re-sum}
\end{eqnarray}
where $\mu_\sigma = \mu_\sigma(m_\pi )$ and $\xi_\sigma = \xi_\sigma(m_\pi )$.
The bare coupling function $g(s,m_\pi^2)$ and the renormalized inverse coupling functions 
$g^{-1}_{R,II}(p,m_\pi)$ of scenario II) are expanded around $p^2=\mu_\sigma^2$. 
At given order $(p^2-\mu_\sigma^2)^n$ the terms can then always be regrouped according 
to (\ref{re-sum}). Note that 
the form of the renormalized coupling functions in (\ref{natural},\ref{re-sum}) avoids a 
fine tuning of parameters which in (\ref{two-scenarios}) is required for the formation 
of the pseudo-bound state pole. 
The parameter $\mu_\sigma $ of scenario II) carries physical significance since it 
will describe the empirical fact that s-wave scattering at relative momentum 
$p\simeq 350 $ MeV is interaction free. This phenomena reflects the well established 
subtle cancellation of attractive $\sigma$-exchange and repulsive $\omega $-exchange 
in phenomenological nucleon-nucleon potentials \cite{Brown-Jackson,Reid68}. 
The expression (\ref{re-sum}) clearly demonstrate that as the value of 
$\mu_\sigma$ decreases the effective chiral power of the first term in $g_{R,II}^{-1}$ 
changes dramatically: for $\mu_\sigma \gg \Lambda_{nat.} $ 
we find  $Q^2$ whereas for $\mu_\sigma \ll \Lambda_{nat.} $ we are led to 
$Q^{0}$. Note however that at momenta $p^2 \ll \mu_\sigma^2$ the 'leading' term is 
suppressed by the factor $z_0^2/\mu_\sigma^2$ and therefore the 'subleading' term, in 
particular the $\xi_0^{(2)}$-term, becomes equally important. Recall that we expect
$\mu_\sigma \sim 350$ MeV and $z_0 \sim z \sim 8 $ MeV. Thus an expansion in ratios
of 'subleading' over 'leading' term is converging for momenta $p^2\sim \mu_\sigma^2$ but 
turns useless for momenta $p^2\ll \mu_\sigma^2$.
Our expansion for $g^{-1}_{R,II}$, in contrast, is expected to converge well for momenta 
$\mu_\sigma^2- \Lambda^{(II)}_{nat.}< p^2< \mu_\sigma^2+\Lambda^{(II)}_{nat.}$ with 
$\Lambda^{(II)}_{nat.} \simeq 600$ MeV (see section 2.2).
This is further support for our claim that $T_{\sigma ,S}$ must not be 
expanded in powers of $T_{\pi ,S}$ (see (\ref{split-3})). 

\subsection{Inefficient chiral expansion (CE) scheme}

In this section we develop the scenario I) in the presence of perturbative pion dynamics. 
The denominator of (\ref{ampl6}) is systematically expanded in chiral powers. We call such an 
approach 'CE' scheme for future reference. The motivation of this section is twofold. On the 
one hand we show the failure of the CE scheme on the other hand we present our
renormalization scheme at work. We will systematically express the scattering amplitude 
at subleading orders in terms of the physical parameters $z_0$, $\zeta_k^{(m)}$ introduced 
in the previous section (see (\ref{z-exp},\ref{ren-bare})) rather than in terms 
of the cutoff and bare coupling parameters. Therewith manifest scheme 
independent results are obtained. In particular we will demonstrate the independence of the 
scattering amplitude on the subtraction function c(s) in a perturbative 
sense: if the scattering amplitude is expanded up to chiral order $Q^n$ with
$c-\bar c \sim Q$ any residual c-dependence contributes at order 
$Q^{n+1}$. The perturbative cancellation of any $c$-dependence demonstrates that physical 
results do not depend on how to distribute strength among $K_{\pi ,S}$ and $K_{\sigma ,S}$ 
in (\ref{split-1},\ref{split-2}). For the full solution of 
the Bethe-Salpeter equation this is of course true by construction. For 
the truncated scattering amplitude this is, however, a non-trivial constraint. 

According to (\ref{ghq-help}) the auxiliary coupling function $g_H^{-1}(s,m_\pi;c)$ carries 
leading chiral order two. This leads to an s-wave scattering amplitude of leading chiral 
order minus one: 
\begin{eqnarray}
T_{[^1S_0]}(p) 
=-\frac{2}{J_R(p,z_0)}+{\mathcal O} \left(Q^0 \right)
=\frac{4\,\pi}{m}\,\frac{i}{p+i\,z_0}
+{\mathcal O} \left(Q^0 \right) \; .
\label{leading-order}
\end{eqnarray}
The result (\ref{leading-order}) is certainly not a 
surprise. One recovers the effective range theory with $r=0$. 
The pole in (\ref{leading-order}) at $p= -i\,z_0$ is the regularization 
scheme dependent result of fine-tuned bare coupling strength and 
intrinsic cutoff parameter. So far the chiral dynamics shows no quantitative predictive power. 
First non-trivial consequence are expected at subleading order. 
Note that the leading order result does not yet reveal the necessity to expand 
around $c=\bar c$ with $\bar c=1/2+{\mathcal O}(Q^2)$. This can first be seen at 
subleading order. 

We proceed and turn to the zeroth order amplitude. At this order the auxiliary 
coupling function $g_H(s,m_\pi;c)$ can be identified with the renormalized coupling 
function $g_R(p,m_\pi)$ (see (\ref{ghq-help})). We collect all terms of chiral order zero:
\begin{eqnarray}
T_{[^1S_0]}(p) &=&-\frac{2}{J_R(p,z)}+2\,g_A^2\,\lambda^2\,R_\pi(p)
-2\,\frac{g_R^{-1}(p,m_\pi)}{J_R^2(p,z_0)} 
\nonumber\\
&+& g_A^2\,\lambda^2\,\Bigg(1 
-4\,\frac{V_{\pi,R}(p)}{J_{R}(p,z_0)}+2\,\frac{J_{\pi ,R}(p,z_0)}{J^2_R(p,z_0)}
\Bigg)+{\mathcal O} \left(Q \right) \;.
\label{result2}
\end{eqnarray}
The result (\ref{result2}) is expressed in terms of the physical parameters in 
$g_R(p,m_\pi)$ and a set of renormalized loop functions given in
(\ref{1-2-a-result},\ref{Rpipi}). We point out that in (\ref{result2}) the explicit 
c-dependence induced by the loop functions $V_{\pi,S}(p) $ and 
$J_{\pi,S}(p)$ (see (\ref{loops-zero})) cancels 
identically upon expansion. Therefore the zeroth order 
scattering amplitude (\ref{result2}) is manifestly independent of the subtraction procedure. 
From the analysis of the effective range theory we expect the scattering amplitude to exhibit 
a single pole at $p=-i\,z$ rather than at $p=-i\,z_0$. We conjecture an 
expansion of the pseudo-bound state pole position $z(m_\pi)$ in powers of the pion mass: 
\begin{eqnarray}
z(m_\pi) = z_0 + \sum_{m \,=\,2}^{\infty } 
\,\bar \zeta_{m}\,m_\pi^m
\label{z-exp}
\end{eqnarray} 
with $\bar \zeta_{m} \sim \Lambda_{nat.}^{-m+1}$ and $z_0 \sim m_\pi \sim  Q$. 
Therefore an expansion scheme which preserves the proper low energy structure of 
the scattering amplitude (\ref{low-rep}) requires a particular summation. In order to comply 
with (\ref{low-rep}) we expand all bare poles in terms of powers of the physical pole term 
and then impose the cancellation of all multiple poles at $p = -i\,z$. The expansion of the 
leading order pole reads:  
\begin{eqnarray}
-\frac{2}{J_R(p,z_0)}
&=&-\frac{2}{J_R(p,z)}+ 2\,\bar \zeta_2\,
\frac{m}{2\,\pi}\,\frac{m_\pi^2\,}{J^2_R(p,z)}
-\Big(\bar \zeta_2 \Big)^2\,\frac{m^2}{2\,\pi^2}\,\frac{m_\pi^4}{J^3_R(p,z)} 
\nonumber\\
&+&2\,\bar \zeta_3\,
\frac{m}{2\,\pi}\,\frac{m_\pi^3\,}{J^2_R(p,z)} + \cdots 
\label{exp-pole}
\end{eqnarray}
where we apply our notation (\ref{z-exp}). At subleading order this leads to 
$\bar \zeta_2=\zeta^{(2)}_0$.

We now turn to the comparison of our result (\ref{result2}) with
the effective range theory. It is instructive to make the analytic 
structure of our result more explicit. We rewrite
(\ref{result2}) and present the central result of this section
in terms of the reduced functions $R^{(1)}_{V,\pi}$ and $R^{(2)}_{J,\pi}$ and the one-pion 
exchange contribution $R_\pi(p)$ (see (\ref{Rpipi})):
\begin{eqnarray}
T_{[^1S_0]}(p)&=&
\frac{4\,\pi}{m}\,\frac{i\,w}{p+i\,z}
+\frac{4\,\pi}{m}\,\zeta^{(0)}_2
\nonumber\\
&+&2\,g_A^2\,\lambda^2\,\Big( R_\pi(p)-2\,R^{(1)}_{V,\pi}(p,z_0)
+R^{(2)}_{J,\pi}(p,z_0)
\Big)+{\mathcal O} \left(Q\right)\;.
\label{result3}
\end{eqnarray}
Here we introduced a convenient notation: for any given loop function $F(p,z)$ we define
the n-th order residual function, $R^{(n)}_{F}(p,z)$, by:
\begin{eqnarray}
R^{(n)}_{F}(p,z) &=& J^{-n}_R(p,z)\,\Big( F(p,z)-\sum_{k=0}^{n-1}\,\frac{1}{k \,!}\,
\big(p+i\,z\big)^k\,\partial^k_p\,F(-i\,z,z) \Big)
\label{def-res}
\end{eqnarray}
Note that for a given loop function $F(p,z)$ which is regular at $p=-i\,z $ also the set of 
residual functions $R^{(n)}_{F}(p,z)$ is regular at $p=-i\,z$ by construction. 
The pole residuum $w$ is:
\begin{eqnarray}
w&=&1+g_A^2\, \lambda^2\,
\Big( 2\,V_{\pi ,R}(-i\,z_0)
-\frac{2\pi}{m}\,\frac{\partial}{\partial \,p}\,J_{\pi,R}(-i\,z_0,z_0)\Big)
\nonumber\\
&-&2\,\zeta_2^{(0)}\,z_0 +{\mathcal O} \left(Q^2\right) \;.
\label{}
\end{eqnarray}
With (\ref{result3}) we arrive at a representation of the scattering amplitude
anticipated in the previous section when discussing the low energy
limit. The amplitude is the sum of the pseudo-bound state pole and
a smooth non-polynomial remainder, which takes care of the
non-trivial pion dynamics. 

\begin{figure}[t]
\epsfysize=10.0cm
\begin{center}
\mbox{\epsfbox{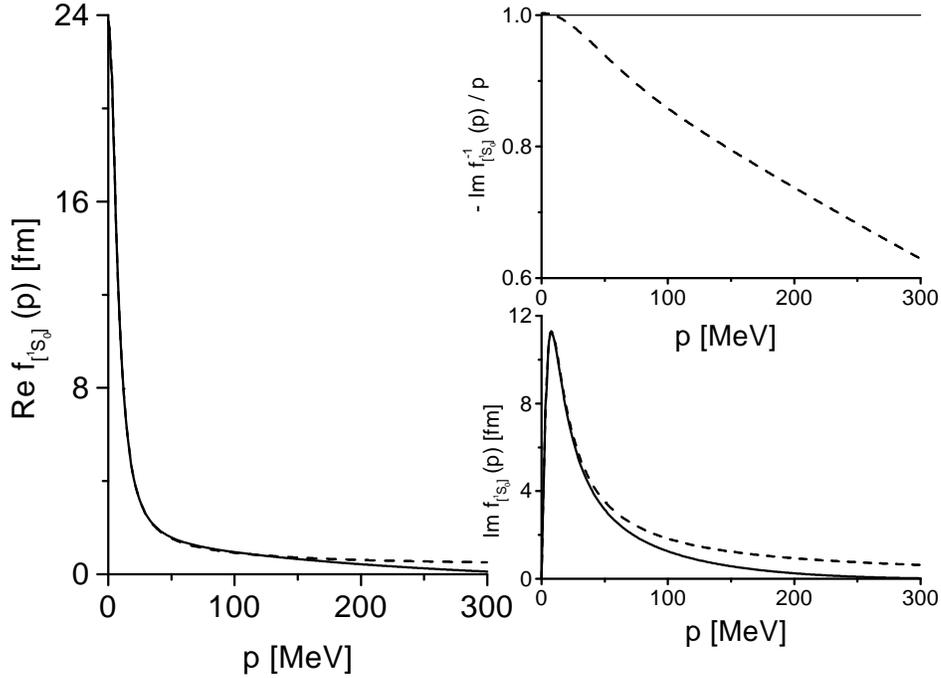}}
\end{center}
\caption{The left hand figure shows the real part of the np-scattering amplitude
$f_{[^1S_0]}(p)=\sqrt{s}\,T_{[^1S_0]}(p)/(8\,\pi)$. The right 
hand figure shows the imaginary part of the np-scattering amplitude
$\Im \,f_{[^1S_0]}(p)$ and the unitarity function $-\Im f_{[^1S_0]}^{-1}(p)/p$, which is one if unitarity 
holds. The solid lines are from the PWA93 solution of 
\cite{Nim} and the dashed lines are the result of the zeroth order CE scheme  with
$\zeta_2^{(0)}= 0.380$ fm, $z_0=8.35$ MeV and $z=7.89$ MeV in (\ref{result3}).}
\label{fig1}
\end{figure}

In Fig. 1) we show the result of a low energy fit to the empirical scattering amplitude. 
For momenta $p< 50$ MeV the scattering amplitude is well reproduced. It is also shown 
that the unitarity condition is violated strongly at momenta $p< 150 $ MeV. This is 
a disappointing result since already the effective range approximation works quite 
accurately up to momenta $p \simeq 150 $ MeV. We conclude that it would be misleading to 
confront the CE result (\ref{result3}) with the empirical scattering phase shift. 

It is instructive to analyze  our result in more detail and apply 
a further numerical expansion to the pole residuum $w$
and the effective scattering length $b_{\mathrm{eff}}^{(0)}$:
\begin{eqnarray}
1-w &=&
 \frac{m\,m_\pi}{4\,\pi}\,g_A^{2}\,\lambda^{2}
\left(2\,\frac{z_0}{m_\pi}+\frac{16}{3}\frac{z^2_0}{m^2_\pi}
+{\mathcal O}\left(\frac{z_0^3}{m_\pi^3}\right)\right)
+2\,\zeta^{(0)}_2 \,z_0 +{\mathcal O}\left( Q^2\right)\; ,
\nonumber\\
b_{\mathrm{eff}}^{(0)} &=&-\zeta^{(0)}_2
-\frac{m}{4\,\pi}\,g_A^2\,\lambda^2
\left(1+\frac{8}{3}\,\frac{z_0}{m_\pi}+\frac{8\,z^2_0}{m^2_\pi}
+{\mathcal O}\left(\frac{z_0^3}{m_\pi^3}\right)
\right)+{\mathcal O}\left( Q\right)
\label{}
\end{eqnarray}

The free parameter $ \zeta^{(0)}_2$ and $z_0 $ are now adjusted to reproduce the 
empirical residuum $w \simeq 0.90 $ and the effective scattering length $ 
b_{\mathrm{eff}}^{(0)} \simeq -1.15$ fm (see (\ref{aeff-def})). We obtain the values 
$\zeta_2^{(0)} \simeq 0.38 $ fm and
$z_0 \simeq 8.35 $ MeV. The scheme dependent bare parameters $\beta_0^{(0)}$,
$\beta_2^{(0)}$ and $\beta_0^{(2)}$ follow with (\ref{ren-bare}). The remaining 
physical parameter $\zeta_0^{(2)}$ follows with $
z = z_0 +\zeta_0^{(2)}\,m_\pi^2+\cdots $ 
which leads to $\zeta_0^{(2)} \simeq -0.0047 $ fm. It is important to inspect also the 
next moment
\begin{eqnarray}
b_{\mathrm{eff}}^{(1)} &=& -\frac{z_0\,m}{\pi\,m^2_\pi}\,g_A^2\,\lambda^2
\left(1+\frac{16}{5}\,\frac{z_0}{m_\pi}
 +{\mathcal O}\left(\frac{z_0^2}{m_\pi^2}\right) \right) +{\mathcal O}\left( Q\right)
\label{result5}
\end{eqnarray}
and compare it with the result from effective range theory (see (\ref{aeff-def})). We conclude
that (\ref{result5}) predicts the wrong sign of this term and therewith confirm our result 
that (\ref{result3}) is in strong conflict with the unitarity condition
\begin{eqnarray}
\Im \, T^{-1}_{[^1S_0]}(p) &=& -\frac{1}{2}\,\Im \, J_{R}(p,z_0)\,.
\label{unitarity-condition}
\end{eqnarray}

\begin{figure}[t]
\epsfysize=10.0cm
\begin{center}
\mbox{\epsfbox{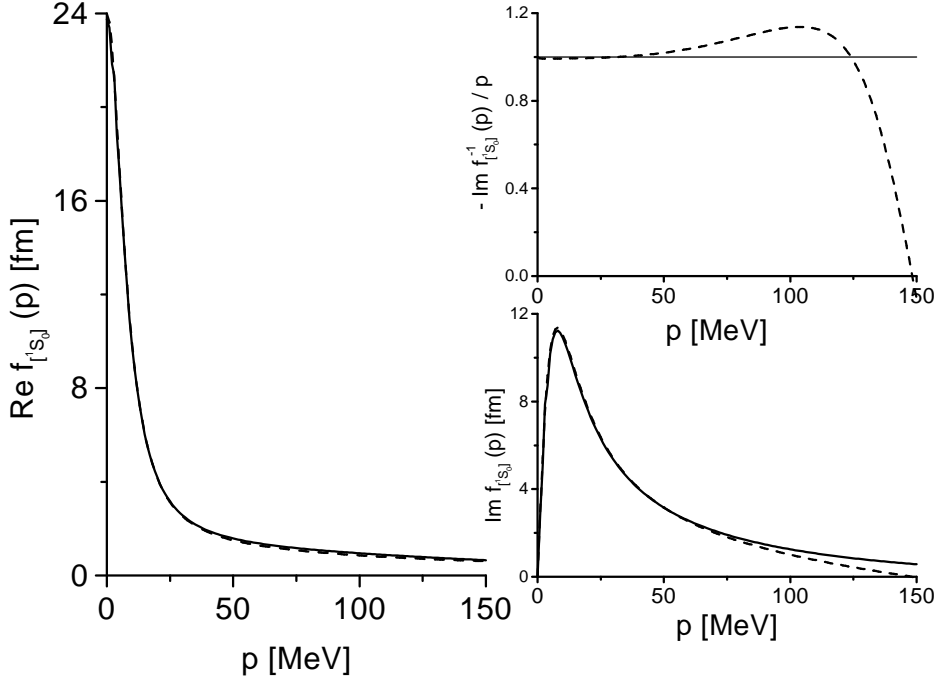}}
\end{center}
\caption{The left hand figure shows real part of the np-scattering amplitude
$f_{[^1S_0]}(p)=\sqrt{s}\,T_{[^1S_0]}(p)/(8\,\pi)$. The right 
hand figure shows imaginary part of the np-scattering amplitude
$\Im \,f_{[^1S_0]}(p)$ and the unitarity function $-\Im f_{[^1S_0]}^{-1}(p)/p$, which is one 
if unitarity holds. The solid lines are from the PWA93 solution of 
\cite{Nim} and the dashed lines are the result of the first order CE scheme  with
$\zeta_2^{(0)}= -1.363$ fm, $\zeta_0^{(2)}=1.771$ fm and $z_0=-45.68$ MeV (see (C.8)).}
\label{fig2}
\end{figure}

In appendix C the reader may find details of the NNLO terms in the CE scheme. In Fig 2 we 
present the result of a low energy fit. The description of the scattering amplitude is 
somewhat improved for $p< 100 $ MeV as compared to the zeroth order result. In particular the 
unitarity condition is met to better accuracy. However the first order calculation clearly 
signals the breakdown of the CE scheme at momenta $p\simeq 150 $ MeV. 
To be explicit: at the chiral orders we are working here (see (\ref{ampl5})) the strength 
of the effective pole at $p=i\,\bar z$ (see (\ref{barz-pole})) must be represented by the 
pion-nucleon box function $R_{\pi \pi }(p)$ (see (\ref{Rpipi})) since it is the only function 
showing a branch  point in the upper complex half plane at $p=i\,m_\pi \sim   i\,\bar z $. 
Recall here that all vertex functions and bubble functions have branch cuts only on the 
lower complex half plane. Since the strength of the cut in $R_{\pi \pi }(p)$ is much 
too weak to represent the strength of the effective pole at $p=i\,\bar z$ the CE scheme 
fails for $p\ll \bar z \simeq 150$ MeV. 
Note also that the parameters $\zeta_0^{(0)}$ and $z_0$ of the zeroth order and first 
order calculation are strongly renormalized. This points at the poor convergence properties of 
the CE scheme even at low momenta $p<150$ MeV. This could have been  anticipated by 
inspecting the ratio of the renormalized functions
\begin{eqnarray}
g_A^2\,\lambda^2\,\frac{J_{\pi, R  }(0,z)}{J_R(0,z)} &&=
g_A^2\,\lambda^2\,\frac{m\,m_\pi }{2\,\pi}\,
+{\mathcal O}\left( z\right)
\simeq 0.9
\label{bad-1}
\end{eqnarray}
at $p=0 $. From our calculation we conclude that in fact we have to change scheme and 
reformulate our effective field theory along the lines already anticipated.  
We emphasize that the result (\ref{bad-1}) does not imply that successive unitary 
iterations of $K_{\pi,S}$ 
need to be summed to all orders. As shown before here
typical ratios are well behaved (see (\ref{ratio-1},\ref{ratio-2})). The 
ratio (\ref{bad-1}) is to be taken as further evidence that the properly renormalized 
denominator (\ref{ampl6}) must not be expanded. 

We close this section by a comment on the so called KSW scheme proposed by
Kaplan, Savage and Wise \cite{KSW}. In the KSW scheme the scattering 
phase $\delta_{[^1S_0]}(p)$ is expanded perturbatively in the pion dynamics  
imposing the correct pseudo-bound state structure. It followed from our CE scheme 
upon a further expansion. For independent calculations in the KSW scheme at subleading 
orders see \cite{Mehennnlo}. In our work we insist on the
CE scheme since the unitarity condition (\ref{unitarity-condition}) provides a 
powerful consistency check for the CE scheme which is obscured in the KSW scheme. If 
the CE scheme showed good convergence properties it would be equivalent to the KSW scheme. 
The rather poor convergence properties of the CE scheme necessarily implies that also the KSW
scheme is poorly convergent. We conclude therefore that neither the CE nor the 
KSW scheme provides an improvement upon the effective range theory. Furthermore 
note that by expanding the phase shift directly rather than the amplitude makes it more 
difficult to detect a potential breakdown of the scheme. This is clearly 
demonstrated by the effective range theory. Here the second low momentum pole 
at $p =i\,\bar z$ (see (\ref{low-rep},\ref{barz-pole})) restricts 
the applicability of the CE and the KSW scheme to momenta $p< |\bar z | $. The breakdown 
is triggered by a 'weak' cut singularity in the KSW scheme as compared to a 'strong' pole 
singularity in the CE scheme:
\begin{eqnarray}
\delta_{[^1S_0]} (p) = \frac{\pi}{2}
-\arctan  \left( -\frac{1}{a\,p}+\frac{1}{2}\,r\,p \right)\; .
\label{branch}
\end{eqnarray}
In the KSW scheme the branch cut of the $\arctan $ function at $\pm \,i$ leads to the 
convergence bound $p<| \bar z |=|2/r|+{\mathcal O}(1/a)$. We conclude that 
in order to uncover the physical significance of non-perturbative strength 
at $p\simeq i\,\bar z$  it is advantageous to choose the CE scheme.

\subsection{Unitary chiral expansion (UCE) scheme }

In the previous sections we collected mounting evidence that the denominator in 
(\ref{split-3}) or (\ref{tpis-one}) must not be expanded further. We have demonstrated that 
the CE scheme, based on such an expansion, is not applicable in the spin singlet channel 
due to poor convergence properties. In this central section of this work our  
new scheme already outlined in section 2.4 is constructed explicitly. It shows much 
improved convergence properties and comprises unitarity exactly. Formally the scheme 
arises if the renormalized coupling, $g^{-1}_R(p,m_\pi)$, changes according to (\ref{re-sum}) 
its chiral power from plus two at 
very small momenta $p\ll \mu_\sigma$ to zero at intermediate small momenta $p\sim \mu_\sigma$. 
As a consequence the denominator in (\ref{tpis-one}) must not be expanded if the effective 
field theory is to be applied for momenta $p <\mu_\sigma+\Lambda_{nat.}$. Keeping 
the denominator on the one hand takes care of the proper chiral power of the renormalized 
coupling at intermediate small momenta $p\sim \mu_\sigma $ and on the other hand induces a 
particular resummation for the low momentum regime which greatly improves its convergence 
property. We refer to this approach as the 'UCE'-scheme since, as we will elaborate on in 
detail below, it in addition leads to exact unitarity. 

The UCE scheme should be viewed as a systematic summation of the CE 
scheme (see (\ref{re-sum})). The renormalization procedure is identical 
to the renormalization procedure of the CE scheme: all vertex functions 
$V_{\pi^n,S}(p)$ are renormalized by appropriate 'loop' subtractions induced 
by a properly constructed local subtraction kernel $\bar K_S$. We emphasize again that this
procedure leads to renormalization scale independent results for the nth order 
vertex functions $V_{\pi^n,R}(p)$ and the nth order box functions $R_{\pi^n}(p)$. 
The remaining divergencies of the nth order bubble functions $J_{\pi^n,S}(p)$ are 
renormalized by a subtraction procedure at $p=-i\, z_0$. Thus the  
renormalized bubble functions $J_{\pi^n,R}(p,z_0)$ show an explicit dependence on 
the subtraction point $z_0$ or more generally they depend on some renormalization scale. 
At strictly $K_S=\bar K_S$ we obtain all together renormalization scale independent results 
since the renormalization scale can be absorbed into the renormalized coupling function 
$g_R(p,m_\pi)$. As advocated in section 2.4 it is advantageous to exploit the 
freedom to not strictly require $K_S=\bar K_S$. For the CE scheme we have 
established that the scattering amplitude evaluated to a given chiral order is independent on 
$(K_S-\bar K_S)^n$ if we count $K_S-\bar K_S \sim Q$. In the UCE scheme this does not hold 
anymore since it is an infinite resummation of the CE scheme. The residual dependence on
$(K_S-\bar K_S)^n$ is fixed as to implement unitarity exactly at given order in the 
expansion. In order not to spoil the renormalization procedure the residual dependence 
on $(K_S-\bar K_S)^n$ 'acts' exclusively on finite parts as defined in (\ref{split-fin-div-1})
and (\ref{split-fin-div-2}). This amounts to dropping all divergent 
terms proportional to $(c-1/2)\,J(-i\,z_0)$ in the loop functions (\ref{loops-zero}).
More explicitly we arrive at:
\begin{eqnarray}
T_{[^1S_0]}(p) &=&2\,g_A^2\,\lambda^2\,\left(R_\pi(p) + c-{\textstyle{1\over2}}\right)
\nonumber\\
&+&\frac{2+4\,g_A^2\,\lambda^2\,\bar V_{\pi,S}(p,z_0)}
{g_R^{-1}(p,m_\pi)-J_R(p,z_0)-g_A^2\,\lambda^2\,\bar J_{\pi,S}(p,z_0)}
\label{bar-t-zero}
\end{eqnarray}
with the renormalized loop functions
\begin{eqnarray}
\bar V_{\pi,S}(p,z_0) &=&
\left(c-{\textstyle{1\over2}}\right) J_R(p,z_0)
+V_{\pi,R}(p )\; , 
\nonumber\\
\bar J_{\pi,S}(p,z_0) &=&  
\left(c-{\textstyle{1\over2}}\right)J^2_R(p,z_0)+J_{\pi,R}(p,z_0) \;.
\label{bar-loops-zero}
\end{eqnarray}
The renormalized coupling $g_R(p,m_\pi)$ is given by (\ref{re-sum}) and the 
loop functions in (\ref{1-2-a-result},\ref{Rpipi}). Note that according to our
definition (\ref{split-fin-div-2}) and (\ref{exp-subtr.-coup}) the finite 'part' of the 
auxiliary coupling function $g_H(s,m_\pi;c(s))$ of (\ref{gren}) must be identified with the 
renormalized coupling $g_R(p,m_\pi)$. The function $c(s)$ is now unambiguously determined by 
the unitarity constraint. According to (\ref{unitarity-construct}) the function $c(s)$, 
representing a separable interaction, is constructed such that the s-wave and on-shell 
projection of $T_{\pi,S} $ vanishes at the given order
\begin{eqnarray}
\langle T_{\pi,S} \rangle 
&=&2\,g_A^2\,\lambda^2\,\left(R_\pi(p) + c-{\textstyle{1\over2}}\right)
=0+{\mathcal O} \left( Q\right) \;.
\label{construct-c-1}
\end{eqnarray}
We find that the unique choice satisfying (\ref{construct-c-1})
\begin{eqnarray}
c(s)= \frac{1}{2}-R_\pi(p)+{\mathcal O} \left( Q\right)
\label{choice-1}
\end{eqnarray}
renders the scattering amplitude (\ref{bar-t-zero}) unitary. Note that the subtraction 
function $c(s)-1/2$ now carries chiral order zero. This does not contradict the L-counting 
rule because the subtraction $c(s)$ is now only active on convergent parts of the loops. 
We emphasize that the construction (\ref{choice-1}) holds irrespective of the form of 
the renormalized coupling $g_R(p,m_\pi)$. 

The mechanism which leads to a unitary scattering amplitude is readily understood. 
The vertex loop function $V_{\pi, R}(p)$, the bubble function 
$J_{\pi, R}(p,z_0)$, and the one-pion exchange contribution $R_\pi(p)$ are not independent 
entities. They are interrelated by causality. Cutkosky's  cutting rules \cite{Cutkosky} 
lead to the identities
\begin{eqnarray}
R_{\pi}(p)\,\Im \, J_{R}(p,z_0) &=& \Im \, V_{\pi,R }(p)\; ,
\nonumber\\
2\,\Big(\Re \, V_{\pi ,R}(p)\Big)\,\Big(\Im \, J_{R}(p,z_0)\Big)
&=& \Im \, J_{\pi , R}(p,z_0) 
\label{causal-1}
\end{eqnarray}
which are consistent with our analytic results (\ref{1-2-a-result},\ref{Rpipi}) and 
directly confirm the unitarity condition (\ref{unitarity-condition}). As a consequence 
the phase shift can now be identified unambiguously
\begin{eqnarray}
p\,\cot \delta_{[^1S_0]}(p) &=&\frac{4\,\pi}{\sqrt{s}}\,
\frac{\bar g_R^{-1}(p,m_\pi)-g_A^2\,\lambda^2 \,\Re\, \bar J_{\pi,S}(p,z_0)}
{1+2\,g_A^2\,\lambda^2 \,\Re \,\bar V_{\pi ,S}(p,z_0 ) }
\label{uce-zero}
\end{eqnarray}
where we introduced
\begin{eqnarray}
\bar g_R^{-1}(p,m_\pi) &=& g_R^{-1}(p,m_\pi)-\Re\, J_{R}(p,z_0)
\label{}
\end{eqnarray}
for notational convenience. 

We point out that any further expansion of (\ref{uce-zero}) is necessarily in conflict 
with unitarity (or even worse with the complicated low energy structure of the 
scattering amplitude) since it would invalidate the derivation of our result (\ref{uce-zero}).
Finally it is instructive to consider the limit of a vanishing 
renormalized coupling function: $g_R(p,m_\pi)=0$ leads to a zero phase shift $\delta_{[^1S_0]}(p)=0$. 
This demonstrates that the renormalized coupling function $g_R$ does not only represent the 
physics of short range repulsion ($\omega-$exchange) and intermediate range attraction 
($\sigma-$exchange) it also encompasses some residual pion dynamics. The important achievement 
lies in the fact that $g_R$ does not pick up pion induced singularities. In other words $g_R$ 
does not exhibit branch points at $p= \pm \,n\,i\,m_\pi/2$
with $n=1,2,3,...$ and therefore $g_R$ is expected to respect a Taylor expansion with the 
typical scale set by a rather large $\Lambda_{nat.}\sim 600$ MeV.

\subsection{NNLO terms of UCE scheme }

We turn to the next terms in our expansion. Again all divergent terms in the 
loop functions (\ref{loops-zero},\ref{2-3-loop}) are dropped according to 
the argument given in the previous section. We derive
\begin{eqnarray}
T_{[^1S_0]}(p) &=&2\,g_A^2\,\lambda^2\,\left(R_\pi(p) + c-{\textstyle{1\over2}}\right)
+2\,g_A^4\,\lambda^4\,\left(c-{\textstyle{1\over 2}}\right)^2 J_R(p,z_0)
\nonumber\\
&+&4\,g_A^4\,\lambda^4\,\left(c-{\textstyle{1\over 2}}\right) 
V_{\pi,R }(p) +2\,g_A^4\,\lambda^4\,R_{\pi \pi }(p)
\nonumber\\
&+&\frac{2\,\Big(1+g_A^2\,\lambda^2\,\bar V_{\pi,S}(p,z_0)\Big)^2+4\,g_A^4\,\lambda^4\,\bar V_{\pi \pi,S}(p,z_0)}
{g_R^{-1}(p)-J_R(p,z_0)-g_A^2\,\lambda^2\,\bar J_{\pi,S}(p,z_0)-g_A^4\,\lambda^4\,\bar J_{\pi \pi,S}(p,z_0)}
\label{bar-t-one}
\end{eqnarray}
with the renormalized loop functions
\begin{eqnarray}
\bar V_{\pi\pi,S}(p,z_0) &=& V_{\pi \pi ,R}(p)
+\left(c-{\textstyle{1\over 2}}\right)^2 J^2_R(p,z_0)
\nonumber\\
&+&\left(c-{\textstyle{1\over 2}}\right)
\left(J_R(p,z_0)\,V_{\pi ,R}(p)+J_{\pi ,R}(p,z_0)
\right) \; ,
\nonumber\\
\bar J_{\pi \pi,S}(p,z_0) &=&
J_{\pi \pi,R }(p,z_0)+\left(c-{\textstyle{1\over 2}}\right)^2 J^3_R(p,z_0)
\nonumber\\
&+&2 \left(c-{\textstyle{1\over 2}}\right) J_R(p,z_0)\,J_{\pi,R }(p,z_0) \; .
\label{bar-loops-one}
\end{eqnarray}
Before constructing the subtraction function $c(s)$ at subleading order we 
collect the relevant causality identities predicted by Cutkosky's \cite{Cutkosky} cutting 
rules:
\begin{eqnarray}
\Im \,R_{\pi \pi}(p) &=&R_\pi^{\,2}(p)\,\Im\,J_R(p,z_0)\; ,
\nonumber\\
\Im \, V_{\pi \pi,R}(p) &=&\Big( \Re \,V_{\pi,R}(p)\Big) \,
\Big(\Im \,V_{\pi ,R}(p)\Big)
+\Big(\Re\,R_{\pi \pi }(p)\Big)\,\Big( \Im \,J_R(p,z_0)\Big)\; ,
\nonumber\\
\Im \, J_{\pi \pi,R }(p,z_0)  &=& 
2\,\Big(\Re \, V_{\pi \pi,R }(p)\Big)\,\Big(\Im \, J_{R}(p,z_0)\Big)
+\Big| V_{\pi ,R}(p)\Big|^2\,\Im \,J_R(p,z_0) \; .
\label{causal-2}
\end{eqnarray}
Again the identities (\ref{causal-2}) are respected by our analytic results 
(\ref{2-3-a-result},\ref{Rpipi}). The subleading term of the subtraction function $c(s)$
can now be constructed by the requirement
\begin{eqnarray}
\langle T_{\pi,S} \rangle 
&=&2\,g_A^2\,\lambda^2\,\left(R_\pi(p) + c-{\textstyle{1\over2}}\right)
+2\,g_A^4\,\lambda^4\,\left(c-{\textstyle{1\over 2}}\right)^2 J_R(p,z_0)
\nonumber\\
&+&4\,g_A^4\,\lambda^4\,\left(c-{\textstyle{1\over 2}}\right) 
V_{\pi ,R }(p) +2\,g_A^4\,\lambda^4\,R_{\pi \pi }(p)
\nonumber\\
&=&0+{\mathcal O} \left( Q^2\right)
\label{construct-c-2}
\end{eqnarray}
that the s-wave and on-shell projection of $T_{\pi,S}$ vanishes to the given 
order (\ref{unitarity-construct}). 
The solution of (\ref{construct-c-2}) is now readily found with
\begin{eqnarray}
c(s)&=& \frac{1}{2}-R_\pi(p)-g_A^2\,\lambda^2\,\Re \,R_{\pi \pi}(p)
-g_A^2\,\lambda^2\,R_\pi^2(p)\,\Re \,J_R(p,z_0)
\nonumber\\
&+&2\,g_A^2\,\lambda^2 \,R_\pi(p)\,\Re \,V_{\pi,R }(p) 
+{\mathcal O} \left( Q^2\right) \; .
\label{choice-2}
\end{eqnarray}
Explicit calculation  indeed confirms that with (\ref{choice-2}) the s-wave scattering 
amplitude (\ref{bar-t-one}) satisfies unitarity. The phase shift can be identified unambiguously 
and we finally arrive at the central result of our work:
\begin{eqnarray}
p\,\cot \delta_{[^1S_0]}(p) &=&\frac{4\,\pi}{\sqrt{s}}\,
\frac{\bar g_R^{-1}(p)-g_A^2\,\lambda^2 \,\Re\, \bar J_{\pi,S}(p,z_0)
-g_A^4\,\lambda^4 \,\Re\, \bar J_{\pi \pi,S}(p,z_0)}
{\Re\,\Big( 1+g_A^2\,\lambda^2 \,\bar V_{\pi ,S}(p,z_0 ) \Big)^2
+2\,g_A^4\,\lambda^4 \,\Re \,\bar V_{\pi \pi,S}(p,z_0 ) }\;.
\nonumber\\
\label{phase-order-1}
\end{eqnarray}
Note that in (\ref{phase-order-1}) the numerator has to be evaluated 
at chiral order three and the denominator at chiral order two.
In  other words in $\bar J_{\pi ,S}(p)$ the subtraction function $c(s)$ has to be included to 
chiral order one as given by (\ref{choice-2}) whereas in $\bar J_{\pi \pi ,S}(p)$ the subtraction 
function $c(s)$ is to be included to chiral order zero as given by (\ref{choice-1}).

\section{Results and discussion}

We confront  our unitary chiral expansion (UCE) with the partial wave analysis (PWA93) of 
the Nijmegen group \cite{Nim}. Our result at order $Q^0$ is given in 
(\ref{bar-loops-zero},\ref{choice-1},\ref{uce-zero}) and 
at order $Q^1$ in (\ref{bar-loops-one},\ref{choice-2},\ref{phase-order-1}). At leading 
chiral orders we encounter the renormalized coupling function
\begin{eqnarray}
g_R^{-1}(p,m_\pi) &=&
\frac{m}{2\,\pi}\,\left(
\xi_\sigma\,\frac{p^2+z_0^2}{p^2-\mu_\sigma^2}
+\xi_2^{(0)} \,\Big(p^2+z_0^2\Big) 
+m_\pi^2\,\xi_0^{(2)}\right)
+{\mathcal O} \left(Q^4 \right) 
\label{ren-coupling2-recall}
\end{eqnarray} 
characterized by five free parameters 
$\xi_\sigma, \xi_2^{(0)}, \xi_0^{(2)},  \mu_\sigma$ and $z_0$. Here we absorb the finite 
contribution $m_\pi^3\,\xi_0^{(3)}$ into $m_\pi^2\,\xi_0^{(2)}$ for convenience. 
Our parameters are adjusted to reproduce the $^1S_0$-partial wave in the 
pn-channel. With $g_\pi =g^2_A\,m\,m_\pi/(16 \pi \,f^2_\pi) = 0.451$, 
$m_\pi=139.0$ MeV and $m=938.9$ MeV we perform a least square fit to the single 
energy $^1S_0$ partial wave  as given in \cite{Nim} with finite error bars. Tab. 1 
presents the resulting parameters together with the $\chi^2$-value of the fit. We observe that 
the $\chi^2$ of the UCE result at order $Q^1$ is significantly improved upon the $\chi^2$ of 
the UCE result at order $Q^0$ although there is no additional free parameter involved. The
parameters are only weakly renormalized indicating good convergence properties and 
consistency with the naturalness assumption (see (\ref{nat-sat})). 
Moreover we confirm that the inclusion of pionic effects leads to a much better description
of the partial wave. The $\chi^2$ is reduced by a factor of 5 from $25$ to $5$. 
The collum in Tab. 1 labeled 'no pions' refers to our calculation with $g_A=0$.

\begin{table}[h,b]
\begin{tabular}{|r||c|c|c|c|c|} \hline
 $$  &  no pions  &UCE-$Q^0$ &UCE-$Q^1$\\ \hline \hline 

$ \chi^2_{\rm PWA93}$   & 25.3 &  12.9 & 5.3  \\ \hline

$ a(^1S_0) [{\rm fm}]$   & -23.33&  -23.48 & -23.59   \\ \hline 

$ r(^1S_0) [{\rm fm}]$   & 2.54 &  2.60 & 2.64  \\ \hline

$\xi_2^{(0)} [{\rm fm}]$   & 0.391 &  0.451 & 0.451  \\ \hline

$\xi_0^{(2)} [{\rm fm}]$   & 0 &  0.6964 & 0.7490  \\ \hline

$\xi_\sigma [{\rm MeV}]$   & -558.9 &  -672.5 & -724.1  \\ \hline

$\mu_\sigma [{\rm MeV}]$   & 354.1 &  353.8 & 353.5  \\ \hline

$z_0 [{\rm MeV}]$   & 8.04&  -45.84 & -37.04  \\ \hline

\end{tabular}
\caption{Set of parameters obtained by fitting the $^1S_0$(np) phase shift of PWA93.} 
\label{tab1}
\end{table}

\begin{table}[t]
\begin{tabular}{|r||c|c|c|c|c|c|} \hline
 $E_{\rm lab} \big[MeV \big]$  & PWA93&  no pions  &UCE-$Q^0$ &UCE-$Q^1$& Bonn& Nijm93\\ \hline \hline 

$ 1$   & 62.07$\pm $ 0.03 &  62.06 & 62.06 & 62.06 &  62.12 & 62.07\\ \hline

$ 5$   & 63.63$\pm $ 0.08 &  63.86 & 63.79 & 63.73 &  63.67 & 63.62\\ \hline 

$10$   & 59.96$\pm $ 0.11 &  60.03 & 60.04 & 60.02 &  59.91 & 59.94 \\ \hline

$25$   & 50.90$\pm $ 0.19 &  50.46 & 50.63 & 50.75 &  50.82 & 50.86\\ \hline

$50$   & 40.54$\pm $ 0.28 &  39.83 & 40.02 & 40.20 &  40.44 & 40.38\\ \hline

$100$  & 26.78$\pm $ 0.38 &  26.43 & 26.45 & 26.53 &  26.27 & 26.17 \\ \hline

$150$  & 16.94$\pm $ 0.41 &  17.08 & 16.98 & 16.96 &  16.17 & 15.71 \\ \hline

$200$  & 8.94$\pm $ 0.39 &  9.36  & 9.24   &  9.15 &   8.14 & 7.07 \\ \hline

$250$  & 1.96$\pm $ 0.37 & 2.35 &  2.27    &  2.19 &   1.40 & -0.45\\ \hline

$300$  & -4.46$\pm $ 0.43 & -4.45 & -4.42 &  -4.44  & -4.45 & -7.18\\ \hline

$350$  & -10.59$\pm $ 0.62& -11.30& -11.13& -11.02  & -9.65 & -13.32\\ \hline

\end{tabular}
\caption{We confront our result (\ref{uce-zero}) and (\ref{phase-order-1}) with the 
PWA93 solution for the $^1S_0(np)$ partial wave.
The $\chi^2$ of (\ref{uce-zero}) with $12.9$ and of (\ref{phase-order-1}) with $5.7$ is 
significantly improved as compared to the $\chi^2$ of $25.3$ which 
follows if pion effects are switched off (i.e. $g_A=0, m_\pi=0$). 
The set of parameters is given in Tab. 1. The $\chi^2$ are $37.0$ for Nijm93 and $12.2$ for 
the Bonn potential. Note that here we omitted the first point at $E_{\rm lab}=1 $ MeV since 
its inclusion would somewhat artificially worsen the $\chi^2$.} 
\label{tab2}
\end{table}

We emphasize that the parameter $z_0$ has a physical interpretation. It represents the 
pseudo-bound state pole position in the chiral limit with $m_\pi =0 $. The chiral limit leads 
in fact to a real bound state at $ p\simeq  i\, 37 $ MeV with binding energy
$\epsilon \simeq z_0^2/m \simeq 1.5$ MeV rather than a pseudo-bound state  obtained with
$m_\pi >0$. Our result demonstrates that the effective range parameter, $r(^1S_0)$, defined in 
(\ref{eff-range}), is driven by the physical scale $\mu_\sigma $ rather than the pion mass 
as one might expect naively. In the language of the $\sigma-, \omega-$meson exchange picture 
we would say that the effective range is mainly determined by the subtle interplay  of 
intermediate attraction and short range repulsion  rather than by the pion exchange.
The parameter $\mu_\sigma $ determines the zero of the phase shift 
and $\zeta_\sigma $ 
\begin{eqnarray}
\delta_{[^1S_0]}(p) \simeq \frac{p^2-\mu_\sigma^2}{\xi_\sigma\,\mu_\sigma}
\label{disc-zero}
\end{eqnarray} 
the behavior close to $p\simeq \mu_\sigma $. In (\ref{disc-zero}) we omitted pion induced
terms for simplicity. We point out that the dynamic chiral power of our renormalized coupling 
function $g_R$ in (\ref{ren-coupling2-recall}) with $g^{-1}_R \sim Q^{2} $ for 
$p\ll \mu_\sigma $ but $g^{-1}_R \sim Q^{-2} $ for $p \simeq \mu_\sigma $ prohibits any 
further expansion of the scattering amplitude beyond (\ref{bar-t-one}) if the theory is to be 
applied for momenta $0< p< m_\omega $.
Our phase shifts are given in Tab. 2. where for the reader's convenience we also 
included  results of two modern potential models as given in \cite{Stoks,Arndt}. 
The latter potentials reproduce the empirical phase shift rather well, where the
Bonn potential shows a somewhat better $\chi^2$ as compared to the 
Nijm93 potential. We conclude that our chiral description of the 
$^1S_0$-phase shift does compete in quality with modern nucleon-nucleon potentials.

\begin{figure}[t]
\epsfysize=10.0cm
\begin{center}
\mbox{\epsfbox{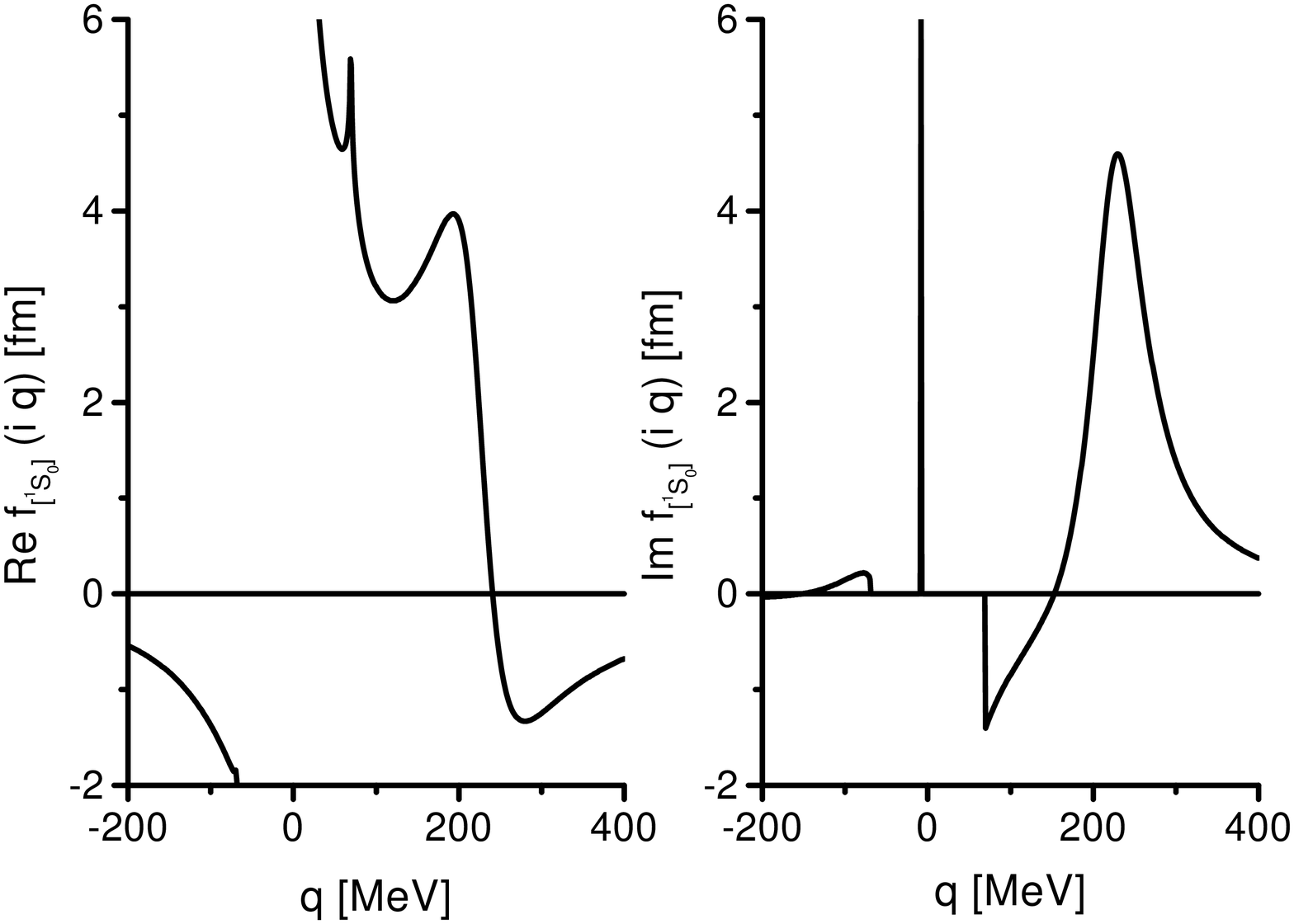}}
\end{center}
\caption{The $^1S_0$-partial wave amplitude $f_{[^1S_0]}(i\,q) =\sqrt{s}\, 
T_{[^1S_0]}(i\,q)/(8\pi)$ of (\ref{bar-t-one}) evaluated with the UCE-$Q^1$ parameters 
as given in Tab. 1.}
\label{fig3}
\end{figure}

It is instructive to discuss the singularity structure of our scattering amplitude 
(\ref{bar-t-one}). 
In Fig. 3 we show the scattering amplitude $f_{[^1S_0]}(i\,q)$ evaluated below threshold 
with $p= i\,q+\epsilon $. The pseudo-bound state pole at $q\simeq -8 $ MeV and 
the anomalous thresholds at $q= \pm \,m_\pi/2 $ are clearly seen. For unphysical negative 
$q\ll -8$ MeV  (2nd Rieman sheet) the amplitude is rather 
smooth whereas for positive $q$ (1st Rieman sheet) the amplitude shows a rather complex 
structure anticipated in our discussion of section 2.3. The figure shows a well defined 
peak structure in the imaginary part of the amplitude and a corresponding zero 
of the real part at $p\simeq i\,225$ MeV, which was roughly predicted by the effective range 
theory (see (\ref{barz-pole})). From an analyticity point of view it is now obvious that a successful 
effective field theory description of the $^1S_0$-partial wave amplitude requires the 
incorporation of this singularity. This explains the failure of the CE scheme 
(see section 3.3) which includes the pseudo-bound state pole but misses to describe 
the complete singularity structure seen in Fig. 3.

In Fig. 4 we show how our UCE-$Q^1$ fit to the $^1S_0$ phase shift extrapolates to higher 
energies. Here we confront our result with Arndt's SP99 solution \cite{Arndt}. We observe 
that in fact our UCE scheme appears better compatible with the PWA93 solution of the Nijmegen 
group \cite{Stoks}. Note that the zeroth order (dashed line) and 
first order (solid line) results give very similar predictions for the phase shift for a given 
set of parameters confirming good convergence properties of our UCE scheme. The main effect of 
the correction terms sit at low momenta $p < 200$ MeV. Most dramatic is the modification at 
low momenta if the pion effects are switched off with $g_A=0$ as demonstrated with the dotted 
line in Fig. 4. Note that of course a set of renormalized parameters leads to a better 
description of the phase shift even with $g_A=0$ as is shown in Tab. 2. Our full result  
(solid line) starts to deviate significantly from  the SP99 solution for momenta $p> 450$ 
somewhat above the pion production threshold at $p\simeq 370$ MeV. 

\begin{figure}[t]
\epsfysize=10.0cm
\begin{center}
\mbox{\epsfbox{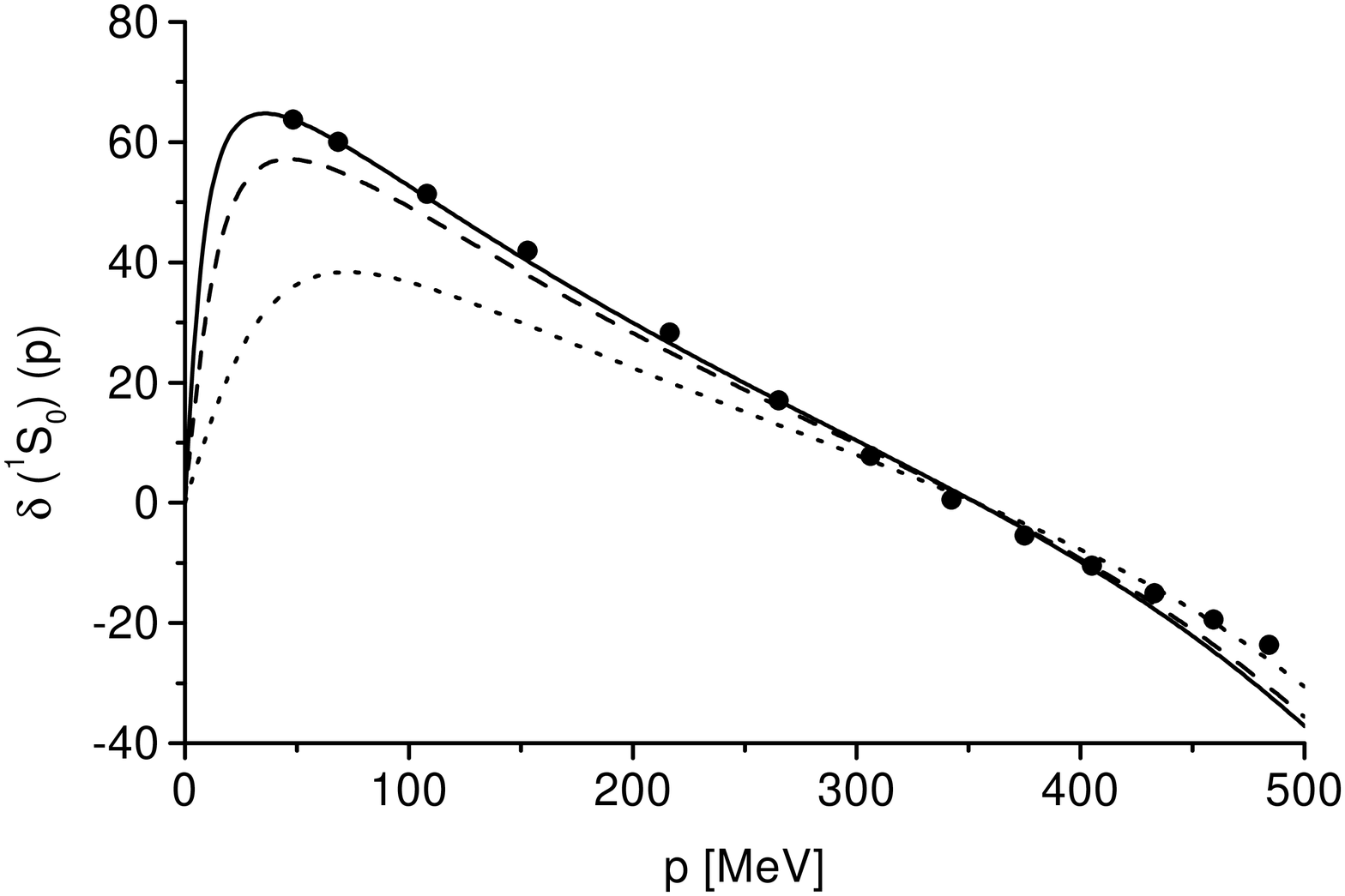}}
\end{center}
\caption{The $^1S_0$-phase shift evaluated with the UCE parameters at order $Q^1$ as given in 
Tab. 1. The solid line is the full result (UCE-$Q^1$) obtained from the fit to the PWA93 
solution. The dashed line is the zero order result of the UCE scheme and the dotted line 
follows from the UCE scheme with $g_A=0$ (no pions). The dots represent the SP99 analysis 
of Arndt \cite{Arndt}.}
\label{fig4}
\end{figure}

We conclude our work with a short explorative discussion of the next terms in our expansion.
At order $Q^2$ one needs to include the irreducible two-pion exchange diagrams in the 
interaction kernel $K_\pi$  studied in \cite{bira1,bira2,brockmann}:
\begin{eqnarray}
K_{\pi,S} &&= 
\parbox{15mm}{
\begin{fmfgraph}(15,10)
\fmfleftn{l}{2}
\fmfrightn{r}{2}
\fmf{fermion}{r2,o,l2}
\fmf{fermion}{r1,o,l1}
\fmfv{d.sh=c,d.filled=0,d.size=3thick}{o}
\end{fmfgraph}}+
\parbox{15mm}{
\begin{fmfgraph*}(15,10)
\fmfleftn{l}{2}
\fmfrightn{r}{2}
\fmf{fermion}{r2,u,l2}
\fmf{fermion}{r1,d,l1}
\fmfforce{(0.0w,0.0h)}{l1}
\fmfforce{(0.5w,0.0h)}{d}
\fmfforce{(1.0w,0.0h)}{r1}
\fmfforce{(0.0w,1.0h)}{l2}
\fmfforce{(0.5w,1.0h)}{u}
\fmfforce{(1.0w,1.0h)}{r2}
\fmf{boson}{d,u}
\fmfdot{u,d}
\end{fmfgraph*}}
\nonumber\\ \nonumber\\ \nonumber\\
&&+
\parbox{20mm}{\begin{fmfgraph*}(20,10)
\fmfleftn{l}{2}
\fmfrightn{r}{2}
\fmf{fermion}{r1,o1,l1}
\fmf{fermion}{r2,o2,l2}
\fmfforce{(0.5w,0.0h)}{o1}
\fmfforce{(0.5w,1.0h)}{o2}
\fmf{boson,left=0.7,tension=0.5}{o1,o2}
\fmf{boson,right=0.7,tension=0.5}{o1,o2}
\fmfv{d.shape=d,d.fill=1,d.size=3thick}{o1,o2}
\end{fmfgraph*}}
+\parbox{20mm}{
\begin{fmfgraph*}(20,10)
\fmfleftn{l}{2}
\fmfrightn{r}{2}
\fmf{fermion}{r1,o11,o12,l1}
\fmf{fermion}{r2,o2,l2}
\fmfforce{(0.33w,0.0h)}{o12}
\fmfforce{(0.66w,0.0h)}{o11}
\fmfforce{(0.5w,1.0h)}{o2}
\fmf{boson}{o11,o2}
\fmf{boson}{o12,o2}
\fmfdot{o11,o12}
\fmfv{d.shape=d,d.fill=1,d.size=3thick}{o2}
\end{fmfgraph*}}
+\parbox{20mm}{
\begin{fmfgraph*}(20,10)
\fmfleftn{l}{2}
\fmfrightn{r}{2}
\fmf{fermion}{r1,o2,l1}
\fmf{fermion}{r2,o11,o12,l2}
\fmfforce{(0.33w,1.0h)}{o12}
\fmfforce{(0.66w,1.0h)}{o11}
\fmfforce{(0.5w,0.0h)}{o2}
\fmf{boson}{o11,o2}
\fmf{boson}{o12,o2}
\fmfdot{o11,o12}
\fmfv{d.shape=d,d.fill=1,d.size=3thick}{o2}
\end{fmfgraph*}}
+
\parbox{20mm}{
\begin{fmfgraph*}(20,10)
\fmfleftn{l}{2}
\fmfrightn{r}{2}
\fmf{fermion}{r2,u2,u1,l2}
\fmf{fermion}{r1,d2,d1,l1}
\fmfforce{(0.0w,0.0h)}{l1}
\fmfforce{(0.30w,0.0h)}{d1}
\fmfforce{(0.70w,0.0h)}{d2}
\fmfforce{(1.0w,0.0h)}{r1}
\fmfforce{(0.0w,1.0h)}{l2}
\fmfforce{(0.30w,1.0h)}{u1}
\fmfforce{(0.70w,1.0h)}{u2}
\fmfforce{(1.0w,1.0h)}{r2}
\fmf{boson}{d2,u1}
\fmf{boson}{d1,u2}
\fmfdot{u1,u2,d1,d2}
\end{fmfgraph*}} \;.
\end{eqnarray}
The scattering amplitude $T_{\pi,S}$ is to be evaluated up to order $Q^2$ applying the L-counting 
rule. This amounts at most a thrice iterated kernel $K_{\pi, S}$. Moreover $1/m$ correction 
terms need to be included. Also the interaction $K_{\sigma ,S}$ receives a contribution
\begin{eqnarray}
K_{\sigma,S} &&=g_{\sigma,S}(s)\,\,\,\,\; 
\parbox{20mm}{
\begin{fmfgraph}(20,15)
\fmfleftn{l}{2}
\fmfrightn{r}{2}
\fmf{fermion}{r2,o,l2}
\fmf{fermion}{r1,o,l1}
\fmfv{d.sh=c,d.filled=1,d.size=3thick}{o}
\end{fmfgraph}}
\nonumber\\&&+
\parbox{20mm}{
\begin{fmfgraph}(20,15)
\fmfleftn{l}{2}
\fmfrightn{r}{2}
\fmfforce{(0.5w,0.5h)}{o}
\fmf{fermion}{r2,or,o,l2}
\fmf{fermion}{r1,o,l1}
\fmfv{d.sh=square,d.filled=1,d.size=3thick}{o}
\fmffreeze
\fmf{boson,left=.7,tension=.5}{or,o}
\fmfdot{or}
\end{fmfgraph}}
+
\parbox{20mm}{
\begin{fmfgraph}(20,15)
\fmfleftn{l}{2}
\fmfrightn{r}{2}
\fmfforce{(0.5w,0.5h)}{o}
\fmf{fermion}{r2,o,or,l2}
\fmf{fermion}{r1,o,l1}
\fmfv{d.sh=square,d.filled=1,d.size=3thick}{o}
\fmffreeze
\fmf{boson,left=.7,tension=.5}{or,o}
\fmfdot{or}
\end{fmfgraph}}
+
\parbox{20mm}{
\begin{fmfgraph}(20,15)
\fmfleftn{l}{2}
\fmfrightn{r}{2}
\fmfforce{(0.5w,0.5h)}{o}
\fmf{fermion}{r2,o,l2}
\fmf{fermion}{r1,or,o,l1}
\fmfv{d.sh=square,d.filled=1,d.size=3thick}{o}
\fmffreeze
\fmf{boson,left=.7,tension=.5}{or,o}
\fmfdot{or}
\end{fmfgraph}}
+
\parbox{20mm}{
\begin{fmfgraph}(20,15)
\fmfleftn{l}{2}
\fmfrightn{r}{2}
\fmfforce{(0.5w,0.5h)}{o}
\fmf{fermion}{r2,o,l2}
\fmf{fermion}{r1,o,or,l1}
\fmfv{d.sh=square,d.filled=1,d.size=3thick}{o}
\fmffreeze
\fmf{boson,left=.7,tension=.5}{or,o}
\fmfdot{or}
\end{fmfgraph}}
\end{eqnarray}
from a new chiral vertex
\begin{eqnarray}
{\mathcal L} &=& \frac{1}{4}\,g^{(\pi )} \,\lambda^4\,\Big( \bar N \,\gamma_5\,C\,\vec \tau \,\tau_2\,\bar N^t \Big)
\, \Big(N^t\,C^{-1}\,\tau_2\, \Big(\partialslash \,\vec \pi \Big) \,N\Big)
+h.c.
\label{pi-prod-vertex}
\end{eqnarray}
part of the chiral Lagrangian. The pion production parameter $g^{(\pi )} $ is 
unconstrained by chiral symmetry and may be estimated by meson-exchange phenomenology.
We point out that the interaction $K_{\sigma ,S}$ receives only 2-nucleon irreducible 
contributions which are also irreducible with respect to pion lines. This is a  
crucial element of our UCE scheme which only then leads to a renormalized inverse coupling 
function $g^{-1}_R(p,m_\pi)$ free of any pionic cut structure. Moreover the two-nucleon propagator 
$G$ in (\ref{bs1},\ref{bs1b}) is corrected by the leading nucleon self energy loop diagrams:
\begin{eqnarray}
G &=& \parbox{20mm}{
\begin{fmfgraph*}(20,10)
\fmfleftn{l}{2}
\fmfrightn{r}{2}
\fmf{fermion}{r1,l1}
\fmf{fermion}{r2,l2}
\fmfforce{(0.33w,0.0h)}{o12}
\fmfforce{(0.66w,0.0h)}{o11}
\end{fmfgraph*}}
+\parbox{20mm}{
\begin{fmfgraph*}(20,10)
\fmfleftn{l}{2}
\fmfrightn{r}{2}
\fmf{fermion}{r1,o11,o12,l1}
\fmf{fermion}{r2,l2}
\fmfforce{(0.33w,0.0h)}{o12}
\fmfforce{(0.66w,0.0h)}{o11}
\fmf{boson,right=.9,tension=.5}{o11,o12}
\fmfdot{o11,o12}
\end{fmfgraph*}}
+\parbox{20mm}{
\begin{fmfgraph*}(20,10)
\fmfleftn{l}{2}
\fmfrightn{r}{2}
\fmf{fermion}{r1,l1}
\fmf{fermion}{r2,o11,o12,l2}
\fmfforce{(0.33w,1.0h)}{o12}
\fmfforce{(0.66w,1.0h)}{o11}
\fmfforce{(0.5w,0.0h)}{o2}
\fmf{boson,left=.9,tension=.5}{o11,o12}
\fmfdot{o11,o12}
\end{fmfgraph*}}\; .
\label{}
\end{eqnarray}
The important new element of the order $Q^2$ calculation are the pion production cuts of 
loop functions. They give rise to a complex phase shift for $p>\mu_\pi $ with the pion 
production scale $\mu^2_\pi =m\,m_\pi+m_\pi^2/4$. We estimate the result of the UCE-Q$^2$
calculation by adding  all pion production contributions of order $Q^2$ to our UCE-Q$^1$ 
result (\ref{bar-t-one}). At the given order there are three different sources for pion production.
First there is the pion production cut of the two-loop bubble diagram 
$J_{\pi, R}^{(\pi)}(p,z_0)$ (see (\ref{sbubbledef})) with
\begin{eqnarray}
J_{\pi, R}^{(\pi)}(p,z_0)&&= \frac{1}{g_A^2\,\lambda^2}\,\left.\parbox{30mm}{
\begin{fmfgraph*}(30,15)
\fmfleftn{l}{2}
\fmfrightn{r}{2}
\fmf{fermion}{ol,l2}
\fmf{fermion}{ol,l1}
\fmf{fermion}{r1,or}
\fmf{fermion}{r2,or}
\fmf{phantom,left=.7,tension=.5,tag=1}{or,ol}
\fmf{phantom,left=.7,tension=.5,tag=2}{ol,or}
\fmfv{d.sh=c,d.filled=1,d.size=4thick}{ol}
\fmfv{d.sh=c,d.filled=1,d.size=4thick}{or}
\fmfposition
\fmfipath{p[]}
\fmfiset{p1}{vpath1(__or,__ol)}
\fmfiset{p2}{vpath2(__ol,__or)}
\fmfi{fermion}{subpath (0,length(p1)/2) of p1}
\fmfi{fermion}{subpath (length(p1)/2,length(p1)) of p1}
\fmfi{fermion}{subpath (length(p2),length(p2)/2) of p2}
\fmfi{fermion}{subpath (length(p2)/2,0) of p2}
\fmfi{boson}{point length(p1)/2 of p1 -- point length(p2)/2 of p2}
\fmfiv{d.sh=c,d.filled=1,d.size=2thick}{point length(p1)/2 of p1}
\fmfiv{d.sh=c,d.filled=1,d.size=2thick}{point length(p2)/2 of p2}
\end{fmfgraph*}} \right|_{\pi-\rm production}
\nonumber\\
&&=\frac{m_\pi^4}{(2\pi)^4}\, \Bigg( 
-\frac{\mu_\pi^2}{p^2}+\frac{(p^2-\mu_\pi^2)^2}{p^4}\,
\ln \Bigg|\frac{\mu_\pi^2}{\mu_\pi^2-p^2} \Bigg|
\nonumber\\
&&-\frac{\mu_\pi^2}{z_0^2}-\frac{(z_0^2+\mu_\pi^2)^2}{z_0^4}\,
\ln \Bigg|\frac{\mu_\pi^2}{\mu_\pi^2+z_0^2} \Bigg|
\nonumber\\
&&+\,i\,\pi\,\Theta (p-\mu_\pi )\,\frac{(p^2-\mu_\pi^2)^2}{p^4}\,
\Bigg)+{\mathcal O}\left(Q^5 \right)
\label{pi-prod-two}
\end{eqnarray}
We wish to emphasize that the chiral expansion of the pion production terms is not performed 
applying the heavy baryon technique since it would destroy the proper pion production threshold. 
Rather we expand the spectral density in powers of $(p^2-\mu_\pi^2)^n$ identifying 
$\mu_\pi \sim Q$ as a further characteristic small scale. This procedure keeps the correct 
pion production threshold. Here we encounter a clear advantage of the 
relativistic chiral Lagrangian over its heavy baryon representation. 

We turn to the second pion production term. It results from the nucleon self energy correction
of the one-loop bubble
\begin{eqnarray}
J_{\Sigma ,R}^{(\pi)}(p,z_0)&&=\frac{1}{g_A^2\,\lambda^2} \left. \left(
\parbox{35mm}{
\begin{fmfgraph*}(35,15)
\fmfleftn{l}{2}
\fmfrightn{r}{2}
\fmf{fermion}{ol,l2}
\fmf{fermion}{ol,l1}
\fmf{fermion}{r1,or}
\fmf{fermion}{r2,or}
\fmfforce{(0.25w,0.5h)}{ol}
\fmfforce{(0.75w,0.5h)}{or}
\fmfforce{(0.35w,0.2h)}{pl}
\fmfforce{(0.65w,0.2h)}{pr}
\fmf{fermion,right=.7,tension=.5}{or,ol}
\fmf{fermion}{or,pr}
\fmf{fermion,left=.3,tension=.7}{pr,pl}
\fmf{boson,right=.7,tension=.5}{pr,pl}
\fmf{fermion}{pl,ol}
\fmfv{d.sh=c,d.filled=1,d.size=4thick}{ol}
\fmfv{d.sh=c,d.filled=1,d.size=4thick}{or}
\fmfdot{pr,pl}
\end{fmfgraph*}}
+
\parbox{35mm}{
\begin{fmfgraph*}(35,15)
\fmfleftn{l}{2}
\fmfrightn{r}{2}
\fmf{fermion}{ol,l2}
\fmf{fermion}{ol,l1}
\fmf{fermion}{r1,or}
\fmf{fermion}{r2,or}
\fmfforce{(0.25w,0.5h)}{ol}
\fmfforce{(0.75w,0.5h)}{or}
\fmfforce{(0.35w,0.8h)}{pl}
\fmfforce{(0.65w,0.8h)}{pr}
\fmf{fermion,left=.7,tension=.5}{or,ol}
\fmf{fermion}{or,pr}
\fmf{fermion,right=.3,tension=.7}{pr,pl}
\fmf{boson,left=.7,tension=.5}{pr,pl}
\fmf{fermion}{pl,ol}
\fmfv{d.sh=c,d.filled=1,d.size=4thick}{ol}
\fmfv{d.sh=c,d.filled=1,d.size=4thick}{or}
\fmfdot{pr,pl}
\end{fmfgraph*}}
\right)\right|_{\pi- \rm prod.}
\label{}
\end{eqnarray}
with its leading contribution proportional to the expression found for the 
first term (\ref{pi-prod-two}) with $J_{\Sigma, R}^{(\pi)}(p,z_0) = 3\,J_{\pi, R}^{(\pi)}(p,z_0)$.
We point out that the dominant term at order $Q^2$ arises from the pion
production vertex (\ref{pi-prod-vertex}) which induces the loop function:
\begin{eqnarray}
J_{V, R}^{(\pi)}(p,z_0)&&=\frac{\lambda^{-5}}{g_A\,g^{(\pi )}}  \left( 
\parbox{35mm}{
\begin{fmfgraph*}(35,15)
\fmfleftn{l}{2}
\fmfrightn{r}{2}
\fmf{fermion}{ol,l2}
\fmf{fermion}{ol,l1}
\fmf{fermion}{r1,or}
\fmf{fermion}{r2,or}
\fmfforce{(0.25w,0.5h)}{ol}
\fmfforce{(0.75w,0.5h)}{or}
\fmf{phantom,left=.7,tension=.5,tag=1}{or,ol}
\fmf{fermion,right=.7,tension=.5,tag=2}{or,ol}
\fmfforce{(0.5w,0.0h)}{md}
\fmf{phantom,right=.7,tension=.4,tag=3}{md,ol}
\fmfv{d.sh=square,d.filled=1,d.size=4thick}{ol}
\fmfv{d.sh=c,d.filled=1,d.size=4thick}{or}
\fmfposition
\fmfipath{p[]}
\fmfiset{p1}{vpath1(__or,__ol)}
\fmfiset{p3}{vpath3(__md,__ol)}
\fmfi{fermion}{subpath (0,length(p1)/2) of p1}
\fmfi{fermion}{subpath (length(p1)/2,length(p1)) of p1}
\fmfi{boson}{point length(p1)/2 of p1--point length(p3)/3 of p3 --
point length(p3) of p3} 
\fmfiv{d.sh=c,d.filled=1,d.size=2thick}{point length(p1)/2 of p1}
\end{fmfgraph*}}
+
\parbox{35mm}{
\begin{fmfgraph*}(35,15)
\fmfleftn{l}{2}
\fmfrightn{r}{2}
\fmf{fermion}{ol,l2}
\fmf{fermion}{ol,l1}
\fmf{fermion}{r1,or}
\fmf{fermion}{r2,or}
\fmfforce{(0.25w,0.5h)}{ol}
\fmfforce{(0.75w,0.5h)}{or}
\fmf{fermion,left=.7,tension=.5,tag=2}{or,ol}
\fmf{phantom,right=.7,tension=.5,tag=1}{or,ol}
\fmfforce{(0.5w,1.0h)}{md}
\fmf{phantom,left=.7,tension=.4,tag=3}{md,ol}
\fmfv{d.sh=square,d.filled=1,d.size=4thick}{ol}
\fmfv{d.sh=c,d.filled=1,d.size=4thick}{or}
\fmfposition
\fmfipath{p[]}
\fmfiset{p1}{vpath1(__or,__ol)}
\fmfiset{p3}{vpath3(__md,__ol)}
\fmfi{fermion}{subpath (0,length(p1)/2) of p1}
\fmfi{fermion}{subpath (length(p1)/2,length(p1)) of p1}
\fmfi{boson}{point length(p1)/2 of p1--point length(p3)/3 of p3 --
point length(p3) of p3} 
\fmfiv{d.sh=c,d.filled=1,d.size=2thick}{point length(p1)/2 of p1}
\end{fmfgraph*}} \right.
\nonumber\\
&&+\left.\left. 
\parbox{35mm}{
\begin{fmfgraph*}(35,15)
\fmfleftn{l}{2}
\fmfrightn{r}{2}
\fmf{fermion}{ol,l2}
\fmf{fermion}{ol,l1}
\fmf{fermion}{r1,or}
\fmf{fermion}{r2,or}
\fmfforce{(0.25w,0.5h)}{ol}
\fmfforce{(0.75w,0.5h)}{or}
\fmf{fermion,left=.7,tension=.5,tag=2}{or,ol}
\fmf{phantom,right=.7,tension=.5,tag=1}{or,ol}
\fmfforce{(0.5w,1.0h)}{md}
\fmf{phantom,right=.7,tension=.4,tag=3}{md,or}
\fmfv{d.sh=c,d.filled=1,d.size=4thick}{ol}
\fmfv{d.sh=square,d.filled=1,d.size=4thick}{or}
\fmfposition
\fmfipath{p[]}
\fmfiset{p1}{vpath1(__or,__ol)}
\fmfiset{p3}{vpath3(__md,__or)}
\fmfi{fermion}{subpath (0,length(p1)/2) of p1}
\fmfi{fermion}{subpath (length(p1)/2,length(p1)) of p1}
\fmfi{boson}{point length(p1)/2 of p1--point length(p3)/3 of p3 --
point length(p3) of p3} 
\fmfiv{d.sh=c,d.filled=1,d.size=2thick}{point length(p1)/2 of p1}
\end{fmfgraph*}}
+
\parbox{35mm}{
\begin{fmfgraph*}(35,15)
\fmfleftn{l}{2}
\fmfrightn{r}{2}
\fmf{fermion}{ol,l2}
\fmf{fermion}{ol,l1}
\fmf{fermion}{r1,or}
\fmf{fermion}{r2,or}
\fmfforce{(0.25w,0.5h)}{ol}
\fmfforce{(0.75w,0.5h)}{or}
\fmf{phantom,left=.7,tension=.5,tag=1}{or,ol}
\fmf{fermion,right=.7,tension=.5,tag=2}{or,ol}
\fmfforce{(0.5w,0.0h)}{md}
\fmf{phantom,left=.7,tension=.7,tag=3}{md,or}
\fmfv{d.sh=c,d.filled=1,d.size=4thick}{ol}
\fmfv{d.sh=square,d.filled=1,d.size=4thick}{or}
\fmfposition
\fmfipath{p[]}
\fmfiset{p1}{vpath1(__or,__ol)}
\fmfiset{p3}{vpath3(__md,__or)}
\fmfi{fermion}{subpath (0,length(p1)/2) of p1}
\fmfi{fermion}{subpath (length(p1)/2,length(p1)) of p1}
\fmfi{boson}{point length(p1)/2 of p1--point length(p3)/3 of p3 --
point length(p3) of p3} 
\fmfiv{d.sh=c,d.filled=1,d.size=2thick}{point length(p1)/2 of p1}
\end{fmfgraph*}} \right) \right|_{\pi- \rm production}
\nonumber\\ \nonumber\\
&&=2\,\frac{m_\pi^4\,p^2}{(2\pi)^4\,m} \Bigg( 
\frac{(p^2-\mu_\pi^2)^2}{p^4}\,
\ln \Bigg|\frac{\mu_\pi^2}{\mu_\pi^2-p^2} \Bigg|
+\frac{(z_0^2+\mu_\pi^2)^2}{z_0^2\,p^2}\,
\ln \Bigg|\frac{\mu_\pi^2}{\mu_\pi^2+z_0^2} \Bigg|
\nonumber\\
&&+\,i\,\pi\,\Theta (p-\mu_\pi )\,\frac{(p^2-\mu_\pi^2)^2}{p^4}\,
\Bigg)+{\mathcal O}\left(Q^7 \right)\; .
\label{}
\end{eqnarray}

Technically the three pion production loop functions are incorporated easily by the formal 
replacement in (\ref{bar-t-one}):
\begin{eqnarray}
g_R^{-1} \rightarrow && g_R^{-1}- g_A^2\,\lambda^2\, 
J_{\pi, R}^{(\pi)}(p,z_0)-g_A^2\,\lambda^2\, 
J_{\Sigma ,R}^{(\pi)}(p,z_0)
\nonumber\\
-&&g^{-1}_R\,g_A\,g^{(\pi)}\lambda^5\,J_{V, R}^{(\pi)}(p,z_0) \; .
\label{pion-inclusion}
\end{eqnarray}
We point out that all loop functions $J_{\pi, R}^{(\pi)}(p,z_0), J_{R}^{(\pi)}(p,z_0)\sim Q^4$ 
and $J_{V, R}^{(\pi)}(p,z_0)\sim Q^6$ contribute 
to the renormalized inverse coupling $g_R^{-1} $ at chiral order $Q^4$: 
we discover here the presence of the enhancement factor $g_R^{-1}\sim Q^{-2}$ which 
implies that at leading chiral order pion production close to threshold is driven by 
the pion production vertex (\ref{pi-prod-vertex}): 
\begin{eqnarray}
\Im \,\delta_{[^1S_0]}(p) \simeq  &&
\frac{g_A\,\zeta_\pi\,m_\pi^4}{4 \pi\,f_\pi}
\,\frac{(p^2-\mu_\sigma^2)\,(p^2-\mu_\pi^2)^2}
{4\,\pi^2\,\zeta_\sigma^2\,\mu_\sigma^4\,p\,m}
\,\Theta \Big( p-\mu_\pi\Big)
\nonumber\\
+&&2\,\frac{g^2_A\,m_\pi^4}{(4 \pi\,f_\pi)^2}
\,\frac{(p^2-\mu_\sigma^2)^2\,(p^2-\mu_\pi^2)^2}
{\zeta^2_\sigma\,\mu_\sigma^4\,p^{3}\,m} \,\Theta \Big( p-\mu_\pi\Big)
\label{pi-prod-approx}
\end{eqnarray}
with:
\begin{eqnarray}
\lambda^4\,g^{(\pi)}\,g^{-1}_R &=& \frac{m}{2\,\pi}\,\frac{\zeta_\pi}{p^2-\mu_\sigma^2}
+\cdots \;.
\label{}
\end{eqnarray}
In (\ref{pi-prod-approx}) we dropped some overall normalization induced by pionic vertex 
corrections for simplicity. Our result clearly demonstrates a subtle interplay of the 
pion production scale $\mu_\pi$ and $\mu_\sigma$. Close to the chiral limit with 
$\mu_\pi  \ll \mu_\sigma $ the second term in (\ref{pi-prod-approx}) is dominant for the 
threshold region with $p\simeq \mu_\pi $. For a realistic pion mass with 
$\mu_\pi \simeq \mu_\sigma $ we find the first term in 
(\ref{pi-prod-approx}) most important since it is suppressed by only one power in 
$p^2-\mu_\sigma^2 $ as compared to the two powers of the second term.

\begin{figure}[t]
\epsfysize=10.0cm
\begin{center}
\mbox{\epsfbox{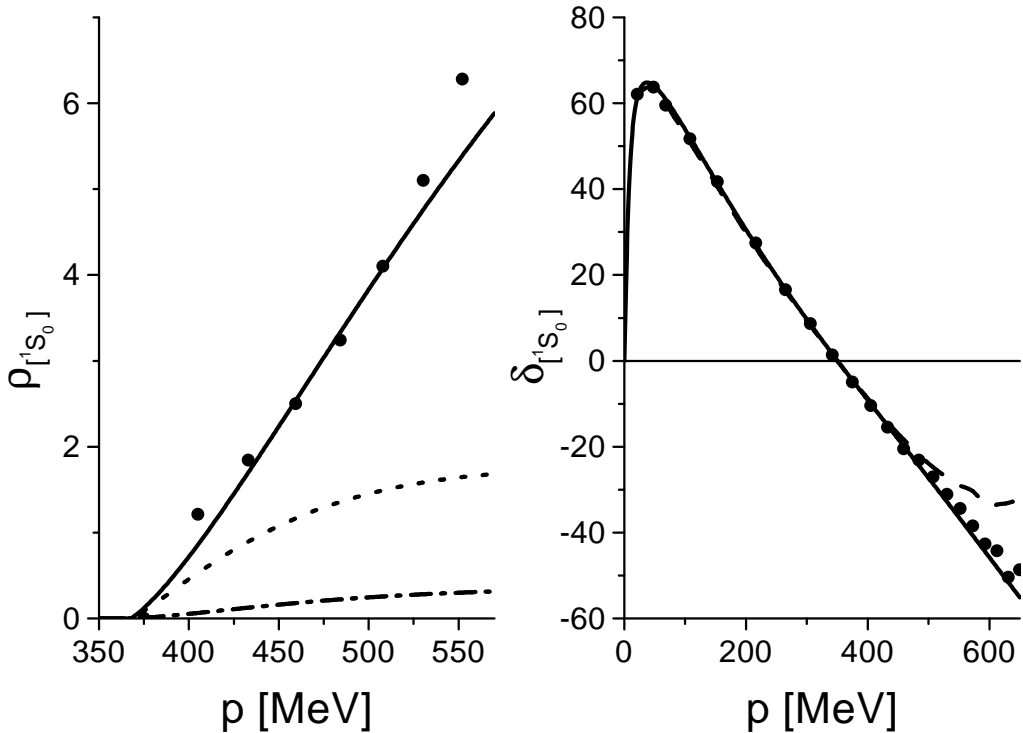}}
\end{center}
\caption{The left figure shows the inelasticity parameter $\rho$ with
$\Im \,\delta =-\ln \cos \rho $ and the right figure the real part of the $^1S_0$-phase 
shift. Our partial UCE-Q$^2$ calculation as described in the text (\ref{pion-inclusion}) 
is given by the solid line. The dashed line is the result of the Bonn potential and the 
dots represent the SP99 analysis \cite{Arndt}. The dotted and dash-dotted lines follow 
from the perturbative expression (\ref{pi-prod-approx}) with $\zeta_\pi =0.301$ fm  
and $\zeta_\pi =0$ fm respectively.}
\label{fig4}
\end{figure}

In Fig. 5 we confront (\ref{pi-prod-approx}) with Arndt's SP99 analysis \cite{Arndt}. The 
inelasticity parameter $\rho $ with $\Im \,\delta =-\ln \cos \rho$ as well as the real part of 
the phase shift are described well with $\xi_0^{(2)}=0.751$ fm, $\xi_2^{(0)}=0.222$ fm,
$\xi_\sigma =-783.5$ MeV, $\mu_\sigma = 350.4$ MeV, $z_0 \simeq -35.27$ MeV, 
and $\zeta_\pi = 0.301$ fm. Note that the set of parameters is rather close to the values 
given in Tab. 1 indicating once again the stability of our expansion. The pion production 
parameter $\zeta_\pi $ turns out natural since it implies $g^{(\pi)} \simeq -0.02$. Fig. 5 
shows also results of the perturbative expression (\ref{pi-prod-approx}) with $\zeta_\pi =0.301$ fm
(dashed line) and  $\zeta_\pi =0$ fm (dashed-dotted line). This demonstrates that  
pion production is driven by the vertex (\ref{pi-prod-vertex}) and furthermore that the production turns  
non-perturbative for momenta $p>400 $ MeV.

The pion production cross section was studied close to threshold by a series of 
recent works \cite{Cohen-pi,Kolck-pi,Park-pi,Hanart-pi,Rocha-pi,Dimitr-pi} in the context of 
chiral perturbation theory. Since the total pion production cross section 
\begin{eqnarray}
\sigma_{\pi-\rm production} &=& \frac{\pi}{2\,p^2}\,\sum_{lJ}\,\Big(2\,J+1 \Big)\,
\Big(1-\exp \left(-4\,\Im \,\delta_{l J}\right) \Big)
\label{pi-total}
\end{eqnarray}
receives contributions also from higher partial waves even close to threshold 
\cite{Green78}, the comparison is not directly possible at this stage. Nevertheless it is 
amusing to observe that our production mechanism contradicts conventional chiral attempts 
to describe pion production insofar that in our scheme the production vertex 
(\ref{pi-prod-vertex}) contributes at leading rather than at subleading 
order \cite{Cohen-pi,Kolck-pi,Park-pi,Hanart-pi,Rocha-pi,Dimitr-pi}. We therewith confirm
phenomenological studies which demonstrate the importance of the $\sigma$-meson 
exchange \cite{Lee-pi,Horowitz-pi}. For a review of one-pion production in nucleon-nucleon 
collisions see \cite{Blankleider}.

We find our result for the $^1S_0$ partial wave encouraging and therefore the unitary chiral 
expansion scheme should be applied also to the spin triplet s-wave and higher partial waves. 
In particular we expect that the zero of the $^3S_1$ and $^3P_0$-phase 
at $E_{\mathrm{lab}} \simeq 300$ MeV and $E_{\mathrm{lab}} \simeq 200$ MeV respectively 
is driven by a mechanism analogous to the mechanism presented for the $^1S_0$ phase.  

\section*{Acknowledgments}

The author acknowledges useful discussions with Bengt Friman and Evgueni Kolomeitsev.

\begin{appendix}

\section{Partial wave representation of invariant amplitudes}

The on-shell nucleon-nucleon scattering amplitude is given in terms of 
five isospin zero amplitudes $a_i^{(0)}(t,u)$ and five isospin one
amplitudes $a_i^{(1)}(u,t)$ as follows:
\begin{equation}
T_c(u,t) ={\mathcal P}_0 \,\sum_{i=0}^5 a_i^{(0)} (u,t)\,  T_i
+ {\mathcal P}_1\,\sum_{i=0}^5 a_i^{(1)} (u,t)\,  T_i
\label{appa1}
\end{equation}
with the Dirac tensors
\begin{eqnarray}
 T_1 &=& \left(   \gamma_5\, C^{-1}\times C\,\gamma_5  \right)\; ,
\nonumber\\
 T_2 &=&  \left(  \gamma_{\alpha }\,C^{-1}\times C\,\gamma^{\alpha }  \right)\; ,
\nonumber\\
 T_3 &=& \left( C^{-1}  \times C \right)\; ,
\nonumber\\
 T_4 &=&  \left(  \gamma_{\alpha }\,C^{-1}\times C\,\gamma^{\alpha }  \right)
-{\textstyle {1\over2}}
\, \left( \sigma_{\alpha \beta }\,C^{-1} \times C \,\sigma^{\alpha \beta }  \right)\; ,
\nonumber\\
 T_5 &=& \left(  \gamma_5\, C^{-1} \times C \,\gamma_5  \right)
+\left(  \gamma_5 \gamma_{\alpha }\,C^{-1}\times C\, \gamma_5\gamma^{\alpha }  \right)
\label{appa2}
\end{eqnarray}
and the isospin projection operators:
\begin{eqnarray}
{\mathcal P}_0 &=& {\textstyle {1\over2}}\left( \tau_2 \times
\tau_2\right)\; ,
\nonumber\\
{\mathcal P}_1 &=& {\textstyle {1\over2}}
\left( \vec \tau \tau_2 \times \tau_2\vec \tau \right) \; .
\label{appa3}
\end{eqnarray}
The tensor product $'\times ' $ in (A.2) is defined via the replacement rule
\begin{eqnarray}
\bar \Gamma \,C^{-1} \times C\, \Gamma  \rightarrow
\Big(\bar u(p_1')\,\bar \Gamma \,C^{-1} \bar u^t(p_2')\Big)
\left(u^t(p_2)\, C \,\Gamma u(p_1) \right) 
\label{}
\end{eqnarray}
with the incoming momenta $p_{1,2}$ and the outgoing momenta $p'_{1,2}$.
The invariant amplitudes $a^{(I)}_i(u,t) $ can be uniquely 
reconstructed from the nucleon phase shifts \cite{Jacob}. It is
convenient first to express the amplitudes in terms of
auxiliary  amplitudes $f_i$ which have a more transparent expansion in partial
wave amplitudes:
\begin{eqnarray}
a_1 &=& 2\frac{m^2}{s}\,f_1 -\frac{m^2}{s}\, f_3
-\frac{m^2}{s}\,\frac{u-t}{u+t}\,f_4
-\frac{m}{\sqrt{s}} \,\frac{u-t}{u+t}\, f_5\; ,
\nonumber\\
a_2 &=&-\frac{m}{\sqrt{s}} \,f_5\; ,
\nonumber\\
a_3 &=& 2\frac{ m^2}{u+t} \,f_2 + m^2 \frac{u-t}{\left(u+t\right)^2}\, f_4
+\frac{m}{\sqrt{s}} \frac{u-t}{\left(u+t\right)^2}\left( s+ 4 m^2\right)\, f_5\; ,
\nonumber\\
a_4 &=&-\frac{m^2}{u+t}\, f_4 - \frac{m\sqrt{s}}{u+t} \,f_5\; ,
\nonumber\\
a_5 &=& -\frac{m^2}{u+t}\,f_3 \; .
\label{appa4}
\end{eqnarray}
For the reader's convenience we also provide the inverse relations
\begin{eqnarray}
f_1 &=&\frac{s}{2\,m^2}\, a_1-\frac{u-t}{2\,m^2}\,a_4-\frac{u+t}{2\,m^2}\,a_5\; ,
\nonumber\\
f_2 &=&2\,\frac{u-t}{u+t}\,a_2+\frac{u+t}{2\,m^2}\,a_3
-\frac{u-t}{2\,m^2}\,a_4\; ,
\nonumber\\
f_3 &=&-\frac{u+t}{m^2}\,a_5\; ,
\nonumber\\
f_4 &=& \frac{s}{m^2}\, a_2-\frac{u+t}{m^2}\,a_4\; ,
\nonumber\\
f_5 &=&-\frac{\sqrt{s}}{m}\,a_2 \; .
\label{appa5}
\end{eqnarray}

The amplitudes $f_i$ are now easily expressed in terms of coupled
$f_{11}^{(j)}$, $f_{22}^{(j)}$, $f_{12}^{(j)}$ and uncoupled
$f_{1}^{(j)}$, $f_{0}^{(j)}$ partial wave amplitudes
\begin{eqnarray}
f_1 &=& \sum_{j=0}^{\infty }
\left( 2 j+1 \right) f_0^{(j)}\, d_{00}^{(j)}(\theta )\; ,
\nonumber\\
f_2 &=& \sum_{j=0}^{\infty }
\left( 2 j+1 \right) f_{11}^{(j)}\, d_{00}^{(j)}(\theta )\; ,
\nonumber\\
f_3 &=& \left( t+u \right) \sum_{j=0}^{\infty }
\left( 2 j+1 \right) f_{1}^{(j)}\,
\left( \frac {d_{11}^{(j)}(\theta )}{2u}+\frac {d_{-11}^{(j)}(\theta )}{2t}\right)
\nonumber\\
&+&\left( t+u \right) \sum_{j=0}^{\infty }
\left( 2 j+1 \right) f_{22}^{(j)}\,
\left( \frac {d_{11}^{(j)}(\theta )}{2u}-\frac {d_{-11}^{(j)}(\theta )}{2t}\right)\; ,
\nonumber\\
f_4 &=& \left( t+u \right) \sum_{j=0}^{\infty }
\left( 2 j+1 \right) f_{1}^{(j)}\,
\left( \frac {d_{11}^{(j)}(\theta )}{2u}-\frac {d_{-11}^{(j)}(\theta )}{2t}\right)
\nonumber\\
&+&\left( t+u \right) \sum_{j=0}^{\infty }
\left( 2 j+1 \right) f_{22}^{(j)}\,
\left( \frac {d_{11}^{(j)}(\theta )}{2u}+\frac {d_{-11}^{(j)}(\theta )}{2t}\right)\; ,
\nonumber\\
f_5 &=& -\left( t+u\right) \sum_{j=0}^{\infty }
\left( 2 j+1 \right) f_{12}^{(j)}\, \frac {d_{10}^{(j)}(\theta )}{2\sqrt{ut}}
\label{appa6}
\end{eqnarray}
where $d^{(j)}_{m'm} (\theta )
=\langle jm'| e^{-i\, \theta J_y}|jm\rangle $. The sum 
over the total angular momentum $j$, in (\ref{appa6}) is constrained by 
proper selection rules. In (\ref{appa6}) we restricted the c.m. scattering angle $ 0<\theta
<\pi $ for technical convenience.

The coupled amplitudes $f^{(j)}_{11},f^{(j)}_{22}$ and $f^{(j)}_{12}$ are expressed
in terms of the amplitudes $f_{l=j+1},f_{l=j-1}$ with well defined
angular momentum $l$,and the transition amplitude $f_{\pm}$ as follows:
\begin{eqnarray}
f^{(j)}_{11} &=& \frac{1+j}{2 j+1} \,f_{l=j+1}
+\frac{j}{2 j+1} \,f_{l=j-1} + 2 \frac {\sqrt{j(j+1)}}{2j+1}\, f_{\pm} \; ,
\nonumber\\
f^{(j)}_{22} &=& \frac{j}{2 j+1}\, f_{l=j+1}
+\frac{1+j}{2 j+1}\, f_{l=j-1} -2 \frac {\sqrt{j(j+1)}}{2j+1}\, f_{\pm}\; ,
\nonumber\\
f^{(j)}_{12} &=& -\frac{\sqrt{j(j+1)}}{2 j+1}\, f_{l=j+1}
+\frac{\sqrt{j(j+1)}}{2 j+1} \,f_{l=j-1} + \frac {1}{2j+1}\, f_{\pm} \; .
\label{appa7}
\end{eqnarray}

The partial wave amplitudes are given in terms of the phase shifts. 
For the uncoupled amplitudes $f_0^{(j)}$ and $f_1^{(j)} $ one defines
\begin{equation}
f^{(j)} = \frac {4 \pi}{m^2}
\sqrt{\frac {-s}{u+t}}\frac{1}{i}
\left(e^{2\,i\,\delta_j }-1 \right)
\label{appa8}
\end{equation}
with the partial wave phase shift $\delta_j $. 
The coupled partial wave amplitudes 
\begin{eqnarray}
f_{l=j\pm 1}
&=& \frac {4 \pi}{m^2} \sqrt{\frac {-s}{u+t}}\frac{1}{i}
\left( \cos \left( 2\,\epsilon_j^S \right)  e^{2\,i\,\delta^S_{l=j\pm 1} }-1
\right)\; ,
\nonumber\\
f_j^{\pm } &=&  \frac {4 \pi}{m^2} \sqrt{\frac {-s}{u+t}}
    \sin \left( 2\, \epsilon_j^S \right)
    e^{i\,\left(\delta^S_{l=j-1}+ \delta^S_{l=j+1}\right)}
\label{appa10}
\end{eqnarray}
are expressed in terms of the phase shift $\delta^S_{l=j\pm 1} $  and the mixing
angle $\epsilon_j $ according to Stapp's convention \cite{Stapp}. Note that the spin 
singlet partial wave phase shift $\delta (^1S_0)$ sits in $f_1$.

\section{Evaluation of loops}

In this appendix we derive analytic results for the one, two and
three-loop function needed for the nucleon scattering amplitude at
leading chiral orders. Since we apply a relativistic version of the
chiral Lagrangian the non-relativistic expansion in $1/m $ is
performed at the level of the loop integrals. Technically this can
be achieved by expanding the spectral density in powers of $1/m$
and then recovering the real part via a dispersion relation.
Equivalently one may  perform the energy integrations in the loop
function first and then expand the integrand before working out the
remaining three momentum integrals. When dealing with fermions a
further complication arises due to the spin structure of fermions.
However, any diagram involving complicated Dirac
structure can be expressed in terms of scalar master loop integrals
which naturally arise in scalar quantum field theory. This
reduction can be achieved purely algebraically relying only on
covariance and the shift invariance of the integration measure $d^4
l $ (as implied for example by dimensional regularization). We
therefore  exemplify our techniques first at hand of these master
loop functions. Consider the scalar functions:
\begin{eqnarray}
I_{2,N}(s)&=&-i\,\int \frac{d^4l}{(2\pi)^4}\,S_N(P-l)\, S_N(K+l)\; ,
\nonumber\\
I_{3,N}(s)&=&i\,\int \frac{d^4l}{(2\pi)^4}\, S_\pi (l+{\textstyle
{1\over2}}\,Q)\, S_N(P-l)\,S_N(K+l)\; ,
\nonumber\\
I_{4,N}(s,t)&=&-i\,\int \frac{d^4l}{(2\pi)^4}\, S_\pi
(l+{\textstyle {1\over2}}\,Q)\, S_\pi (l-{\textstyle
{1\over2}}\,Q)\, S_N(P-l)\,S_N(K+l)
\label{master-loops-I}
\end{eqnarray}
with $S_i(p)=1/(p^2-m_i^2+i\,\epsilon )$. For technical convenience we introduce
the four vectors $P,K$ and $Q$ as follows:
\begin{eqnarray}
\begin{array}{ll}
p_1=K+{\textstyle {1\over2}} Q \; ,\;\;\;\;\;\;& p_1' = K-{\textstyle
{1\over2}} Q\; ,\\ p_2=P-{\textstyle {1\over2}} Q \; ,\;\;\;\;\;\;\;\;\;&
p_2' = P+{\textstyle {1\over2}} Q \; .
\end{array}
\end{eqnarray}
The master functions are easily calculated by means of dispersion
techniques:
\begin{eqnarray}
I_{2,N}(s) &=&
\int_{4\, m^2}^{4\left( m^2+\Lambda^2\right)}
\frac{d\, s'}{\pi } \, \frac{\rho_{2,N }(s')}{s'-s}\;\; ,
\;\;\;\;
I_{3,N}(s) =
\int_{4\, m^2}^{4\left( m^2+\Lambda^2\right)}
\frac{d\, s'}{\pi } \, \frac{\rho_{3,N }(s')}{s'-s}\; ,
\nonumber\\
I_{4,N}(s,t)&=&
\int_{4\, m^2}^{4\left( m^2+\Lambda^2\right)}
\frac{d\, s'}{\pi } \, \frac{\rho_{4,N}(s',t)}{s'-s}
\label{disp-all}
\end{eqnarray}
with $(P+K)^2=s$. The finite cutoff $\Lambda < m $ is kept at this
intermediate stage in order to mathematically justify the $1/m$ expansion and provide 
an instructive guide how to define a suitable regularization scheme. The 
cutoff $\Lambda $ will ultimately be removed and may be replaced by an appropriate 
scale arising from dimensional regularization. We derive the s-channel spectral densities:
\begin{eqnarray}
\rho_{2,N}(s) &=&\frac{1}{8\,\pi}\,\sqrt{\frac{1}{4}-\frac{m^2}{s}}\; ,
\nonumber\\
\rho_{3,N}(s)&=&\frac{1}{16\,\pi }\frac{1}{\sqrt{4\, m^2 -u-t}}\,
\frac{1}{\sqrt{-u-t}}\,\ln
\left( \frac{m_\pi^2-u-t}{ m_\pi^2 }\right)\; ,
\nonumber\\
\rho_{4,N}(s,t) &=&\frac{1}{4\, \pi}\,\frac{1}{\sqrt{-t}}\,
\frac{1}{\sqrt{4\, m^2-u-t}}\, \frac{1}{\sqrt{b(s,t)}}\,
\arctanh \sqrt{\frac{t(u+t)}{b(s,t)}}
\label{spec-all}
\end{eqnarray}
with $u+t=4\,m^2-s$ and
\begin{eqnarray}
b(s,t) =4\,m_\pi^4
-4\,m_\pi^2 \left(u+t\right) + t \left( u+t\right)\; .
\label{}
\end{eqnarray}

In the algebraic reduction of the pion nucleon box diagram we will
encounter further master loop functions:
\begin{eqnarray}
I_2(p^2) &=&-i\,\int
\frac{d^4l}{(2\pi)^4}\,S_N(l+p)\, S_\pi (l)\; ,
\nonumber\\
I_{2,\pi }(t) &=&i\,\int
\frac{d^4l}{(2\pi)^4}\, S_\pi (l+{\textstyle {1\over2}}\,Q)\, S_\pi
( l-{\textstyle {1\over2}}\,Q) \; ,
\nonumber\\
I_{3,\pi }(t) &=&i\,\int
\frac{d^4l}{(2\pi)^4}\, S_\pi (l+{\textstyle {1\over2}}\,Q)\, S_\pi
( l-{\textstyle {1\over2}}\,Q)\, S_N(l+K) \; .
\label{master-loops-II}
\end{eqnarray}
We point out that the master loops of (\ref{master-loops-II}) are not independent of
the master loops introduced in (\ref{master-loops-I}). They are interrelated by 
exact algebraic identities required for the cancellation of kinematical singularities:
\begin{eqnarray}
&&I_{2}(m^2)= I_{2,N}(4\,m^2)-m_\pi^2\,I_{3,N}(4\,m^2) \; ,
\nonumber\\
&&I_{3,\pi}(t) =I_{3,N}(4\,m^2-t)
-\left( m_\pi^2-{\textstyle {1\over2}}\,t\right) I_{4,N}(4\,m^2-t,t )\;,
\nonumber\\
&&\left(2\,m_\pi^2-t\right)^2\,
{\textstyle {\partial\over \partial s}}\,I_{4,N}(s,t)\Big|_{s=4\,m^2-t}
=I_{3,N}(4\,m^2-t) 
\nonumber\\
&&\;\;\;\;\;\;\;\;\;\;\;\; -
{\textstyle {1\over2}}\left(4\,m_\pi^2-t\right)\,I_{4,N}(4\,m^2-t,t)
+2\,\left(2\,m_\pi^2-t\right)I'_{3,N}(4\,m^2-t)
\nonumber\\
&&\;\;\;\;\;\;\;\;\;\;\;\;-{\textstyle {2\over t}}\,\Big( m_\pi^2\,I_{3,N}(4\,m^2-t)-I_{2,N}(4\,m^2-t)+I_2(m^2) \Big)
\nonumber\\
&&\;\;\;\;\;\;\;\;\;\;\;\;-
{\textstyle {2\over 4\,m^2-t}} \,\Big(\left(m_\pi^2-{\textstyle {1\over2}}\,t\right)\,I_{3,\pi}(t)
+ I_{2,\pi}(t)-I_2(m^2)\Big)\;.
\label{algebra-constraint}
\end{eqnarray}
The identities (\ref{algebra-constraint}) can be verified for example in dimensional 
regularization\footnote{More conveniently the identities (\ref{algebra-constraint}) can be 
derived algebraically if one requires the reduced loop function $b_{1,3}(s,t) $ 
of the pion-nucleon box diagram not to be singular at $u=0$ and 
$u+t=0$. See appendix B.6.}. Any consistent regularization procedure must respect 
(\ref{algebra-constraint}) order by order in the $1/m$-expansion. 
Once we specify our regularization for $I_{i,N}(s)$ in terms of the cutoff $\Lambda$ the 
regularized loops $I_2(p^2), I_{2,\pi}(t)$ and $I_{3,\pi}(t)$ must be compatible with  
(\ref{algebra-constraint}). 

One may conjecture that the triangle loops $I_{2,\pi }(t)$ and $I_{3,\pi }(t)$ could  
be evaluated by means of a dispersion integral:
\begin{eqnarray}
I_{2,\pi }(t) &=&\int_{4\, m_\pi^2}^{4\left(
m_\pi^2+\bar \Lambda^2\right)}
\frac{d\, t'}{\pi } \, \frac{\rho_{2,\pi }(t')}{t'-t}\;\;,\;\;\;\;
I_{3,\pi }(t) =\int_{4\, m_\pi^2}^{4\left(
m_\pi^2+\bar \Lambda^2\right)}
\frac{d\, t'}{\pi } \, \frac{\rho_{3,\pi }(t')}{t'-t}
\label{disp-t-channel}
\end{eqnarray}
with the t-channel spectral densities:
\begin{eqnarray}
\rho_{2,\pi }(t)&=&\frac{\Theta \Big( t-4\,m_\pi^2\Big)}{8\,\pi}\,
\sqrt{\frac{1}{4}-\frac{m_\pi^2}{t}}\; ,
\nonumber\\
\rho_{3,\pi }(t)&=&\frac{1}{8\, \pi }\, \frac{\Theta \Big( t-4\,m_\pi^2\Big)}
{\sqrt{t(4\, m^2-t)}}\,
\arctan \left( \sqrt{4\, m^2-t}\,
\frac{\sqrt{t-4\, m_\pi^2}}{t-2\, m_\pi^2}\right)
\label{rho-3-pi}
\end{eqnarray}
for $t>4\, m_\pi^2 $. The pion bubble $I_{2,\pi}(t)$ is rendered  finite by one subtraction.
Performing the dispersion integral for $I_{2,\pi}(t)$ in (\ref{disp-t-channel}) one finds:
\begin{eqnarray}
I_{2,\pi}(4\,m_\pi^2) &=& 
\frac{1}{8\,\pi^2}\,\arcsinh \left( \frac{\bar \Lambda}{m_\pi} \right)
\;,
\nonumber\\
I_{2,\pi}(t) &=& I_{2,\pi}(0)
-\frac{1}{8\,\pi^2}\, \left( \sqrt{\frac{4\,m_\pi^2-t}{t}}
\,\arctan \sqrt{\frac{t}{4\,m_\pi^2-t}} -1 \right)
\label{i2-pi}
\end{eqnarray}
where we applied the large $\bar \Lambda$-cutoff limit in the subtracted loop. We point out,
however, that the cutoff $\bar \Lambda$ in $I_{2,\pi}(t)$ and $I_{3,\pi}(t)$ must not be 
identified with $\Lambda $. The $1/m$ expansion of the loop integral $I_{3,\pi }(t) $ 
is subtle. The regularization of $I_{3,\pi}(t)$ by means 
of the cutoff $\bar \Lambda \ll m$ is incompatible with the $1/m$ expansion. 
This follows from the fact that the $1/m$ expansion of the t-channel spectral 
density leads, at subleading orders, to singularities in the dispersion
integral at $t'=4\,m_\pi^2 $ whereas $I_{3,\pi}(t)$ given via (\ref{algebra-constraint}) 
leads to well defined moments in the $1/m$ expansion. 

At the one-loop level we complete the collection of master loops with the pionic and 
nucleonic tadpole diagrams $I_{1,\pi}$ and $I_{1,N}$:
\begin{eqnarray}
I_{1,N} &=& i\,\int
\frac{d^4l}{(2\pi)^4}\,\frac{1}{l^2-m_N^2+i\,\epsilon }=
\int_{4\,m^2}^{4\,(m^2+\Lambda^2)}\frac{d s}{2\,\pi}\,\Im \,I_{2,N}(s)
\nonumber\\
&=&\frac{1}{12\,\pi^2}\,\frac{\Lambda^3}{m}\left( 1
+{\mathcal O}\left( \frac{\Lambda^2}{m^2}\right)\right)
\nonumber\\
I_{1,\pi} &=& i\,\int
\frac{d^4l}{(2\pi)^4}\,\frac{1}{l^2-m_\pi^2+i\,\epsilon }
=\int_{4\,m_\pi^2}^{4\,(m_\pi^2+\bar \Lambda^2)}\frac{d t}{2\,\pi}\,\Im \,I_{2,\pi}(t)
\nonumber\\
&=& \frac{1}{8\,\pi^2}\left( \bar \Lambda\,\sqrt{m_\pi^2+\bar \Lambda^2}
-m_\pi^2\,\arcsinh \left(\frac{\bar \Lambda}{m_\pi} \right)\right)
\label{tadpole}
\end{eqnarray}
where consistency of the cutoff regularization relates $I_{1,N}$ with  
$\Im \,I_{2,N}(s)$ and $I_{1,\pi}$ with $\Im \,I_{2,\pi}(t)$. 
It is crucial to observe that the proper $\Lambda $ dependence of the pionic 
tadpole $I_{1,\pi}$ follows from a further exact algebraic identity
\begin{eqnarray}
I_{1,\pi}-I_{1,N}&=& 
6\,m^2\,m_\pi^2\,
\left( m_\pi^2\,\frac{d}{d\,s}\,I_{3,N}(4\,m^2)-\frac{d}{d\,s}\,I_{2,N}(4\,m^2) \right)
\nonumber\\
&+& \Big( 2\,m^2\,m_\pi^2+m_\pi^4\Big)\,I_{3,N}(4\,m^2)
+\Big( m^2-m_\pi^2\Big) \,I_{2,N}(4\,m^2)\; 
\label{more-algebra}
\end{eqnarray}
which may be verified in dimensional regularization. Again we insist that (\ref{more-algebra})
must be respected by the $1/m$-expansion. In particular (\ref{more-algebra}) implies a 
non-trivial matching of the s-channel cutoff parameters $\Lambda$ and the t-channel 
cutoff $\bar \Lambda$. 

We point out that all $t$-channel loops $I_{3,\pi}(t)$, $I_{2,\pi}(t)$ and $I_{1,\pi}$ can  
in fact be defined via (\ref{algebra-constraint}) and (\ref{more-algebra}) in terms 
of the s-channel loops $I_{4,N}(s,t)$, $I_{3,N}(s)$ and $I_{2,N}(s)$. 
It is, however, convenient to simplify our regularization scheme and derive more 
explicit expressions for the loop functions in order to facilitate the renormalization 
procedure. First we assume that the diagrams are already partly $1/m$ 
renormalized via the construction (\ref{1/m-ren}). This implies that the factor 
$1/\sqrt{s'}$ in the spectral functions $\rho_{i,N}(s')$ can be pulled out of the dispersion 
integral (\ref{disp-all}). This procedure moves a large class of power divergencies of the 
form $(\Lambda/m)^n$ into the interaction kernel where they are absorbed into the local 
interaction terms. Next we systematically drop nucleonic tadpole diagrams since they  
also involve multiple powers of the cutoff $\Lambda^n$ with $n\geq 3$. Such terms are 
ill-defined and are ultimately canceled from contributions induced by the properly 
constructed cutoff dependence of the path-integral measure \cite{cutoff-theory}. Finally we 
apply the large cutoff limit for all finite parts. Our regularization scheme leads now to at 
most linear divergent one-loop functions. 

In more formal terms our regularization scheme is defined  by writing the 
one-loop functions $I_{2,N}(s), I_{3,N}(s)$ and $I_{4,N}(s,t)$ as follows:
\begin{eqnarray}
I_{2,N}(s) &=& \frac{1}{8\,\pi
\,\sqrt{s}}\,\Bigg(\frac{2}{\pi}\,\Lambda+i\, p \Bigg)+\Delta I_{2,N}(s) \; ,
\nonumber\\
I_{3,N}(s) &=&\frac{1}{16\,\pi\,\sqrt{s}}\,\frac{i}{p}\,
\ln \left( 1-2\,i\,\frac{p}{m_\pi}\right)+\Delta I_{3,N}(s)\; ,
\nonumber\\
I_{4,N}(s,t) &=&
\frac{1}{4\,\pi \sqrt{s}}\,\frac{1}{\sqrt{-t\,b(s,t)}}
\Bigg( i\,\arctanh  \left(\sqrt{\frac{t(u+t)}{b(s,t)}}\,\right)
\nonumber\\
&+&\frac{1}{2}\,\arctan
\left(  \frac{\sqrt{-t\,b(s,t) }}{4\,m_\pi^3}
-\frac{\sqrt{-u-t}}{m_\pi } \left(1-\frac {t}{4\,m_\pi^2} \right) \right)
\nonumber\\
&+&\frac{1}{2}\,\arctan
\left(  \frac{\sqrt{-t\,b(s,t) }}{4\,m_\pi^3}
+\frac{\sqrt{-u-t}}{m_\pi } \left(1-\frac {t}{4\,m_\pi^2} \right) \right)
\Bigg)
\nonumber\\
&+&\Delta I_{4,N}(s,t)\;.
\label{master-loops-results}
\end{eqnarray}
where 
\begin{eqnarray}
\Delta I_{2,N}(s)&=&
-\int_{4\,m^2}^{4\,(m^2+\Lambda^2)}\frac{d s'}{\pi}\,
\frac{\Im \,I_{2,N}(s')}{s+\sqrt{s\,s'}}\;,
\nonumber\\
\Delta I_{3,N}(s)&=& -\int_{4\,m^2}^{4\,(m^2+\Lambda^2)}\frac{d s'}{\pi}\,
\frac{\Im \,I_{3,N}(s')}{s+\sqrt{s\,s'}} \;,
\nonumber\\
\Delta I_{4,N}(s,t)&=& -\int_{4\,m^2}^{4\,(m^2+\Lambda^2)}\frac{d s'}{\pi}\,
\frac{\Im \,I_{4,N}(s', t)}{s+\sqrt{s'\,s}} \;.
\label{delta-i4n-a}
\label{}
\end{eqnarray}
Note that for example the separation of the term $\Delta I_{2,N}(s)$ from $I_{2,N}(s)$ in
(\ref{master-loops-results}) is induced uniquely by our construction 
(\ref{tilde-G},\ref{1/m-ren}). The divergent structure $\Delta I_{2,N}(s)$  can systematically 
be absorbed into local counter terms in accordance with (\ref{1/m-ren}). This follows from 
the representation in (\ref{delta-i4n-a}) which shows that $\Delta I_{2,N}(s)$ can be 
expanded in a Taylor series around $s=4\,m^2$. In our simplified regularization scheme 
such terms are from now on simply dropped. The loop functions $I_{3,N}(s)$ and $I_{4,N}(s,t)$ 
require special attention since $\Delta I_{3,N}(s)$ leads to chiral log terms and 
$\Delta I_{4,N}(s,t)$ depends on $t$ in a non-polynomial fashion. It is instructive to
consider first $\Delta I_{3,N}(s)$ and $\Delta I_{4,N}(s,t)$ as defined via 
(\ref{delta-i4n-a}) in terms of the finite cutoff $\Lambda $ and inspect the point 
$s=4\,m^2\,, t=0$:
\begin{eqnarray}
\Delta I_{3,N}(4\,m^2)&=&
\frac{\Lambda }{64\,\pi^2\,m^3}
\left(1-\ln \left(\frac{2\,\Lambda }{m_\pi}\right)-\frac{\pi}{2}\,\frac{m_\pi}{\Lambda}+
{\mathcal O}\left(\frac{\Lambda^2}{m^2} \right)\right)\;,
\nonumber\\
\Delta I_{4,N}(4\,m^2,0)&=&
-\frac{\Lambda }{64\,\pi^2\,m^3\,m^2_\pi}
\left(1+{\mathcal O}\left(\frac{\Lambda^2}{m^2} \right)\right)
\label{delta-i4n-0}
\end{eqnarray}
where $\Lambda < m$. Note that $\Delta I_{3,N}(4\,m^2)$ must not be considered as a counter 
term since it is non-analytic in the squared pion mass. Also the term 
$\Lambda/m^2_\pi $ cannot be absorbed into local interaction terms of the chiral Lagrangian. 
The result (\ref{delta-i4n-0}) signals an infrared singularity induced by the $1/m$ expansion 
and asks for a more careful regularization procedure. In fact the solution to this puzzle 
follows simply if one evaluates $\Delta I_{4,N}(4\,m^2,0)$ for an infinite cutoff 
$\Lambda >m$:
\begin{eqnarray}
\Delta I_{3,N}(4\,m^2)&=&
\frac{-1}{16\,\pi^2\,m^2}
\Bigg( \ln \Bigg| \frac{m}{m_\pi}\Bigg|+\pi\,\frac{m}{m_\pi}
\nonumber\\
&&\;\;\;\;\;\;\;\;\;\;\;\;\;\;-\sqrt{\frac{4\,m^2}{m_\pi^2}-1}\,
\arctan \left( \sqrt{\frac{4\,m^2}{m_\pi^2}-1}\,\right)\Bigg)
\nonumber\\
&=& \frac{-1}{16\,\pi^2\,m^2}\left( 
1+\ln \Bigg| \frac{m}{m_\pi}\Bigg|
+\frac{\pi}{8}\,\frac{m_\pi}{m}-\frac{1}{12}\,\frac{m_\pi^2}{m^2}
+{\mathcal O}\left(\frac{m_\pi^3}{m^3}\right)\right)\;.
\nonumber\\
\Delta I_{4,N}(4\,m^2,0)&=& \frac{-1}{16\,\pi^2\,m\,m_\pi^3}
\left( \frac{\pi}{2}-\frac{2\,m}{\sqrt{4\,m^2-m_\pi^2}}
\arctan \left( \sqrt{\frac{4\,m^2}{m_\pi^2}-1}\,\right)\right)
\nonumber\\
&=& \frac{-1}{32\,\pi^2\,m^2\,m_\pi^2}\left( 
1-\frac{\pi}{8}\,\frac{m_\pi}{m}+\frac{1}{6}\,\frac{m_\pi^2}{m^2}
+{\mathcal O}\left(\frac{m_\pi^3}{m^3}\right)\right)\;.
\label{i4-resum}
\end{eqnarray}
This is a surprising result since it demonstrates that the chiral log term $\ln (m_\pi) $ in 
(\ref{i4-resum}) results in part from modes probing large nucleon virtuality with 
$\Lambda > m$.

In order to perform the $1/m$ expansion of $\Delta I_{4,N}(s,t)$ and derive explicit results 
it is convenient to perform the angle average in (\ref{bs4}) after the $s'$-integration 
in (\ref{delta-i4n-a}). Now with $\Lambda =\infty$ in (\ref{delta-i4n-a}) we obtain:
\begin{eqnarray}
\Delta I_{4,N}(4\,m^2,t) &=&\frac{-1}{8\,\pi^2} \int_{-1/2}^{1/2} \,\frac{d x}{m_\pi^4}\,
\left( \frac{\pi}{2\,\sqrt{a(x,t)}} 
-\frac{\arctan \left(\sqrt{-1+a(x,t)}\, \right)}{\sqrt{-1+a(x,t)}}
\right)\; ,
\nonumber\\
a(x,t)&=& \frac{m^2}{m_\pi^2}\left( 4 +\frac{t}{m_\pi^2}\left( 4\,x^2-1\right) \right)\;.
\label{delta-i4n-d}
\end{eqnarray}
The $1/m$ expansion of $\Delta I_{4,N}(s,t)$ follows upon expanding the integrand 
of (\ref{delta-i4n-d}) in powers of $1/a$:
\begin{eqnarray}
\Delta I_{4,N}(s,t) &=& \frac{-1}{8\,\pi^2\,m^2}\,\left(
\frac{\arctanh \sqrt{\frac{-t}{4\,m_\pi^2-t}}}{\sqrt{4\,m_\pi^2-t}\,\sqrt{-t}}\,
-\frac{\pi\,m_\pi}{8\,m}\,\frac{1}{4\,m_\pi^2-t}\right)
+{\mathcal O}\left( \frac{1}{m^4}\right) \;,
\label{delta-i4n-c}
\end{eqnarray}
where $t< 4\,m_\pi^2 $. Note that the $1/m$ expansion of $\Delta I_{4,N}(s,t)$ stops 
converging at $t= 4\,m_\pi^2$. 

We turn to the master loop function $I_{3,\pi}(t)$. It is instructive to  
provide the loop including next-to-leading-order contributions
\begin{eqnarray}
I_{3,\pi }(t) &=&\frac{1}{16\,\pi\,m}\left( 1+\frac{t}{8\,m^2}\right)\frac{1}{\sqrt{-t}}\,
\arctan \left(\frac{\sqrt{-t}}{2\,m_\pi} \right)
-\frac{m_\pi}{128\,\pi\,m^3}\,\frac{2\,m_\pi^2-t}{4\,m_\pi^2-t}
\nonumber\\
&+&\frac{1}{16\,\pi^2\,m^2}\,\frac{2\,m_\pi^2-t}{ \sqrt{-t\,(4\,m_\pi^2-t)}}
\,\arctanh \sqrt{\frac{-t}{4\,m_\pi^2-t}}
\nonumber\\
&+&\Delta I_{3,N}(4\,m^2)+{\mathcal O}\left(\frac{1}{m^4} \right) 
\label{i3pi-sub-lead}
\end{eqnarray}
which follow with (\ref{algebra-constraint},\ref{master-loops-results},\ref{delta-i4n-c}). 
The result (\ref{i3pi-sub-lead}) is in agreement with the leading expression derived in the 
heavy-baryon framework \cite{Mojzis,Tang} and with 
(\ref{master-loops-II},\ref{master-loops-results}).  The subleading term of order 
$1/m^2$ agrees with a corresponding expression given in \cite{Becher}. Our result 
(\ref{i3pi-sub-lead}) confirms that the triangle loop $I_{3,\pi}(t)$ and the scalar 
pion nucleon box function $I_{4,N}(s,t)$ show a singular branch point at $t=4\,m_\pi^2$ 
if expanded in $1/m$. A systematic resummation leading to a smooth
behavior at $t=4\,m_\pi^2$ is proposed in \cite{Becher}. 

The function $\Delta I_{4,N}(s,t)$ can alternatively be deduced from 
(\ref{algebra-constraint}). Imposing the functional form for 
$I_{2,N}(s)$, $I_{3,N}(s)$ and $I_{4,N}(s,t)$ as given in (\ref{master-loops-results}) 
with $\Delta I_{2,N}(s)=0$ we derive:
\begin{eqnarray}
\Delta I_{4,N}(s,t)
&=&\frac{2\,I_{2,\pi}(t)-I_{2}(m^2)+2\,m^2\,\Delta I_{3,N}(4\,m^2)}
{2\,m^2\,(4\,m_\pi^2-t)}+{\mathcal O}\left( \frac{1}{m^4}\right)\,.
\label{delta-i4n-b}
\end{eqnarray}
Our results (\ref{delta-i4n-b}) and (\ref{i4-resum}) determine the divergent 
subtraction $I_{2,\pi}(0)$ in terms of $I_{2,N}(4\,m^2)$. This is a surprising 
result since it shows that the $1/m$ expansion converts the $\log$-divergence in 
$I_{2,N}(s)$ {\it and } $I_{2,\pi}(t)$ into a power divergence. In fact with 
(\ref{more-algebra}) and (\ref{algebra-constraint}) we find that also the pionic tadpole 
diagram $I_{1,\pi}$ and $I_2(m^2)$ must not be dropped and assigned a linear divergence:
\begin{eqnarray}
I_{1,\pi} &=&\left(m^2-\frac{1}{4}\,m_\pi^2\right) I_{2,N}(4\,m^2)
+4\,m^2\,m_\pi^2\,\Delta I_{3,N}(4\,m^2)+{\mathcal O}\left( \frac{m^3_\pi}{m} \right)\;,
\nonumber\\
I_{2,\pi}(0) &=& 
\frac{1}{2}\,I_{2,N}(4\,m^2)-2\,m^2\,\Delta I_{3,N}(4\,m^2)
-\frac{1}{8\,\pi^2}-\frac{m_\pi}{64\,\pi\,m}
+{\mathcal O}\left( \frac{m^2_\pi}{m^2} \right) \;,
\nonumber\\
I_2(m^2) &=&I_{2,N}(4\,m^2)-m_\pi^2\,\Delta I_{3,N}(4\,m^2)-\frac{m_\pi}{16\,\pi\,m}\;.
\label{div-terms}
\end{eqnarray}
We emphasize that our result (\ref{div-terms}) follows from the convenient prescription to 
systematically drop the divergent terms $I_{1,N}$ and $\Delta I_{2,N}(s)$.
This defines our simplified regularization procedure. The regularization of the remaining loop 
functions is then implied by the algebraic constraints 
(\ref{algebra-constraint},\ref{more-algebra}). The latter follow for example in 
dimensional regularization and must be satisfied  by any consistent 
regularization scheme. To this extend it is not surprising to find that for example the 
pionic tadpole in (\ref{div-terms}) acquires an explicit dependence on the nucleon mass. 
The only divergent structure left at the one-loop level is $I_{2,N}(4\,m^2)$. It may 
be written in terms of the cutoff parameter $\Lambda $ (see (\ref{master-loops-results})) or 
equivalently in terms of a scale resulting from dimensional regularization. Consistent
power counting rules arise if we count $I_{2,N}(4\,m^2)\sim \Lambda/m \sim Q$. We conclude 
that the scalar nucleon bubble is in fact the only master loop where we have to strictly 
insist on $\Lambda < m$ since for $\Lambda \gg m$ one finds 
$I_{2,N}\sim \ln \Lambda \sim Q^0$. All finite loop functions $I_{3,\pi}(t)$, $I_{3,N}(s)$ and 
$I_{4,N}(s,t)$ must be evaluated with $\Lambda \gg m$ in order to generate the proper 
chiral log content and avoid infrared singularities (see also \cite{Gegelia-2}).

\subsection{Static pion limit}

Here we would like to confront our relativistic scheme with the potential approach. 
Inherent of the potential approach is the use of the static pion exchange. 
The pion propagator is expanded
\begin{eqnarray}
S_\pi(l_0,\vec l\,) &=& \frac{-1}{ l^2+m_\pi^2} -\frac{l^2_0}{(l^2+m_\pi^2)^2}+
{\mathcal O} \left(l_0^4 \right)
\label{}
\end{eqnarray}
where it is commonly argued that terms involving the energy transfer $l^2_0 $ are 
suppressed by $1/m $. Mathematically it is not quite obvious that such an expansion is 
justified and in accordance with covariance since the expansion is highly divergent due to 
multiple powers of the energy transfer $l_0$. We evaluate the master pion nucleon box 
function $I_{4,N}(u,t)$ in the limit of static pions
\begin{eqnarray}
I^{(static )}_{4,N} (\vec p,\vec p\,' ) &=& \frac{1}{4\,m}
\int \frac{d^3l}{(2\pi)^3}\,\frac{1}{(\vec l+\vec p\,)^2+m_\pi^2}
\,\frac{1}{(\vec l+\vec p\,'\,)^2+m_\pi^2} \,\frac{1}{l^2-p^2-i\,\epsilon }
\nonumber
\label{static-result}
\end{eqnarray}
where we chose the center of mass frame with $p^2=p'^2$ and 
$\vec p\cdot \vec p' = p^2\,\cos \theta $. Note that we dropped terms suppressed by $1/m^3$.
We compare the static pion result (\ref{static-result}) with the 
proper non-relativistic expanded master loop function of (\ref{master-loops-results})
and find agreement.

\subsection{One-loop bubble}

Start with the one-loop bubble function
\begin{eqnarray}
J(s )&=&-i\,\tr \int \frac {d^4 l}{(2\pi)^4} \,C\,\gamma_5
\,S_F\left(P+l \right)\gamma_5\, C^{-1}
\,S^t_F\left(K-l \right)\;,
\nonumber\\
J(s,\Lambda ) &=&
\frac{1}{\pi} \int_{0}^{\Lambda^2 }
\frac {d\,\bar p^2\,\rho_J (\bar p )}{\bar p^2-p^2-i\,\epsilon} \; ,
\;\; \rho_J (p)=\frac{m\,p}{2\pi}\,\sqrt{1+\frac{p^2}{m^2}} 
\label{1-loop2}
\end{eqnarray}
with the nucleon propagator $S_F(p) = 1/(\pslash-m+i\,\epsilon)$. The loop function may be 
regularized by a cutoff $\Lambda $ introduced in the s-channel dispersion relation
with $s=(P+K)^2=4\,(m^2+p^2)$. Equivalently the loop may be expressed in terms of the master
functions:
\begin{eqnarray}
J(s) &=& 2\,s\,I_{2,N}(s)+ 4\,I_{1,N} \; .
\label{1-loop4}
\end{eqnarray}
The expressions (\ref{1-loop2}) and (\ref{1-loop4}) agree if the cutoff regularized 
tadpole is given by (\ref{tadpole}). Our simplified regularization
procedure leads to a linear divergent one-loop function $J(s,\Lambda)$:
\begin{eqnarray}
J(s, \Lambda) &=& \frac{m}{2\pi}\,\sqrt{1+\frac{p^2}{m^2}} 
\left( \frac{2}{\pi}\,\Lambda +i\,p \right) \;.
\label{}
\end{eqnarray}
At given order in the $1/m$ expansion the renormalized loop function is introduced
by the minimal number of subtractions at $p=-i\,z$. At leading chiral orders this 
leads to:
\begin{eqnarray}
J_R(p,z)&=&\frac{m}{2\,\pi} \Big(-z+i\,p\Big)
\nonumber\\
&+&\frac{1}{4\,\pi\,m} \Big(-\frac{3\,z}{2}\,p^2 -\frac{z^3}{2}+i\,p^3\Big)
+{\mathcal O}\left(Q^4 \right)\; .
\label{1-loop9}
\end{eqnarray}
Note that the leading term is renormalized via $J_R(-i\,z,z)=0$ whereas the 
subleading term requires a further renormalization condition.

\subsection{One-loop vertex }

The one-loop vertex function is introduced with
\begin{eqnarray}
&&{\mathcal V}_{ \pi }^{(i)}(p_2,p_1) = \frac{i}{2}\,\int \frac{d^4
l}{(2\pi)^4} \left( S_F(P+l)\, \vec \tau
\,\gamma_5\,(\lslash+{\textstyle {1\over2}}\, \Qslash )\right)^t
\tau_2\,\tau_i \,C \,\gamma_5\,
\nonumber\\
&&\;\;\;\;\;\;\;\;\;\;\;\;\;\;\;\;\;\;
\cdot S_\pi (l+{\textstyle {1\over2}} \,Q )\, S_F(K-l)\, \vec
\tau\,\gamma_5\, (\lslash+{\textstyle {1\over2}}\, \Qslash )\;,
\nonumber\\
&&u^t(p_2)\,{\mathcal V}_{ \pi }^{(i)}(p_2,p_1)\,u(p_1) =  V_\pi
(s)\, u^t(p_2)\, C\, \gamma_5 \, \frac{1}{2}\, \tau_2\, \tau_i
\,u(p_1)
\label{oneloopvertex}
\end{eqnarray}
where $u(p_{1,2})$ are the incoming nucleon spinors. The loop
function is expressed in terms of the master functions
\begin{eqnarray}
V_{\pi }(s) &=& 4\,m^2\,\Big( m_\pi^2\,I_{3,N}(s)-I_{2,N}(s) \Big)
+I_{1,\pi}-2\,I_{1,N}+2\,m_\pi^2\,I_2(m^2) \; .
\label{v-result-red}
\end{eqnarray}
According to our renormalization procedure we first transform to the on-shell irreducible 
interaction kernel. This leads to the pion-exchange with pseudo-scalar coupling which is 
known to be on-shell equivalent to the pseudo-vector coupling used in (\ref{oneloopvertex}).
The vertex evaluated with pseudo-scalar coupling agrees with the pseudo-vector result 
(\ref{v-result-red}) up to the divergent terms $I_{1,\pi}-2\,I_{1,N}+2\,m_\pi^2\,I_2(m^2)$. 
The latter terms are included in the local part of the on-shell irreducible kernel 
$\hat K=\hat K_\pi+\hat K_\sigma $ according to (\ref{matching-xx}):
\begin{eqnarray}
\hat K_\sigma &=& \Bigg(g(s) +\frac{g_A^2\,g(s)}{2\,f_\pi^2}\,
\Big( I_{1,\pi}-2\,I_{1,N}+2\,m_\pi^2\,I_2(m^2) \Big)\Bigg) \,T_1\,{\mathcal P}_1+\cdots\,.
\label{}
\end{eqnarray}
with the bare coupling function $g(s)$ of (\ref{mut0}). Similarly the $\Delta I_{3,N}(s)$ 
and $\Delta I_{2,N}(s)$ terms in (\ref{v-result-red}) may be absorbed into 
$\tilde K=\tilde K_\pi+\tilde K_\sigma $ 
according to (\ref{1/m-ren}):
\begin{eqnarray}
\tilde K_\sigma &=& \Bigg(g(s) +\frac{g_A^2\,g(s)}{2\,f_\pi^2}\,
\Big( I_{1,\pi}-2\,I_{1,N}+2\,m_\pi^2\,I_2(m^2) \Big)
\nonumber\\
&+& \frac{g_A^2\,g(s)}{2\,f_\pi^2}\,4\,m^2\,
\Big(m_\pi^2\, \Delta I_{3,N}(s)-\Delta I_{2,N}(s) \Big)\Bigg)\, T_1\,{\mathcal P}_1
+\cdots \;.
\label{}
\end{eqnarray}

Note that in our regularization prescription the nucleonic tadpole $I_{1,N}$ 
is dropped whereas the pionic tadpole (see (\ref{div-terms})) contributes at chiral 
order $Q$. The renormalized vertex function $V_{\pi,R}(s)$ follows 
upon a proper loop subtraction according to (\ref{ren-condition-1},\ref{barc-def}) with
\begin{eqnarray}
V_{\pi }(s) &=& V_{\pi ,R} (s)-\bar c(s)\,J(s)
+I_{1,\pi}-2\,I_{1,N}+2\,m_\pi^2\,I_2(m^2) \;, 
\nonumber\\
V_{\pi,R} (s)&=&\frac{m^2\,m_\pi}{8\,\pi\,\sqrt{m^2+p^2}}\,
\frac{m_\pi}{p}\,i\,\ln \left(1-i\,\frac{2\,p}{m_\pi}\right)
\;,\;\;\;
\bar c(s) = \frac{1}{2}\,\frac{m^2}{m^2+p^2} \;.
\label{ren-vpi}
\end{eqnarray}
Note that the loop function $V_{\pi,R} (s)$ is most easily derived by first evaluating 
its imaginary part. The real part is then constructed with ease since 
$\sqrt{m^2+p^2}\,V_{\pi,R} (4\,m^2+4\,p^2)$ is an analytic function in the upper 
complex half $p$-plane.

\subsection{Two-loop bubble}

The two-loop bubble integral is
\begin{eqnarray}
J_{\pi}(s)\,   &=&-\frac{i}{2}\,\tr
\int \frac{d^4 l}{(2\pi)^4}
\tau_j \,\tau_2 \,\gamma_5\,C^{-1} S^t_F(P+l)\,{\mathcal V}_{ \pi }^{(j)}(P+l,K-l)\,S_F(K-l)
\nonumber\\
\label{twoloopbubble}
\end{eqnarray}
defined in terms of the full off-shell one-loop vertex function
(\ref{oneloopvertex}). Here we implicitly assume that the vertex (\ref{oneloopvertex}) is 
evaluated with pseudo-scalar pion nucleon coupling according to (\ref{matching-xx}). 
We evaluate $J_{\pi }(s) $ by constructing its imaginary 
part and then recovering its real part by means of dispersion relations. The loop is written 
as  a sum of three terms
\begin{eqnarray}
J_{\pi }(s) =\bar J_{\pi }(s)-\bar c(s)\,J^2(s)+ J^{(\pi )}_{\pi }(s)
\label{def-jpi-s}
\end{eqnarray}
where $J^{(\pi )}_{\pi }(s)$ is the unique contribution due to the pion production cut. The
square of the one-loop bubble $J(s)$ in  (\ref{def-jpi-s}) takes care of the 
overlapping divergence structure of the two-loop function $J_{\pi }(s) $. 

We first evaluate the loop-subtracted bubble function $\bar J_{\pi }(s)$ 
by imposing a dispersion relation on $s\,\bar J_{\pi }(s)$. Here we pull out the factor
$1/s'$ in the dispersion relation in order to comply with our regularization scheme 
and (\ref{1/m-ren}). With $\bar c (s)$ as given in (\ref{ren-vpi}) the appropriate spectral 
density is readily derived
\begin{eqnarray}
\Im \,\bar J_{\pi}(s) =2\, \rho_J(s)\, \Re \, V_{\pi,R} (s)
\label{}
\end{eqnarray}
in terms of  the renormalized vertex function $V_{\pi ,R} (s)$ of (\ref{ren-vpi}).
The following representation can now be established
\begin{eqnarray}
\bar J_\pi(p)-\bar
J_\pi(0)&&=i\,\frac{m\,p}{\pi}\,V^{(1)}_{\pi,R }(p) 
=-2\,\frac{m^2\,m^2_\pi}{(4\,\pi)^2}\,
\ln \left(1-i\,\frac{2\,p}{m_\pi}\right) \;
\label{jpi-identity}
\end{eqnarray}
where $\sqrt{s}\,V_{\pi ,R} (s)  =m\,V^{(1)}_{\pi,R} (p)$. Note that (\ref{jpi-identity}) 
shows that all $1/m$ correction terms for $\bar J_\pi(p)-\bar J_\pi(0)$ cancel identically. 
The identity (\ref{jpi-identity}) follows from the Kramers-Kronig relations:
\begin{eqnarray}
i\left( V^{(1)}_{\pi,R}(p)-V^{(1)}_{\pi ,R}(0)\right)
&=&\int_{-\infty }^\infty \frac{d\,\bar p}{\pi}\,\frac{p}{\bar p}\,
\frac{\Re\, V^{(1)}_{\pi,R}(\bar p)}{\bar p-p-i\epsilon}
\nonumber\\
&=& \frac{\pi}{m\,p}\,\Big( \bar J_{\pi}(p)-\bar J_{\pi}(0)\Big)-i\,V^{(1)}_{\pi ,R }(0)\; ,
\nonumber\\
\bar J_\pi(p)-\bar J_\pi(0) &=&
\int_0^{\infty } \frac{d \,\bar p^2}{\pi^2 }\,\frac{p^2}{\bar p^2}
\frac{m\,\bar p}{\bar p^2 -p^2-i\, \epsilon }
\, \Re \, V^{(1)}_{\pi ,R } (\bar p) \; .
\label{}
\end{eqnarray}
This is an amazingly simple result.
The two-loop bubble function is given in terms of the
one-loop vertex function $V^{(1)}_{\pi,R }(p) $. Note that the loop function $\bar J_\pi (p)$ 
shows a $\log$-divergence. In analogy to (\ref{1-loop2}) the subtraction constant 
$\bar J_{\pi}(0)$  may be reconstructed by a cutoff regularized dispersion relation:
\begin{eqnarray}
\bar J_\pi(0) &=& m^2\,
\int_0^{\Lambda^2 } \frac{d \,\bar p^2}{\pi^2 }\,
\left( \frac{m}{\bar p }+\frac{\bar p}{m}\right)
\Re \, V^{(1)}_{\pi ,R } (\bar p)
\nonumber\\
&=&\frac{m^2\,m_\pi^2}{(4\,\pi)^2}\left( \ln \left(
\frac{\Lambda^2}{m^2_\pi}\right) +{\mathcal O}\left(\frac{m_\pi
}{\Lambda },\frac{\Lambda^2}{m^2}\right) \right) \;.
\label{}
\end{eqnarray}
The renormalized loop follows now with 
$J_{\pi,R}(p,z) =\bar J_{\pi}(p)-\bar J_{\pi}(-i\,z )$ according to 
(\ref{ren-condition-2}).

We turn to the evaluation of the pion production term $J^{(\pi )}_{\pi }(s)$. This contribution
is conceptually interesting since it involves a further small scale 
\begin{eqnarray}
\mu^2_\pi  = m\,m_\pi +\frac{m_\pi^2}{4} 
\label{}
\end{eqnarray}
characterizing the  the pion production process at $p> \mu_\pi$.
The spectral density is systematically expanded in powers of $(p^2-\mu_\pi^2)^n$
\begin{eqnarray}  
\rho^{(\pi )}_{J, \pi }(p) &=& m_\pi^4\,\Theta (p-\mu_\pi) \,
\frac{(p^2-\mu_\pi^2)^2}{2\,(2\pi)^3\,p^4}+ {\mathcal O} \left( Q^5\right)
\label{}
\end{eqnarray}
where we count $p\sim \mu_\pi \sim Q$. We point out that 
our expansion of the pion production spectral function is in full analogy to the 
non relativistic expansion of the two nucleon unitarity cut where one expands in powers of  
the relative momentum $p$ also around the threshold value $p=0$. The real part is derived from 
the dispersion relation:
\begin{eqnarray}
J^{(\pi )}_{\pi ,R }(p,z)  &=&\frac{m_\pi^4}{(2\pi)^4}\,\Bigg(
-\frac{\mu_\pi^2}{p^2}+\frac{(p^2-\mu_\pi^2)^2}{p^4}\,
\ln \Bigg|\frac{\mu_\pi^2}{\mu_\pi^2-p^2} \Bigg|
\nonumber\\
&-&\frac{\mu_\pi^2}{z^2}-\frac{(z^2+\mu_\pi^2)^2}{z^4}\,
\ln \Bigg|\frac{\mu_\pi^2}{\mu_\pi^2+z^2} \Bigg|
\nonumber\\
&+&i\,\pi\,\Theta (p-\mu_\pi )\,\frac{(p^2-\mu_\pi^2)^2}{p^4}\,
\Bigg)+{\mathcal O}\left(Q^5 \right)
\label{jpi-prod}
\end{eqnarray}
properly subtracted at $p^2=-z^2$. Note that the chiral order 
$J^{(\pi )}_{\pi ,R }(p,z) \sim Q^4 $ is deduced by applying simply Weinberg's counting rules. 
In other words due to the absence of an intermediate nucleon pair state  there is no 
enhancement factor for the pion production spectral density.

\subsection{Pion production loops}

We consider the self energy correction to the one-loop bubble diagram
\begin{eqnarray}
J_\Sigma(s)
&=&-i\,\tr \int \frac {d^4 l}{(2\pi)^4} \,C^{-1}\gamma_5
\,S_F\left(l+{\textstyle {1\over2}} w \right)\,\Sigma\left(l+{\textstyle {1\over2}} w \right)
S_F\left(l+{\textstyle {1\over2}} w \right)
\nonumber\\
&&\;\;\;\;\;\;\gamma_5 \,C
\,S^t_F\left(-l+{\textstyle {1\over2}} w \right)
\nonumber\\
&-&i\,\tr \int \frac {d^4 l}{(2\pi)^4} \,C^{-1}\gamma_5
\,S_F\left(l+{\textstyle {1\over2}} w \right)\gamma_5 \,C
\,S^t_F\left(-l+{\textstyle {1\over2}} w \right)
\nonumber\\
&&\;\;\;\;\;\;\Sigma^t \left(-l+{\textstyle {1\over2}} w \right)
S^t_F\left(-l+{\textstyle {1\over2}} w \right)
\label{mut2-recall-1}
\end{eqnarray}
with $w^2=s$ and the self energy 
\begin{eqnarray}
\Sigma(p) &=&\Sigma_{\pi N} - 3\,g^2_A\,\lambda^2 \,i\int \frac{d^4l}{(2 \pi)^4}\,
\gamma_5\,\lslash \,S_F(p+l)\,\gamma_5\,\lslash\, S_\pi(l) +c.t.
\label{}
\end{eqnarray}
receiving contributions from the pion nucleon sigma term $\Sigma_{\pi N}\simeq 45 $ MeV and 
suitable counter terms. We also consider the two-loop function, $J_V(s)$, induced by the pion 
production vertex in (\ref{pi-prod-vertex}):
\begin{eqnarray}
J_V(s)
&=&-2\,i\,\tr \int \frac {d^4 l}{(2\pi)^4} \,C^{-1}
\,\,V\left(l+{\textstyle {1\over2}} w \right)
S_F\left(l+{\textstyle {1\over2}} w \right)
\gamma_5 \,C \,S^t_F\left(-l+{\textstyle {1\over2}} w \right)
\nonumber\\
&-&2\,i\,\tr \int \frac {d^4 l}{(2\pi)^4} \,C^{-1}\gamma_5
\,S_F\left(l+{\textstyle {1\over2}} w \right)
\bar V \left(-l+{\textstyle {1\over2}} w \right)
C\,S^t_F\left(-l+{\textstyle {1\over2}} w \right)\;,
\nonumber\\
V(p) &=& - i\int \frac{d^4l}{(2 \pi)^4}\,\lslash \,S_F(p+l)\,
\gamma_5\,\lslash\, S_\pi(l) \;.
\label{mut2-recall-2}
\end{eqnarray}
with the vertex function $\bar V(p) = \gamma_0\,V^\dagger (p) \,\gamma_0$.
Note that in (\ref{mut2-recall-1},\ref{mut2-recall-2}) we suppress the isospin space 
but include isospin factors effectively. 

The two-loop functions $J_\Sigma(s)$ and $J_V(s)$ have two types of s-channel cuts. 
Cuts which do not separate the pion line simply give rise to a renormalization of the 
nucleon mass in the one-loop bubble $J(s)$ in case of $J_\Sigma(s)$ and a renormalization
of the bare coupling in case of $J_V(s)$. 
They are effectively taken into account by using the physical nucleon mass 
in $J(s)$ and a renormalized bare coupling. 
The more interesting cuts with the pion line separated reflect the pion production
process. We derive the leading contribution:
\begin{eqnarray}
\rho^{(\pi )}_{J, \Sigma }(p) &=& 3\,m_\pi^4\,\Theta (p-\mu_\pi) \,
\frac{(p^2-\mu_\pi^2)^2}{2\,(2\pi)^3\,p^4}+ {\mathcal O} \left( Q^5\right) \;,
\nonumber\\
\rho^{(\pi )}_{J, V }(p) &=& m_\pi^4\,\Theta (p-\mu_\pi) \,
\frac{(p^2-\mu_\pi^2)^2}{(2\pi)^3\,p^2\,m}+ {\mathcal O} \left( Q^7\right)
\label{}
\end{eqnarray}
and conclude $J^{(\pi)}_{\Sigma, R}(p,z)= 
3\,J^{(\pi )}_{\pi ,R }(p,z)+{\mathcal O} \left( Q^5\right)$ 
(see (\ref{jpi-prod})). Again we confirm that Weinberg's counting rules can be applied for 
pion production cuts.

\subsection{Pion nucleon box}

Consider the pion nucleon box diagram
\begin{eqnarray}
B =-i\,\int \frac{d^4l}{(2\pi)^4} && S_\pi (l-{\textstyle
{1\over2}}\, Q)\,
\tau_i\,\gamma_5 \left(\lslash +{\textstyle {1\over2}}\,\Qslash \right)\,
S_F(P-l)\, \tau_j\,\gamma_5 \left(-\lslash +{\textstyle
{1\over2}}\,\Qslash \right)
\nonumber\\
\otimes \,
\tau_i \,\gamma_5 && \left(\lslash +{\textstyle {1\over2}}\,\Qslash \right)\,
S_F(K+l)\,
\tau_j\,\gamma_5 \left(-\lslash +{\textstyle {1\over2}}\,\Qslash \right)\,
S_\pi (l+{\textstyle {1\over2}}\, Q) \; .
\label{pv-box}
\end{eqnarray}
where the tensor product $'\otimes ' $ is defined via the replacement rule
\begin{eqnarray}
\bar \Gamma  \otimes \Gamma  \rightarrow
\Big(\bar u(p_2')\,\bar \Gamma \, u(p_2)\Big)
\Big(\bar u(p_1')\, \Gamma \,u(p_1) \Big) 
\label{}
\end{eqnarray}
with the incoming momenta $p_{1,2}$ and the outgoing momenta $p'_{1,2}$ of the nucelons.
On-shell projection simplifies the diagram considerably. The isospin one component, 
$B^{(1)}$, follows with:
\begin{eqnarray}
B^{(I=1)}&=& X^{\mu \nu} \, \left( \gamma_\mu
\otimes \gamma_\nu\right) +  \left( \gamma_\mu \otimes X^\mu +
\bar X^\mu \otimes\gamma_\mu  \right) +X
\label{apbx8}
\end{eqnarray}
where
\begin{eqnarray}
X_{\mu \nu } &=& 16\,m^4 \,i\int \frac{d^4l}{(2\pi)^4} \,l_\mu \,l_\nu\,
S_\pi (l-{\textstyle {1\over2}}\, Q)\,S_N(P-l)\,
S_N(K+l)\,S_\pi (l+{\textstyle {1\over2}}\, Q)\; ,
\nonumber\\
\bar X_{\mu } &=& 4\,m^2 \,i\int \frac{d^4l}{(2\pi)^4} \,l_\mu \,\Big( \lslash+2\,m\Big)
S_\pi (l-{\textstyle {1\over2}}\, Q)\,S_N(K+l)\,S_\pi (l+{\textstyle {1\over2}}\, Q)\; ,
\nonumber\\
X_{\mu } &=& 4\,m^2 \,i\int \frac{d^4l}{(2\pi)^4} \,l_\mu \,\Big( \lslash-2\,m\Big)
S_\pi (l-{\textstyle {1\over2}}\, Q)\,S_N(P-l)\,S_\pi (l+{\textstyle {1\over2}}\, Q)\; ,
\nonumber\\
X &=& -i\int \frac{d^4l}{(2\pi)^4} \,\Big( 2\,m-\lslash\Big) 
\otimes \Big( 2\,m+\lslash \Big)
S_\pi (l-{\textstyle {1\over2}}\, Q)\,S_\pi (l+{\textstyle {1\over2}}\, Q) \; .
\label{}
\end{eqnarray}
It is useful to introduce auxiliary loop functions $J_i(t)$ and $b_i(s,t) $ which exhaust 
the structure of $X_{\mu \nu }, X_\mu, \bar X_\nu $ and $X$: 
\begin{eqnarray}
X_{\mu \nu} &=& b_1(s,t)\,g^{\mu \nu}
+\frac{b_2(s,t)}{m^2}\,\Big(P+K \Big)^\mu \,\Big( P+K\Big)^\nu
\nonumber\\
&+&\frac{b_3(s,t)}{m^2}\,\Big(P-K \Big)^\mu \,\Big( P-K\Big)^\nu\; ,
\nonumber\\
\bar X_\mu &=&-{\textstyle {1\over2}}\,J^{(\pi )}_1(t)\,  \gamma_\mu
- \left({\textstyle {1\over2}}\,J^{(\pi )}_2(t)+2\, J^{(\pi )}_4(t)\right) \frac{K_\mu}{m}\; ,
\nonumber\\
X_\mu &=&-{\textstyle {1\over2}}\,J^{(\pi )}_1(t)\,  \gamma_\mu
- \left({\textstyle {1\over2}}\,J^{(\pi )}_2(t)+2\, J^{(\pi )}_4(t)\right) \frac{P_\mu}{m}\; ,
\nonumber\\
X&=& {\textstyle {1\over3}}\, J^{(\pi )}_0(t)\,\left(  \gamma_\mu \otimes \gamma^\mu \right)
-2\,J^{(\pi )}_3(t)\,\left( 1\otimes 1\right) \; .
\label{aux-def}
\end{eqnarray}
Here we drop terms proportional to $Q_\mu$ or $Q_\nu $ since they would not
contribute to the on-shell scattering matrix. Note that transforming to the on-shell 
irreducible one-pion exchange with pseudo-scalar coupling 'moves' all $J^{(\pi)}_i$ 
functions into the on-shell irreducible kernel. 
The invariant functions $b_i(s,t)$ follow easily after applying the substitution rule
\begin{eqnarray}
l_\mu l_\nu && \rightarrow
g_{\mu \nu}\left(
l^2-\frac{\left(l\cdot Q\right)^2}{Q^2}
-\frac{\left(l\cdot W\right)^2}{W^2}
-\frac{\left(l\cdot (K-P)\right)^2}{(K-P)^2}\right)
\\
&+& \frac{W_\mu W_\nu}{W^2}
\left(
\frac{\left(l\cdot Q\right)^2}{Q^2}
+2\,\frac{\left(l\cdot W\right)^2}{W^2}
+\frac{\left(l\cdot (K-P)\right)^2}{(K-P)^2}
-l^2\right)
\nonumber\\
&+& \frac{(K-P)_\mu (K-P)_\nu}{(K-P)^2}
\left(
\frac{\left(l\cdot Q\right)^2}{Q^2}
+\frac{\left(l\cdot W\right)^2}{W^2}
+2\,\frac{\left(l\cdot (K-P)\right)^2}{(K-P)^2}
-l^2\right)
\nonumber
\label{apbx13}
\end{eqnarray}
in (\ref{aux-def}) with $W=P+K$ which is justified by covariance. 
The reduced box functions, $b_i(u,t)$, can now be 
expressed in terms of the master loop functions systematically:
\begin{eqnarray}
b_1(s,t)  &=&16\, m^4\,
\left(\frac{\left(m_\pi^2-{\textstyle {t\over2}}\right)^2}{u}
- m_\pi^2+{\textstyle {t\over4}}  \right)\,
I_{4,N}(s,t)
\nonumber\\
&+&16\, m^4\, \left(
\frac{1}{2}-
\frac{m_\pi^2-{\textstyle {t\over2}}}{u}
\right)\, I_{3,N}(s)
+16\, m^4\,\frac{m_\pi^2-{\textstyle {t\over2}}}{u}\, I_{3,\pi }(t)\; ,
\nonumber\\
b_3(s,t)  &=&16\, \frac{m^6}{u}\,
\left( m_\pi^2-{\textstyle {t\over4}}
-2\,\frac{\left(m_\pi^2-{\textstyle {t\over2}}\right)^2}{u}
\right)\,
I_{4,N}(s,t)
\nonumber\\
&-&16\, \frac{m^6}{u}\, \left(\frac{1}{2}-\frac{m_\pi^2}{u+t}-
2\,\frac{m_\pi^2-{\textstyle {t\over2}}}{u}
\right)\, I_{3,N}(s)
\nonumber\\
&-&32\, \frac{m^6}{u}\,
\,\frac{m_\pi^2-{\textstyle {t\over2}}}{u}\, I_{3,\pi }(t)
-16\, \frac{m^6}{u(u+t)}\Big( I_{2,N}(s)-I_2(m^2) \Big)
\nonumber\\
&-&16\, \frac{m^6}{u}\,
\,\frac{m_\pi^2-{\textstyle {t\over2}}}{s+u}\, I_{3,\pi }(t)
-16\, \frac{m^6}{u(s+u)}\Big( I_{2,\pi}(t)-I_2(m^2) \Big)  
\label{b-1-3-result}
\end{eqnarray}
with $u=4\,m^2-s-t$. The box function $b_2(s,t)$ can in fact be written in terms of 
$b_1(s,t)$ and $J^{(\pi)}_4(t)$ as follows:
\begin{eqnarray}
b_2(s,t) = -\frac{m^2}{s}\, b_1(s,t)+2\, \frac{m^2}{s}\, J_4^{(\pi)}(t) \; .
\label{}
\end{eqnarray}
It remains the evaluation of the reducible loops $J^{(\pi)}_i(t)$. Applying 
similar methods we derive:
\begin{eqnarray}
J^{(\pi )}_0(t)&=& \frac{1}{2}\, I_{1,\pi}-
\left( m_\pi^2-\frac {t}{4} \right)\, I_{2, \pi}(t)\; ,
\nonumber\\
J^{(\pi )}_1(t) &=&4\, m^2 \left(
\frac{\left(m_\pi^2-{\textstyle {1\over2}}\,t\right)^2}{s+u}
-m_\pi^2+\frac{t}{4}\right)\, I_{3,\pi}(t)
\nonumber\\
&+&\frac{4\, m^2}{s+u}\, \left(m_\pi^2-\frac{t}{2} \right)
\Big(I_{2,\pi }(t)-I_2(m^2) \Big)
+2\, m^2\, I_2(m^2)\; ,
\nonumber\\
J^{(\pi )}_2(t) &=&\frac{16\, m^4}{s+u}
\left( m_\pi^2-\frac{t}{4}
-3\,\frac{\left(m_\pi^2-{\textstyle {1\over2}}\,t\right)^2}{s+u}
\right)\, I_{3,\pi}(t)
\nonumber\\
&-&3\,\frac{16\, m^4}{(s+u)^2}\,
\left(m_\pi^2-\frac{t}{2} \right)\, I_{2,\pi }(t)
\nonumber\\
&+&\frac{16\, m^4}{s+u}\,
\left(3\,\frac{m_\pi^2-{\textstyle {1\over2}}\,t}{s+u}-
\frac{1}{2}+ \frac{m_\pi^2}{2\, m^2} \right)\, I_2(m^2)
+\frac{8\, m^2}{s+u}\Big(  I_{1,\pi}-I_{1,N}\Big)\; ,
\nonumber\\
J^{(\pi )}_3(t)&=&-2\, m^2\,I_{2,\pi }(t)
\nonumber\\
J^{(\pi )}_4(t)&=&8\, \frac{m^4}{s+u}
\left( m_\pi^2-{\textstyle {1\over2}}\, t\right)\, I_{3,\pi}(t)
+8\, \frac{m^4}{s+u}\,\Big( I_{2,\pi}(t)-I_2(m^2) \Big) 
\label{}
\end{eqnarray}
with $s+u=4\,m^2-t$.
Finally it is convenient to decompose the on-shell projected pion nucleon box into 
our Dirac tensors $T_i$, with simple exchange properties
\begin{eqnarray}
B(p_2',p_2;p_1',p_1) ={\mathcal P}_0\, \sum_{i=1}^5 \,
B_i^{(0)}(u,t)\, T_i +{\mathcal P}_1\, \sum_{i=1}^5 \,
B_i^{(1)}(u,t)\, T_i \; .
\label{}
\end{eqnarray}
The exchange box diagram is now deduced by properly exchanging
$u$ and $t$ in the reduced functions $B_i(u,t)$ (see (\ref{ampl4})). Note that 
the isospin zero amplitudes are $B_i^{(0)}(u,t) = 9\,B_i^{(1)}(u,t) $. Here we 
provide the necessary on-shell tensor identities (see \cite{Brown-Jackson}):
\begin{eqnarray}
\left( \Kslash-m \right) \otimes \left(m- \Pslash \right)
&=&\left({\textstyle {1\over4}}\, t -{\textstyle {3\over4}}\, u  \right)
 T_3
+{\textstyle {1\over4}} \left( u+t\right) T_4
+{\textstyle {1\over4}} \left( u+t\right) T_5\; ,
\nonumber\\
\Kslash \otimes m+ m\otimes \Pslash &=&
\left( m^2-{\textstyle {1\over4}} \left( u-t\right) \right)  T_1
+\left( m^2-{\textstyle {1\over4}} \left(u+t\right) \right)  T_2
\nonumber\\
&-&\left( {\textstyle {1\over2}}\, m^2
       +{\textstyle {1\over4}} \left(u-t\right) \right)  T_3
-\left( {\textstyle {1\over2}}\, m^2
       -{\textstyle {1\over4}} \left(u+t\right) \right)  T_4
\nonumber\\
&+&{\textstyle {1\over2}} \,m^2 \, T_5\; ,
\nonumber\\
\Kslash \otimes \Pslash &=&
\left({\textstyle {1\over2}}\, m^2
      -{\textstyle {1\over4}} \left( u-t\right) \right)  T_1
+\left({\textstyle {1\over2}} \,m^2
      -{\textstyle {1\over4}} \left( u+t\right) \right)  T_2
\nonumber\\
&-&\left({\textstyle {3\over4}} \,m^2
      -{\textstyle {1\over2}}\, u \right)  T_3
- {\textstyle {1\over4}}\, m^2 \, T_4
\nonumber\\
&+&\left({\textstyle {3\over4}}\, m^2
      -{\textstyle {1\over4}} \left(u+t\right) \right)  T_5\; ,
\nonumber\\
\left( \Kslash+m \right) \otimes \left( m+\Pslash \right)
&=&
\left(2\, m^2
      -{\textstyle {1\over2}} \left( u-t\right) \right)  T_1
+\left(2 \,m^2
      -{\textstyle {1\over2}} \left( u+t\right) \right)  T_2
\nonumber\\
&-&\left(m^2
      -{\textstyle {1\over4}}\, (u+t) \right)
\left(  T_3+  T_4- T_5\right) \; .
\label{}
\end{eqnarray}
We collect the leading terms of chiral order one:
\begin{eqnarray}
B_1^{(1)}(u,t) &=&
-2\,m^2\left(m_\pi^2-{\textstyle {t\over 2}}\right) I_{3,\pi}(t)
-{\textstyle {t\over 4\,m^2}}\, b_1(s,t)
\nonumber\\
&+&2\,m^2\,I_2(m^2)-{\textstyle {11\over 12}}\,I_{1,\pi}+\cdots\; ,
\nonumber\\
B_2^{(1)}(u,t) &=&
-2\,m^2\left(m_\pi^2-{\textstyle {t\over 2}}\right) I_{3,\pi}(t)
+2\,m^2\,I_2(m^2)-{\textstyle {11\over 12}}\,I_{1,\pi}+\cdots\; ,
\nonumber\\
B_3^{(1)}(u,t) &=&
-{\textstyle {3\over 4}}\, b_1(s,t)
+{\textstyle {t-3\,u\over 4\,m^2}}\,b_3(s,t)+\cdots\; ,
\nonumber\\
B_4^{(1)}(u,t) &=& {\textstyle {1\over 4}}\, b_1(s,t) +{\textstyle
{u+t\over 4\,m^2}}\,b_3(s,t)+\cdots\; ,
\nonumber\\
B_5^{(1)}(u,t) &=& {\textstyle {1\over 4}}\, b_1(s,t) +{\textstyle
{u+t\over 4\,m^2}}\,b_3(s,t)+\cdots
\label{leading-box}
\end{eqnarray}
where the dots represent further terms with chiral power $\sim Q^2 $ 
not relevant in this work. 

It is instructive to consider also the pion-nucleon box diagram defined with a 
pseudo-scalar pion-nucleon coupling vertex (PS). In our calculation  
the pseudo-scalar box follows formally from the pseudo-vector box diagram 
defined in (\ref{pv-box}) by dropping the $X_\mu, \bar X_\mu$ and $X$ terms in (\ref{apbx8}). 
We observe that the difference of pseudo-vector and pseudo-scalar box 
diagram carries leading chiral order $Q^0$:
\begin{eqnarray}
B_1^{(1)}(u,t) -B_1^{(1,PS)}(u,t) &=& 
-4\,m^2\left(m_\pi^2-{\textstyle {t\over 2}}\right) I_{3,\pi}(t)-2\,m^2\,I_{2,\pi}(t)
\nonumber\\
&+&4\,m^2\,I_2(m^2) -{\textstyle {11\over 12}}\,I_{1,\pi}+{\mathcal O} \left( Q^2 \right)\;,
\nonumber\\
B_2^{(1)}(u,t) -B_2^{(1,PS)}(u,t) &=& 
-4\,m^2\left(m_\pi^2-{\textstyle {t\over 2}}\right) I_{3,\pi}(t)-2\,m^2\,I_{2,\pi}(t)
\nonumber\\
&+&4\,m^2\,I_2(m^2) -{\textstyle {11\over 12}}\,I_{1,\pi}+{\mathcal O} \left( Q^2 \right)\;.
\label{PV-PS}
\end{eqnarray}
This implies that the on-shell irreducible kernel 
$\hat K =\sum_{i=1}^5\,\hat g_i(u,t)\,T_i\,{\mathcal P}_1$ of
(\ref{matching-xx}) receives a non-local contribution. For example:
\begin{eqnarray}
\hat g_1(u,t) &=& \frac{1}{2}\,g_A^4\,\lambda^4\Big(  B_1^{(1)}(u,t) -B_1^{(1,PS)}(u,t) \Big)
\nonumber\\
&+&\frac{1}{2}\,g_A^4\,\lambda^4\Big(  B_1^{(1)}(t,u) -B_1^{(1,PS)}(t,u) \Big) \;.
\label{non-local}
\end{eqnarray}
The non-local terms of (\ref{non-local}) are then included in $\hat K_\pi$ where 
$\hat K =\hat K_\pi +\hat K_\sigma$. The pionic tadpole in (\ref{PV-PS}) is included 
in $\hat K_\sigma$ and absorbed into the coupling function $g(s)$. Since both $K_\pi$ and
$\hat K_\pi$ are treated in perturbation theory our procedure (\ref{matching-xx}) 
defines a systematic procedure to absorb divergent terms, which are not associated with the 
s-channel unitartity cut, into the local bare coupling function $\hat K_\sigma $. At subleading 
order it leads to the inclusion of the pionic tadpole $I_{1,\pi}$ into $g(s)$.

We close our essay on the pion-nucleon box diagram with the relevant combination for 
s-wave scattering in the spin singlet channel:
\begin{eqnarray}
T_{[^1S_0]}(u,t)&=&
a^{(1)}_1(u,t)-\frac{u-t}{s}\,a^{(1)}_4(u,t)-\frac{u+t}{s}\,a^{(1)}_5(u,t)
\nonumber\\
&=&2\,g_A^4\,\lambda^4\,m_\pi^4\,m^2\,I_{4,N}(s,t)
-4\,g_A^4\,\lambda^4\,m_\pi^2\,m^2\,I_{3,N}(s)
\nonumber\\
&+&2\,g_A^4\,\lambda^4\,m^2\,I_{2,N}(s) +m^2\,{\mathcal O }
\left(\frac{1}{m^2} \right)+\Big( u \leftrightarrow t\Big) \; 
\label{}
\end{eqnarray}
where we now dropped the pionic tadpole contribution.

\subsection{Two-loop vertex}

The two-loop vertex function is
\begin{eqnarray}
&&{\mathcal V}^{(i)}_{\pi \pi }(p_2,p_1) = i\,\int \frac{d^4
l}{(2\pi)^4} \left( S_F(P+l)\, \vec \tau
\,\gamma_5\,(\lslash+{\textstyle {1\over2}}\, \Qslash )\right)^t
{\mathcal V}_{ \pi }^{(i)}(P+l,K-l)
\nonumber\\
&&\;\;\;\;\;\;\;\;\;\;\;\;\;\;\;\;\;\;\cdot S_\pi (l+{\textstyle {1\over2}} \,Q )\, S_F(K-l)\, \vec
\tau\,\gamma_5\, (\lslash+{\textstyle {1\over2}}\, \Qslash ) \;,
\nonumber\\
&&u^t(p_2)\,{\mathcal V}^{(i)}_{\pi \pi }(p_2,p_1)\, u(p_1)= V_{\pi
\pi }(s)\, u^t(p_2)\left( C\,
\gamma_5 \, \frac{1}{2}\,\tau_2\, \tau_i \right)u(p_1) \; .
\label{twoloopvertex}
\end{eqnarray}
We evaluate $V_{\pi \pi}(s) $ by imposing a dispersion relation on the renormalized 
vertex $s\,V_{\pi \pi ,R}(s) $ with
\begin{eqnarray}
V_{\pi \pi}(s) &=& V_{\pi \pi ,R}(s)
-\bar c(s)\,J(s)\,V_{\pi ,R}(s)
-\bar c(s)\,\Big( \bar J_\pi(s)+J_\pi^{(\pi)}(s) \Big)
\nonumber\\
&+&\bar c^2(s)\,J^2(s)+V_{\pi \pi,R}^{(\pi)}(s)\; .
\label{}
\end{eqnarray}
Our renormalization via loop subtraction keeps track of the overlapping divergency structure
of the vertex function $V_{\pi \pi }(s) $. The unique contribution from the one-pion 
production cut is denoted by $V_{\pi \pi,R}^{(\pi)}(s)$. With $\bar c(s)$ given in 
(\ref{ren-vpi}) the renormalized loop functions $V_{\pi \pi,R}(s) $ and 
$V^{(\pi)}_{\pi \pi ,R}(s)$ are finite. Note that the pion production contribution 
$V^{(\pi)}_{\pi \pi ,R}(s)$ is renormalized by the proper subtraction of the pion 
production loop $J_{\pi}^{(\pi)}(s)$. From (\ref{jpi-prod}) we expect 
$V^{(\pi)}_{\pi \pi ,R}(s)$ to carry chiral order $Q^4$.
The spectral function of $V_{\pi \pi,R}(s)$ follows in terms of the renormalized 
one-loop vertex $V_{\pi ,R}(s)$ and an effective pion-nucleon box $B_{\pi \pi}(s)$:
\begin{eqnarray}
\Im \,V_{\pi \pi,R}(s) &=&\Big(\Im \,V_{\pi ,R} (s)\Big)\,\Big(\Re\,V_{\pi,R}(s)\Big)
+\Big(\Im\,J(s)\Big)\,\Big(\Re\,B_{\pi \pi}(s)\Big) \; ,
\nonumber\\
B_{\pi \pi}(s)&=&2\,m^2\,m_\pi^4\,
\int_{-1}^{1} \frac{d\,x}{2}\,I_{4,N}(s,t)\Bigg|_{t=-2\,p^2\,(1-x)}
+{\mathcal O} \left( Q^2\right)
\label{imvpipi}
\end{eqnarray}
with $s=4\,(m^2+p^2)$. Note that the function $B_{\pi \pi}(s) $ coincides with 
the s-wave projected pion-nucleon box diagram $R_{\pi \pi}(p)$ (see (\ref{Rpipi}))
as should be expected. The renormalized loop is decomposed into its chiral moments:
\begin{eqnarray}
V_{\pi \pi,R}(s) = \sum_{n=2}^\infty\,V_{\pi \pi,R}^{(n)}(p)
\label{}
\end{eqnarray}
where $V_{\pi \pi,R}^{(n)}(p)$ carries chiral power $n$. 
The real part of the two-loop function $V_{\pi \pi,R}^{(2)}(p)$ can be easily derived without 
performing further integrations. Recall that $V_{\pi \pi,R}^{(2)}(p)$ 
satisfies a dispersion relation in $p$: 
\begin{eqnarray}
V_{\pi\pi,R}^{(2)}(p) &=& \int_{4\,m^2}^\infty \frac{d s'}{\pi}\, 
\frac{\Im\,V_{\pi\pi,R}^{(2)}(p')}{s'-s-i\,\epsilon}
= \int_{-\infty}^\infty \frac{d p'}{\pi}\, 
\frac{\Im\,V_{\pi\pi,R}^{(2)}(p')}{p'-p-i\,\epsilon} \; .
\label{}
\end{eqnarray}
Therefore, if considered as a function of $p$ with $s=4\,(m^2+p^2)$, the chiral moment 
$V_{\pi \pi,R}^{(2)}(p)$ is analytic in the upper complex half plane. We note that 
the full vertex function $V_{\pi \pi}(s)$ on the other hand is not analytic in the upper 
complex half plane if considered as a function of p. It shows branch points at $p=\pm i\,m$ 
which are systematically eliminated performing the $1/m$ expansion. 

Start with the first term in (\ref{imvpipi}). The result can immediately be deduced  
from the simple identity:
\begin{eqnarray}
\Im \,\ln^2 \left( 1-2\,i\,\frac{p}{m_\pi}\right) &=&
-\arctan \left(2\,\frac{p}{m_\pi} \right)\,
\ln  \left( 1+4\,\frac{p^2}{m_\pi^2}\right) \; .
\label{}
\end{eqnarray}
The second term follows with the somewhat more involved identity:
\begin{eqnarray}
\Re \,R_{\pi \pi}(p)&=&
\frac{m\,m_\pi}{4\,\pi }
\,\frac{m_\pi^3}{8\,p^3} \,
\Im\,\Bigg(\Li\left(\frac{1}{2}+\frac{i\,p}{m_\pi}\right)
-2\,\Li\left(\frac{1}{2}\,\frac{m_\pi}{m_\pi-i\,p}\right)
+\frac {\pi^2}{12}
\nonumber\\
&&\;\;\;\;\;\;\;\;\;\;\;\;\;\;
-\ln^2 \left(\frac{m_\pi-2\,i\,p}{2\,m_\pi-2\,i\,p }\right)
+\frac{1}{2}\,\ln^2\left(\frac{1}{2}-\frac{i\,p}{m_\pi } \right)
 \Bigg) \; .
\label{}
\end{eqnarray}
For both terms we succeed in writing the spectral function as the imaginary part of 
an analytic function with branch cuts only on the lower complex half plane. Thus it is
straightforward to reconstruct the full analytic function. 
We collect all terms and display our result\footnote{Note the identity:
$\Li\left({\textstyle{1\over 2}}\right)
={\textstyle{\pi^2\over 12}}-{\textstyle{1\over 2}}\,\ln^2 2 $}:
\begin{eqnarray}
V_{\pi \pi,R }^{(2)}(p)&=&
\frac{m^2\,m^2_\pi}{(4\,\pi)^2 }
\,\frac{m_\pi^2}{4\,p^2} \,
\Bigg(\Li\left(\frac{1}{2}+\frac{i\,p}{m_\pi}\right)
-2\,\Li\left(\frac{1}{2}\,\frac{m_\pi}{m_\pi-i\,p}\right)
+\Li\left(\frac{1}{2}\right)
\nonumber\\
&&\;\;\;\;\;\;\;\;\;
-\ln^2 \left(\frac{m_\pi-2\,i\,p}{2\,m_\pi-2\,i\,p }\right)
-\ln \left(\frac{1}{2}-\frac{i\,p}{m_\pi } \right)\,\ln \,2  \Bigg) \; .
\label{}
\end{eqnarray}

\subsection{Three-loop bubble }

We turn to the three-loop bubble
\begin{eqnarray}
J_{\pi \pi}(s) &=&-\frac{i}{2}\,\tr
\int \frac{d^4 l}{(2\pi)^4}
\tau_j \,\tau_2 \,\gamma_5\,C^{-1} S^t_F(P+l)\,{\mathcal V}^{(j)}_{\pi \pi }(P+l,K-l)\,S_F(K-l)
\; .
\nonumber\\
\label{}
\end{eqnarray}
which is defined in terms of the full off-shell two-loop vertex of
(\ref{twoloopvertex}). We introduce the loop-subtracted bubble function,
$\bar J_{\pi \pi } (s) $, with
\begin{eqnarray}
J_{\pi \pi}(s) =\bar J_{\pi \pi}(s)-2\,\bar c(s)\,J(s)\,\bar J_{\pi }(s)
+\bar c^{\,2}(s)\,J^3(s)+\cdots
\label{}
\end{eqnarray}
where we dropped terms of suppressed chiral orders induced by one and two-pion production.
The loop $\bar J_{\pi \pi } (s) $ is decomposed into its chiral moments:
\begin{eqnarray}
\bar J_{\pi \pi } (s) =\sum_{n=3}^\infty\,\bar J_{\pi \pi }^{(n)} (p)
\label{}
\end{eqnarray}
where $\bar J_{\pi \pi}^{(n)}(p)$ carries chiral power $n$. 
The spectral function is expressed in terms of the renormalized one and two-loop vertex 
functions $V_{\pi ,R}(s)$ and $V_{\pi \pi,R}(s)$
\begin{eqnarray}
\Im \,\bar J_{\pi \pi }(s) &=& 2\, \Big(\Im\,J(s) \Big)\,
\Big( \Re \, V_{\pi \pi,R} (s)\Big)
+\Big(\Im\,J(s)\Big)\, \Big( V_{\pi ,R }(s)\, V^*_{\pi ,R}(s) \Big) \;.
\label{}
\end{eqnarray}
For the leading moment $\bar J^{(3)}_{\pi \pi }(p)$ we find in analogy to (\ref{jpi-identity}) 
the relation:
\begin{eqnarray}
\bar J^{(3)}_{\pi \pi }(p)-\bar J^{(3)}_{\pi \pi  }(0) &=&i\,\frac{m\,p}{2\,\pi}
\left(2\, V^{(2)}_{\pi \pi ,R } (p)
+\Big|V_{\pi ,R}^{(1)}\Big|^2_{c}(p) \right)
\Big)
\label{}
\end{eqnarray}
where we introduced the proper analytic continuation $|V_{\pi ,R}^{(1)}|^2_{c}(p)$ of 
$|V_{\pi ,R}^{(1)}(p)|^2$ with:
\begin{eqnarray}
V_{\pi,R}^{(1)}(p)\,V_{\pi,R }^{(1),*}(p)
&=&\Re \, \Big|V_{\pi ,R}^{(1)}\Big|^2_{c}(p)
\nonumber\\
&=&\frac{m^2\,m_\pi^2}{(4\,\pi )^2}
\,\frac{m_\pi^2}{2\,p^2}\;
\;\Re \,\Bigg(
\Li\left(\frac{1}{2}\right)
-\Li \left(\frac{1}{2}+\frac{i\,p}{m_\pi} \right)
\nonumber\\
&+&\ln \left(2\right)\,\ln\left(1-2\,\frac{i\,p}{m_\pi} \right)
+\frac{i\,p}{m_\pi}\,\ln \left(16 \right)
\Bigg) \; .
\label{}
\end{eqnarray}
We collect all terms and present the leading chiral order contribution
\begin{eqnarray}
\bar J^{(3)}_{\pi \pi }(p)-\bar J^{(3)}_{\pi \pi}(0) &=&-i\,
\frac{m^3\,m^3_\pi}{(4\,\pi)^3 }
\,\frac{m_\pi}{p} \,
\Bigg(
2\,\Li\left(\frac{1}{2}\,\frac{m_\pi}{m_\pi-i\,p}\right)
\nonumber\\
&-&\frac{i\,p}{m_\pi}\,\ln\left( 16\right)
+\ln^2 \left(\frac{m_\pi-2\,i\,p}{2\,m_\pi-2\,i\,p }\right)
-\frac{\pi^2}{6}
\Bigg)\; .
\label{jpipi-result}
\end{eqnarray}
Note that the loop function $\bar J^{(3)}_{\pi \pi}(p) $ is convergent with
\begin{eqnarray}
\bar J^{(3)}_{\pi \pi}(0) =
\int_0^{\infty } \frac{d \,\bar p^2}{\pi }\,
\frac{\Im \,\bar J^{(3)}_{\pi \pi }(\bar p)}{\bar p^2 }
=\ln (16)\,\frac{m^3\,m_\pi^3}{(4\,\pi)^3} \;.
\label{jpipi-0}
\end{eqnarray}
The integral (\ref{jpipi-0}) follows from (\ref{jpipi-result}) and the observation 
$\bar J^{(3)}_{\pi \pi}(\infty ) =0 $. The leading order renormalized three-loop function is 
$J_{\pi \pi,R}(p,z)=\bar J^{(3)}_{\pi \pi}(p)-\bar J^{(3)}_{\pi \pi}(-i\,z)$.

\section{NNLO terms in CE scheme}

In this appendix we systematically collect all terms of chiral order one in CE scheme. 
We emphasize that again explicit calculation shows that no $c$-dependence 
survives at this order. The auxiliary coupling function $g_H^{-1}$ receives the contributions
\begin{eqnarray}
g_H^{-1}(s,m_\pi; c ) &&= g_{\sigma ,S}^{-1}(s,m_\pi;c)-J(-i\,z_0)
- g_A^2\,\lambda^2\,J_{\pi,S}(-i\,z_0 )
\nonumber\\
&& -g_A^4\,\lambda^4\,J_{\pi \pi ,S}(-i\,z_0)
-\Big( c-{\textstyle{1\over 2}}\,\Big)^3 \,g_A^6\,\lambda^6\,J^4(-i\,z_0)
+\cdots  
\label{gren-app}
\end{eqnarray}
where we included the anomalous contribution from $J_{\pi \pi \pi ,S}$ according 
to (\ref{all-order}). It is convenient to resolve the structure of the subtraction 
constants $J_{\pi,S}(-i\,z_0)$ and $J_{\pi \pi ,S}(-i\,z_0)$ with:
\begin{eqnarray}
g_A^4\,\lambda^4\,J_{\pi,S}(-i\,z_0) &=& 
\left(c-{\textstyle{1\over2}}\right) \bar \kappa_J^2(z_0)
+g_A^2\,\lambda^2\,\Big( \bar \kappa_{J,\pi}(z_0) +\kappa^{(\pi )}_{J,\pi}(z_0) \Big)\; ,
\nonumber\\
g_A^6\,\lambda^6\,J_{\pi \pi,S}(-i\,z_0)&=&
\left(c-{\textstyle{1\over 2}}\right)\,\bar \kappa_J(z_0)
\left(\left(c-{\textstyle{1\over 2}}\right)\,\bar \kappa^2_J(z_0)+
2\, g_A^2\,\lambda^2\,\bar \kappa_{J,\pi}(z_0) \right)
\nonumber\\
&+&g_A^2\,\lambda^2\,\bar \kappa_{J,\pi \pi}(z_0)
\label{}
\end{eqnarray}
where we introduce $\bar \kappa_J(z_0) =g_A^2\,\lambda^2\,J(-i\,z_0)$,
$\bar \kappa_{J,\pi}(z_0) =g_A^2\,\lambda^2\,\bar J_{\pi }(-i\,z_0)$ and 
$\bar \kappa_{J, \pi \pi}(z_0) =g_A^4\,\lambda^4\,\bar J_{\pi \pi}(-i\,z_0)$.
For later convenience it is appropriate to generalize here our notation and introduce 
further dimensionless coefficients $\bar \kappa^{(l)}_{J,\pi^k}(z)$ and 
$\bar \kappa^{(l)}_{V,\pi^k}(z)$:
\begin{eqnarray}
\bar \kappa^{(l)}_{J,\pi^k}(z) &=&g_A^{2\,k}\,\lambda^{2\,k} \, \left(\frac{2\,\pi }{i\,m}\right)^l\,\frac{1}{l\,!}\,
\partial^l_p\,J_{\pi^k,R}(-i\,z,z) \; ,
\nonumber\\
\bar \kappa^{(l)}_{V,\pi^k}(z) &=&g_A^{2\,k}\,\lambda^{2\,k} \, \left(\frac{2\,\pi }{i\,m}\right)^l\,\frac{1}{l\,!}\,
\partial^l_p\,V_{\pi^k,R}(-i\,z) \; .
\label{def-kappa-general}
\end{eqnarray}
The auxiliary coupling $g_H$ does show an explicit $c$-dependence at chiral 
order $Q^3$. However such terms are canceled by a contribution to the scattering 
amplitude with an explicit $c$-dependence 
\begin{eqnarray}
T_{[^1S_0]}(u,t) &=& 4\,\bar \kappa_J\,\Bigg( c-\frac{1}{2}
\Bigg)\,\frac{g_H^{-1}(s)}{J_R^2(p,z)}+\cdots \; .
\label{}
\end{eqnarray}
The cancellation follows from the intermediate result
\begin{eqnarray}
g_A^2\,\lambda^2\,g_H^{-1}(s) -2\,\bar \kappa_J(z_0) 
\,\left(c-{\textstyle{1\over 2}}\right)\,g_H^{-1}(s) &=& 
g_A^2\,\lambda^2\,g_R^{-1}(p) 
+{\mathcal O}\left( Q^4\right)  
\label{gren-sub}
\end{eqnarray}
which considers the relevant combination. Again our result (\ref{gren-sub}) relies on the 
leading order expression (\ref{ren-bare}). We decompose the amplitude in its chiral moments 
$T_{[^1S_0]}(p)=\sum_{n=-1}^\infty\,T^{(n)}_{[^1S_0]}(p)$ and display our complete 
collection of terms with chiral power one:
\begin{eqnarray}
T^{(1)}_{[^1S_0]}(u,t) &=&
2\,\frac{\bar \kappa_{J,\pi \pi}(z_0)}{J_R^2(p,z)} 
+2\, \bar \zeta_3 \,\frac{m}{2\,\pi}\,\frac{m_\pi^3}{J^2_R(p,z)} 
- 2\,
\bar \zeta_2\,m_\pi^2\,\frac{m}{2\,\pi}\,\frac{\bar \kappa^{(1)}_{J, \pi}(z_0)}{J_R^2(p,z)} 
\nonumber\\
&-& 2\,\frac{g^{-2}_R(p)}{J_R^3(p,z)}-2\,\Big( 
\bar \zeta_2\,m_\pi^2 \Big)^2\,\frac{m^2}{4\,\pi^2}\,\frac{1}{J^3_R(p,z)} 
\nonumber\\
&+& 4\,\bar \zeta_2\,m_\pi^2\,\frac{m}{2\,\pi}\,\frac{g^{-1}_R(p)}{J_R^3(p,z)} 
-2\,g_A^4\,\lambda^4
\left(2\,\frac{V_{\pi \pi ,R}(p)}{J_R(p,z_0)}
-\frac{J_{\pi \pi ,R}(p,z_0)}{J^2_R(p,z_0)} \right)
\nonumber\\
&-&4\,g_A^2\,\lambda^2 \,\frac{g_R^{-1}(p)}{J_R(p,z)}
\left( \frac{V_{\pi ,R}(p)}{J_R(p,z_0) }
-\frac{J_{\pi ,R}(p,z_0)}{J^2_R(p,z_0) }\right)
\nonumber\\
&-&2\,\frac{g_A^4\,\lambda^4}{J^3_R(p,z_0)}
\Big( J_R(p,z_0)\, V_{\pi ,R}(p)- J_{\pi,R}(p,z_0) \Big)^2
\nonumber\\
&+&g_A^4\,\lambda^4\,\Big( B_{\pi \pi }(u,t) + B_{\pi \pi
}(t,u)\Big)  \; .
\label{result6}
\end{eqnarray}
Our result (\ref{result6}) involves the previously given
renormalized one-loop vertex $V_{\pi ,R}(p) \sim Q $ and two
loop bubble $J_{\pi ,R}(p,z_0) \sim Q^2 $. It further requires the renormalized two-loop
vertex $V_{\pi \pi,R}(p) \sim Q^2 $ and the three-loop bubble
$J_{\pi \pi ,R}(p,z_0) \sim Q^3$. Again all higher order pole terms must cancel. 
This requirement confirms $\bar \zeta_2 =\zeta_2^{(0)}$  and determines the hitherto unknown
coefficients $\bar \zeta_3$ 
\begin{eqnarray}
0 &=& \frac{\bar \kappa_{J,\pi \pi}(z_0)}{m_\pi^3} 
+\bar \zeta_2 \,\frac{\bar \kappa^{(1)}_{J, \pi}(z_0)}{m_\pi} \,\frac{m}{2\,\pi}
+\bar \zeta_3 \,\frac{m}{2\,\pi}
\label{}
\end{eqnarray}
in terms of the known parameters $\bar \zeta_2$ and 
$ \bar \kappa^{(1)}_{J,\pi }(z_0),\bar \kappa_{J,\pi \pi}(z_0)$
\footnote{Note that here we suppress a finite renormalization of 
$\zeta_0^{(2)}\rightarrow \zeta_0^{(2)} \,(1-2\,z_0\,\zeta_2^{(0)})$. This renormalization
leads effectively to $ g_R^{-1} \sim \zeta_2^{(0)} \,(p^2+z^2)$.}.

We make the analytic structure of our result (\ref{result6}) more explicit by 
expressing our result in terms of residual functions $R^{(n)}_{J,\pi^k }(p,z)$ 
and $R^{(n)}_{V,\pi^k }(p,z)$ which are defined 
\begin{eqnarray}
g_A^{2\,k}\,\lambda^{2\,k} \,J_{\pi^k, R}(p,z) &=&
g_A^{2\,k}\,\lambda^{2\,k} \,J_R^{n}(p,z)\,R^{(n)}_{J,\pi^k }(p,z)
+\sum_{l=0}^{n-1} \,\bar \kappa^{(l)}_{J,\pi^k}(z)\,J_R^{l}(p,z) 
\nonumber\\
g_A^{2\,k}\,\lambda^{2\,k} \,V_{\pi^k, R}(p) &=&
g_A^{2\,k}\,\lambda^{2\,k} \,J_R^{n}(p,z)\,R^{(n)}_{V,\pi^k }(p,z)
\nonumber\\
&+&\sum_{l=0}^{n-1} \,\bar \kappa^{(l)}_{V,\pi^k}(z)\,J_R^{l}(p,z) 
\label{}
\end{eqnarray} 
in  terms of the coefficients $\bar \kappa^{(l)}_{J,\pi^k}(z)$ and 
$\bar \kappa^{(l)}_{V,\pi^k}(z)$ of (\ref{def-kappa-general}). 
This leads to the central result of this appendix 
\begin{eqnarray}
T^{(1)}_{[^1S_0]}(p) &=&
\frac{4\,\pi}{m}\,\frac{i\,w^{(2)}}{p+i\,z}
+\frac{4\,\pi }{m}\,\zeta^{(0)}_2\, \Big( 
2\,\bar \kappa_{V,\pi}^{(0)}(z_0)-2\,\bar \kappa^{(1)}_{J,\pi}(z_0)-3\,\zeta^{(0)}_2 \,z_0\Big)
\nonumber\\
&-&\frac{4\,\pi }{m}\,\Big(\zeta^{(0)}_2 \Big)^2\,i\,p+2\,g_A^4\,\lambda^4\,R_{\pi \pi }(p)
\nonumber\\
&+&4\,g_A^2\,\lambda^2\,
\zeta^{(2)}_0\,m_\pi^2\,\frac{m}{2\,\pi}\,
\Big(R^{(3)}_{J,\pi }(p,z_0)-R^{(2)}_{V,\pi }(p,z_0)\Big)
\nonumber\\
&-&4\,g_A^2\,\lambda^2\,\Big(R^{(2)}_{J,\pi}(p,z_0)-R^{(1)}_{V,\pi}(p,z_0)\Big)\,\zeta^{(0)}_2 
\,\Big(z_0 +i\,p \Big)
\nonumber\\
&+&4\,g_A^2\,\lambda^2
\Big(\bar \kappa_{V,\pi}^{(0)}(z_0) -\bar \kappa^{(1)}_{J,\pi}(z_0) \Big)
\Big(R^{(2)}_{J,\pi }(p,z_0)-R^{(1)}_{V,\pi}(p,z_0)\Big)
\nonumber\\
&+&2\,g_A^4\,\lambda^4 \Big(R^{(2)}_{J,\pi \pi }(p,z_0)-2\,R^{(1)}_{V,\pi \pi }(p,z_0)\Big)
\nonumber\\
&+&2\,g_A^4\,\lambda^4 \, \Big(z_0-i\,p \Big)\,\frac{m}{2\,\pi}\,
\Big(R^{(1)}_{V,\pi }(p,z_0)-R^{(2)}_{J,\pi }(p,z_0)\Big)^2 \; .
\label{result7}
\end{eqnarray}
where we decomposed the residuum $w=\sum_{n=0}^\infty \,w^{(n)}$ into its chiral moments.
In (\ref{result7}) we encounter the reduced functions $R^{(1)}_{V,\pi \pi}$, 
$R^{(2)}_{J,\pi \pi}$
and $R^{(2)}_{V,\pi }$, $R^{(3)}_{J,\pi }$ defined according to (\ref{def-res}). Note that
all terms in (\ref{result7}) are regular at $p= -i\,z $ and $p=-i\,z_0 $ except the first one.
The contribution from the pion nucleon box diagram sits in $R_{\pi
\pi }(p)$ (see (\ref{Rpipi})). 
With (\ref{result7}) we arrive at the desired representation of the
scattering amplitude. The chiral correction terms of order one
renormalize the pole residuum $w$ and add further smooth remainder
terms. The pole residuum is:
\begin{eqnarray}
1-w &=&
\bar \kappa^{(1)}_{J, \pi }(z_0) -2\,\bar \kappa_{V, \pi }^{(0)}(z_0)
+2\,\zeta^{(0)}_2 \,z_0
\nonumber\\
&+&\bar \kappa^{(1)}_{J, \pi \pi }(z_0) -2\,\bar \kappa_{V, \pi \pi }^{(0)}(z_0)
-\Big( \bar \kappa_{V, \pi }^{(0)}(z_0)-\bar \kappa^{(1)}_{J, \pi }(z_0)\Big)^2 
\nonumber\\
&+& 4\,\Big(\bar \kappa_{V, \pi }^{(0)}(z_0)- \bar \kappa^{(1)}_{J,\pi }(z_0) -\zeta^{(0)}_2\,z_0 \Big) \,\zeta^{(0)}_2\,z_0
\nonumber\\
&-& 2\,\Big(\bar \kappa^{(1)}_{V, \pi }(z_0)- \bar \kappa^{(2)}_{J,\pi }(z_0) \Big) \,
\frac{m}{2\,\pi}\,\zeta^{(2)}_0\,m_\pi^2 +{\mathcal O}\left( Q^3\right) \; .
\label{}
\end{eqnarray}

We close this appendix with a more detailed analysis of the low energy behavior 
of (\ref{result7}).  A systematic $z_0/m_\pi$ expansion is worked out. 
In particular we show results for the low energy coefficients $b^{(0)}_{\mathrm{eff}}$ and 
$b^{(1)}_{\mathrm{eff}}$:
\begin{eqnarray}
b^{(0)}_{\mathrm{eff}} &=&   -\zeta^{(0)}_2
-\frac{m}{4\,\pi}\,g_A^2\,\lambda^2 \,\Big(1-4\,R^{(1)}_{V, \pi }(0,z_0)+2\,R^{(2)}_{J, \pi }(0,z_0) \Big)
\nonumber\\
&-&\zeta^{(0)}_2\, \Big( 
2\,\bar \kappa_{V,\pi}^{(0)}(z_0)-2\,\bar \kappa^{(1)}_{J,\pi}(z_0)-3\,\zeta^{(0)}_2 \,z_0\Big)
\nonumber\\
&-& g_A^2\,\lambda^2\,
\zeta^{(2)}_0\,m_\pi^2\,\frac{m^2}{2\,\pi^2}\,
\Big(R^{(3)}_{J,\pi }(0,z_0)-R^{(2)}_{V,\pi }(0,z_0)\Big)
\nonumber\\
&-&g_A^2\,\lambda^2
\Big(\bar \kappa_{V,\pi}^{(0)}(z_0) -\bar \kappa^{(1)}_{J,\pi}(z_0)-\zeta^{(0)}_2\,z_0 \Big)
\frac{m}{\pi}\,\Big(R^{(2)}_{J,\pi }(0,z_0)-R^{(1)}_{V,\pi}(0,z_0)\Big)
\nonumber\\
&-&g_A^4\,\lambda^4 \,\frac{m}{2\,\pi}\,\Big(R^{(2)}_{J,\pi \pi }(0,z_0)-2\,R^{(1)}_{V,\pi \pi }(0,z_0)\Big)
\nonumber\\
&-&g_A^4\,\lambda^4 \,\frac{m^2\,z_0}{4\,\pi^2}\,
\Big(R^{(1)}_{V,\pi }(0,z_0)-R^{(2)}_{J,\pi }(0,z_0)\Big)^2
\nonumber\\
&-&g_A^4\,\lambda^4\,\frac{m}{2\,\pi}\,R_{\pi \pi }(0)
+{\mathcal O} \left( Q^2\right)
\nonumber\\
b^{(1)}_{\mathrm{eff}}&=&
-g_A^2\,\lambda^2 \,\frac{m}{2\,\pi}\,(-i\,\partial_p )\,
\Big(R^{(2)}_{J, \pi }(0,z_0)-2\,R^{(1)}_{V, \pi }(0,z_0) \Big)
+\Big(\zeta^{(0)}_2\Big)^2\, 
\nonumber\\
&-& g_A^2\,\lambda^2\,\frac{m^2}{2\,\pi^2}\,
\zeta^{(2)}_0\,m_\pi^2\,(-i\,\partial_p )\,
\Big(R^{(3)}_{J,\pi }(0,z_0)-R^{(2)}_{V,\pi }(0,z_0)\Big)
\nonumber\\
&+&g_A^2\,\lambda^2\,\frac{m}{\pi}\,\zeta^{(0)}_2\,
\Big(R^{(2)}_{J,\pi }(0,z_0)-R^{(1)}_{V,\pi }(0,z_0)\Big)
\nonumber\\
&+&g_A^2\,\lambda^2\, 
\frac{m}{\pi}\,\zeta^{(0)}_2\,z_0\,(-i\,\partial_p) \,\Big(R^{(2)}_{J,\pi }(0,z_0)-R^{(1)}_{V,\pi}(0,z_0)\Big)
\nonumber\\
&-&g_A^2\,\lambda^2
\Big(\bar \kappa_{V,\pi}^{(0)}(z_0) -\bar \kappa^{(1)}_{J,\pi}(z_0)\Big)
\frac{m}{\pi}\,(-i\,\partial_p) \,\Big(R^{(2)}_{J,\pi }(0,z_0)-R^{(1)}_{V,\pi}(0,z_0)\Big)
\nonumber\\
&-&g_A^4\,\lambda^4 \,\frac{m}{2\,\pi}\,(-i\,\partial_p) \,\Big(R^{(2)}_{J,\pi \pi }(0,z_0)-2\,R^{(1)}_{V,\pi \pi }(0,z_0)\Big)
\nonumber\\
&-&g_A^4\,\lambda^4 \,\frac{m^2\,z_0}{4\,\pi^2}\,(-i\,\partial_p )\,
\Big(R^{(1)}_{V,\pi }(0,z_0)-R^{(2)}_{J,\pi }(0,z_0)\Big)^2
\nonumber\\
&+&g_A^4\,\lambda^4 \,\frac{m^2}{4\,\pi^2}
\Big(R^{(1)}_{V,\pi }(0,z_0)-R^{(2)}_{J,\pi }(0,z_0)\Big)^2
\nonumber\\
&-&g_A^4\,\lambda^4\,\frac{m}{2\,\pi}\,(-i\,\partial_p ) \,R_{\pi \pi }(0)
+{\mathcal O} \left( Q^2\right)
\label{}
\end{eqnarray}
introduced when discussing the structure of the scattering amplitude in (\ref{low-rep}). 
First the  coefficients $\bar \kappa^{(l)}_{J,\pi^k}(z)$ and 
$\bar \kappa^{(l)}_{V,\pi^k}(z)$ are systematically expanded in the small ratio $z/m_\pi$ 
with the leading moments
\begin{eqnarray}
\bar \kappa^{(0)}_{V,\pi}(z) &=& g_A^2\,\lambda^2\, \,
\frac{m\,m_\pi}{4\,\pi}
\left( 1+ \frac{z}{m_\pi}+\frac{4}{3}\,\frac{z^2}{m_\pi^2}+\frac{2\,z^3}{m_\pi^3}
+{\mathcal O}\left(\frac{z^4}{m^4_\pi} \right) \right)\; ,
\nonumber\\
\bar \kappa^{(1)}_{V, \pi } (z) &=& g_A^2\,\lambda^2\,\Bigg(
\frac{1}{2} +\frac{4\,z}{3\,m_\pi}+ \frac{3\,z^2}{m^2_\pi}+\frac{32\,z^3}{5\,m^3_\pi}+
{\mathcal O}\left(\frac{z^4}{m^4_\pi} \right) \Bigg) \; ,
\nonumber\\
\bar \kappa^{(0)}_{J,\pi} (z)&=&
g_A^2\,\lambda^2\,\frac{m^2\,m_\pi^2}{(4\,\pi)^2}\,\ln \left( \frac{\Lambda^2}{m^2_\pi}\right)
\left(1+{\mathcal O}\left(\frac{m_\pi }{\Lambda },\frac{z}{m_\pi}\right)\right)\; ,
\nonumber\\
\bar \kappa^{(1)}_{J,\pi}(z) &=&g_A^2\,\lambda^2\,\frac{m\,m_\pi}{4\,\pi}\left(
2+\frac{4\,z}{m_\pi}+\frac{8\,z^2}{m_\pi^2} +\frac{16\,z^3}{m_\pi^3}+{\mathcal
O}\left(\frac{z^4}{m_\pi^4}\right) \right)\; ,
\nonumber\\
\bar \kappa^{(2)}_{J, \pi } (z) &=&g_A^2\,\lambda^2\,\Bigg( 1 +\frac{4\,z}{m_\pi}+ 
\frac{12\,z^2}{m^2_\pi}+\frac{32\,z^3}{m^3_\pi}+ 
{\mathcal O}\left(\frac{z^4}{m^4_\pi} \right) \Bigg) \; ,
\nonumber\\
\bar \kappa^{(0)}_{V,\pi\pi}(z)&=&g_A^4\,\lambda^4\,\frac{m^2\,m_\pi^2}{(4\,\pi)^2}
\left(\ln \left(2 \right)+\frac{3}{2}\,\frac{z}{m_\pi}
+\left( \frac{3}{2}+\ln \left( 4\right) \right)
\frac{z^2}{m_\pi^2}
+{\mathcal O}\left(\frac{z^3}{m_\pi^3} \right)\right)\; ,
\nonumber\\
\bar \kappa^{(1)}_{J,\pi \pi}(z) 
&=&g_A^4\,\lambda^4\frac{m^2\,m_\pi^2}{(4\,\pi)^2}\Bigg(
1+2\,\ln\left(2\right)+
\left(\frac{14}{3}+\frac{16}{3}\,\ln \left(2\right) \right)
\frac{z}{m_\pi}
\nonumber\\
&&\;\;\;\;\;\;\;\;\;\;\;\;\;\;\;\;\;\;\;\; +\Big(14+12\,\ln
\left(2\right)\Big) \frac{z^2}{m_\pi^2} +{\mathcal
O}\left(\frac{z^3}{m_\pi^3}\right) \Bigg) \; .
\label{k-exp}
\end{eqnarray}
Next we provide the leading moments of the residual functions
\begin{eqnarray}
R^{(1)}_{V,\pi} (p,z) &=&\left( \frac{1}{2}+\frac{2\,z}{3\,m_\pi}+\frac{z^2}{m_\pi^2} \right)
\nonumber\\
&+&i\,\frac{p}{m_\pi}\left(\frac{2}{3}+\frac{z}{m_\pi}
+\frac{8\,z^2}{5\,m_\pi^2} \right) +{\mathcal
O}\Bigg(\frac{z^3}{m_\pi^3},\frac{p^2}{m_\pi^2}\Bigg)\; ,
\nonumber\\
R^{(2)}_{V,\pi} (p,z) &=&\frac{2 \,\pi}{m\,m_\pi}
\left(\frac{2}{3} +\frac{2\,z}{m_\pi}+\frac{24\,z^2}{5\,m_\pi^2} \right)
\nonumber\\
&+&i\,\frac{2 \,\pi}{m\,m_\pi}\,\frac{p}{m_\pi}
\left(1+\frac{16\,z}{5\,m_\pi}
+\frac{8\,z^2}{m_\pi^2} \right) +{\mathcal
O}\Bigg(\frac{z^3}{m_\pi^3},\frac{p^2}{m_\pi^2}\Bigg)\; ,
\nonumber\\
R^{(2)}_{J,\pi} (p,z) &=&\left(1+\frac{8\,z}{3\,m_\pi}+\frac{6\,z^2}{m_\pi^2}  \right)
\nonumber\\
&+&i\,\frac{p}{m_\pi}\left(\frac{4}{3}+\frac{4\,z}{m_\pi}
+\frac{48\,z^2}{5\,m_\pi^2} \right) +{\mathcal
O}\Bigg(\frac{z^3}{m_\pi^3},\frac{p^2}{m_\pi^2}\Bigg)\; ,
\nonumber\\
R^{(3)}_{J,\pi} (p,z) &=&\frac{2 \,\pi}{m\,m_\pi}
\left(\frac{4}{3} +\frac{6\,z}{m_\pi}+\frac{96\,z^2}{5\,m_\pi^2} \right)
\nonumber\\
&+&i\,\frac{2 \,\pi}{m\,m_\pi}\,\frac{p}{m_\pi}
\left(2+\frac{48\,z}{5\,m_\pi}
+\frac{32\,z^2}{m_\pi^2} \right) +{\mathcal
O}\Bigg(\frac{z^3}{m_\pi^3},\frac{p^2}{m_\pi^2}\Bigg)\; ,
\nonumber\\
R^{(1)}_{V,\pi \pi}(p,z) &=&\frac{m\,m_\pi}{4\,\pi}\,
\Bigg(\frac{3}{4}+\left(\frac{3}{4}+\ln (2) \right) \frac{z}{m_\pi}
+\frac{65}{24}\,\frac{z^2}{m^2_\pi} \Bigg)
\nonumber\\
&+&i\,\frac{m\,m_\pi}{4\,\pi}\left(
\frac{3}{4}+\ln (2) +\frac{65}{24}\,\frac{z}{m_\pi}
+\left(\frac{77}{24}+\frac{8}{3}\,\ln (2)\right)
\frac{z^2}{m^2_\pi} \right)\frac{ p}{m_\pi}
\nonumber\\
&+&{\mathcal
O}\left(\frac{z^3}{m_\pi^3} , \frac{p^2}{m_\pi^2}\right)\; ,
\nonumber\\
R^{(2)}_{J,\pi \pi }(p,z) &=&\frac{m \,m_\pi}{4\,\pi}
\Bigg(\frac{7}{6}+\frac{4}{3}\,\ln \left(2\right)
+\left(\frac{14}{3}+4\,\ln\left(2\right)\right)\frac{z}{m_\pi}
\nonumber\\
&&\;\;\;\;\;\;\;
+\left(\frac{269}{20}+\frac{48}{5}\,\ln \left(2\right)\right)
\frac{z^2}{m^2_\pi} \Bigg)
\nonumber\\
&+&i\,\frac{m \,m_\pi}{4\,\pi}
\Bigg(\frac{7}{3}+2\,\ln \left(2\right)
+\left( \frac{269}{30}+\frac{32}{5}\,\ln \left(2\right)
\right) \frac{z}{m_\pi}
\nonumber\\
&&\;\;\;\;\;\;\;
+\left(\frac{1531}{60}+16 \,\ln \left(2 \right)
\right) \frac{z^2}{m_\pi^2}
\Bigg)\,\frac{p}{m_\pi}
+{\mathcal O}\left(\frac{z^3}{m_\pi^3} ,\frac{p^2}{m_\pi^2}\right) \; .
\label{r-exp}
\end{eqnarray}
Finally note the low energy characteristics of the s-wave projected box function
\begin{eqnarray}
R_{\pi \pi } (p) &=& \frac{m\,m_\pi}{4\,\pi}\left( 
\frac{1}{4}+\frac{i\,p}{2\,m_\pi} + {\mathcal O}\left( \frac{p^2}{m_\pi^2}\right)
\right) \; .
\label{}
\end{eqnarray}

Applying the our results (\ref{k-exp}) and (\ref{r-exp}) we derive
\begin{eqnarray}
1-w
&=& \frac{m\,m_\pi}{4\,\pi}\,g_A^{2}\,\lambda^{2}
\left(2\,\frac{z_0}{m_\pi}+\frac{16}{3}\frac{z^2_0}{m^2_\pi}
+{\mathcal O}\left(\frac{z_0^3}{m_\pi^3}\right)\right)
+2\,\zeta^{(0)}_2 \,z_0
\nonumber\\
&+&\left(g_A^2\,\lambda^2\,\frac{m\,m_\pi}{4\,\pi}\right)^2
\Bigg(\left(-\frac{13}{3}+\frac{16}{3}\,\ln \left( 2\right)\right)
\frac{z_0}{m_\pi}
\nonumber\\
&&\;\;\;\;\;\;\;\;\;\;\;\;\;\;\;\; +\left( -\frac{34}{3}+8\,\ln
\left(2\right) \right)\frac{z_0^2}{m_\pi^2} +{\mathcal
O}\left(\frac{z_0^3}{m_\pi^3} \right)
\Bigg)
\nonumber\\
&-& 4\,g_A^2\,\lambda^2\,\frac{m\,m_\pi}{4\,\pi}
\left(1+\frac{3\,z_0}{m_\pi}  +{\mathcal O}\left(\frac{z_0^2}{m_\pi^2} \right)\right)
\,\zeta^{(0)}_2 \,z_0
\nonumber\\
&+&2\,g_A^2\,\lambda^2\,\frac{m\,m_\pi}{4\,\pi}\left( 1+\frac{16\,z_0}{3\,m_\pi} +\frac{18\,z^2_0}{m^2_\pi} 
+{\mathcal O}\left(\frac{z_0^3}{m_\pi^3} \right)
\right)\,\zeta^{(2)}_0\,m_\pi
\nonumber\\
&-& 4\,\zeta^{(0)}_2\,z_0  \,\zeta^{(0)}_2\,z_0
+{\mathcal O}\left( Q^3\right)\;,
\nonumber\\
b^{(0)}_{\mathrm{eff}} &=&-\zeta^{(0)}_2
-\frac{m}{4\,\pi}\,g_A^2\,\lambda^2
\left(1+\frac{8}{3}\,\frac{z_0}{m_\pi}+\frac{8\,z^2_0}{m^2_\pi} +{\mathcal O}\left(\frac{z_0^3}{m_\pi^3}\right)
\right)
\nonumber\\
&+&g_A^4\,\lambda^4\,\frac{m^2\,m_\pi}{(4\,\pi)^2}
\Bigg( \frac{13}{6}-\frac{8}{3}\,\ln (2)
+\left(\frac{20}{3}-4\,\ln \left(2\right) \right)\frac{z_0}{m_\pi}
\nonumber\\
&&\;\;\;\;\;\;\;\;\;\;\;\;\;\;\;\;\;\;\;
+\left( \frac{499}{15}-\frac{96}{5} \,\ln \left(2\right)\right)
\frac{z^2_0}{m^2_\pi}
+{\mathcal O}\left(\frac{z^3_0}{m_\pi^3} \right)
\Bigg)
\nonumber\\
&-& g_A^2\,\lambda^2 \,\frac{m\,m_\pi}{4\,\pi}\,\zeta_0^{(2)}
\left( \frac{8}{3} +\frac{16\,z_0}{m_\pi}+\frac{288\,z^2_0}{5\,m^2_\pi}
+{\mathcal O}\left(\frac{z^3_0}{m_\pi^3} \right)
\right)
\nonumber\\
&+& g_A^2\,\lambda^2 \,\frac{m\,m_\pi}{4\,\pi}\,\zeta_2^{(0)}
\left( 2 +\frac{8\,z_0}{m_\pi}+\frac{64\,z^2_0}{3\,m^2_\pi}
+{\mathcal O}\left(\frac{z^3_0}{m_\pi^3} \right)
\right)
\nonumber\\
&+&3\,z_0 \,\Big(\zeta_2^{(0)}\Big)^2 +{\mathcal O} \left( Q^2\right)\;,
\nonumber\\
b^{(1)}_{\mathrm{eff}} &=&
-\frac{z_0\,m}{\pi\,m^2_\pi}\,g_A^2\,\lambda^2
\left(1+\frac{16}{5}\,\frac{z_0}{m_\pi}
 +{\mathcal O}\left(\frac{z_0^2}{m_\pi^2}\right) \right)
\nonumber\\
&+&g_A^4\,\lambda^4\,\frac{m^2}{(4\,\pi)^2}
\Bigg( 1
+\left(\frac{547}{30}-\frac{64}{5}\,\ln \left(2\right) \right)\frac{z_0}{m_\pi}
\nonumber\\
&&\;\;\;\;\;\;\;\;\;\;\;\;\;\;\;\;\;\;\;
+\left( \frac{2741}{45}-\frac{64}{3} \,\ln \left(2\right)\right)
\frac{z^2_0}{m^2_\pi}
+{\mathcal O}\left(\frac{z^3_0}{m_\pi^3} \right)
\Bigg)
\nonumber\\
&-& g_A^2\,\lambda^2 \,\frac{m}{4\,\pi}\,\zeta_0^{(2)}
\left( 4 +\frac{128\,z_0}{5\,m_\pi}+\frac{96\,z^2_0}{m^2_\pi}
+{\mathcal O}\left(\frac{z^3_0}{m_\pi^3} \right)
\right)
\nonumber\\
&+& g_A^2\,\lambda^2 \,\frac{m}{4\,\pi}\,\zeta_2^{(0)}
\left( 2 +\frac{32\,z_0}{3\,m_\pi}+\frac{32\,z^2_0}{m^2_\pi}
+{\mathcal O}\left(\frac{z^3_0}{m_\pi^3} \right)
\right)
\nonumber\\
&+&\Big(\zeta_2^{(0)}\Big)^2 +{\mathcal O} \left( Q^2\right) \; .
\label{}
\end{eqnarray}
 
\end{appendix}

\end{fmffile}
\end{document}